\documentclass[12pt]{article}
\usepackage{amsmath}
\usepackage{amssymb}
\usepackage{tabularx} 
\begin{document}
\begin{center}
{\bf {\large{St$\ddot u$ckelberg-Modified Massive Abelian 3-Form Theory: Constraint Analysis, Conserved Charges and BRST Algebra}}} 

\vskip 3.4cm

{\sf  A. K. Rao$^{(a)}$, R. P. Malik$^{(a,b)}$}\\
$^{(a)}$ {\it Department of Physics, Institute of Science,}\\
{\it Banaras Hindu University, Varanasi - 221 005, (U.P.), India}\\

\vskip 0.1cm

$^{(b)}$ {\it DST Centre for Interdisciplinary Mathematical Sciences,}\\
{\it Institute of Science, Banaras Hindu University, Varanasi - 221 005, India}\\
{\small {\sf {e-mails: amit.akrao@gmail.com;  rpmalik1995@gmail.com}}}
\end{center}

\vskip 1.2 cm

\noindent
{\bf Abstract:}
For the St$\ddot u$ckelberg-modified massive Abelian 3-form theory in any arbitrary D-dimension 
of spacetime, we show that its classical gauge symmetry transformations are generated by the first-class constraints. 
We establish that the Noether conserved charge (corresponding to the local gauge symmetry transformations) is same as the
standard form of the generator for the underlying local gauge symmetry transformations (expressed in terms of the 
first-class constraints).
We promote these {\it classical} local, continuous and infinitesimal
 gauge symmetry transformations to their {\it quantum} counterparts Becchi-Rouet-Stora-Tyutin
(BRST) and anti-BRST symmetry transformations which are respected by the coupled (but equivalent) 
Lagrangian densities. We derive the conserved (anti-)BRST charges by exploiting the 
theoretical potential of Noether's theorem. However, these charges turn out to be non-nilpotent. 
Some of the highlights of our present investigation are (i) the derivation of the off-shell nilpotent versions of the  (anti-)BRST charges
from the standard  non-nilpotent Noether conserved (anti-)BRST charges, (ii) the appearance of the
operator forms of the first-class constraints at the {\it quantum} level through the physicality criteria w.r.t. the nilpotent versions of the
(anti-)BRST charges, and (iii) the deduction of the CF-type restrictions from the 
straightforward equality  of the coupled (anti-)BRST invariant Lagrangian densities as well as 
 from the requirement of the absolute anticommutativity of the off-shell nilpotent versions of the
 conserved (anti-)BRST charges.



\newpage
\section{Introduction}
The modern developments in the domain of (super)string theories have brought {\it together}, 
in an unprecedented manner, the top-class mathematicians and 
theoretical physicists of the highest calibre on a single platform which has led to the confluence of thoughts as well as
the pollination  of ideas that have been specifically responsible for  
many interesting developments in the realm of theoretical high energy physics. In this context, mention can be made of the 
AdS/CFT correspondence, gauge-gravity  duality, classification of the spacetime manifolds, 
higher-spin gauge theories, higher $p$-form ($p = 2, \, 3, \, 4,...$) gauge theories, etc. 
The central objective of our present endeavor is connected with the study of the St$\ddot u$ckelberg-modified 
massive {\it higher} $p$-form (i.e. $ p = 3)$ Abelian gauge theory within the framework of Becchi-Rouet-Stora-Tyutin (BRST) formalism
where the symmetry considerations (i.e. continuous symmetry transformations, conserved Noether 
 charges, algebra obeyed by the conserved charges, etc.) are given utmost importance.  Our present endeavor is
essential in view of the fact that it is connected with the BRST-quantization of a D-dimensional {\it modified} massive Abelian
3-form theory which is beyond the purview of the theoretical potential and power of the standard model of particle physics which is based on
the non-Abelian 1-form (i.e.  $p = 1 $) {\it interacting} gauge theory that incorporates into its folds the
matter as well as the  gauge fields {\it together} with a  well-defined  coupling on the basis of the principle of local gauge
invariance and ensuing interaction.

A deeper understanding of the  quantum field theory of the {\it massive} and {\it massless} higher $p$-form ($p = 2,\, 3,\,  4,...$) 
gauge theories has become essential and important because the higher $p$-form gauge fields appear in the 
excitations of the (super)strings which are one of the most promising candidates $(i)$ to provide a 
consistent theory of quantum gravity, $(ii)$ to address the question of the  {\it complete} unification of all the {\it four} fundamental 
interactions of nature, and $(iii)$ to go beyond
the realm of the standard model of particle physics which is based on the 
non-Abelian {\it interacting} 1-form gauge theory (see, e.g. [1-5] for details).
The {\it latter} issue is important because the experimental
evidence of the masses of the neutrinos has led to the conclusion that, even though very successful, the
standard model of particle physics is {\it not} a complete theory.
The purpose of our present investigation is to study some field-theoretic aspects of the 
St$\ddot u$ckelberg-modified  {\it massive} Abelian 3-form gauge theory in any arbitrary D-dimension of spacetime. 
In our earlier works (see, e.g. [6-8]), we have studied the  BRST quantization of the
$p$-form ($p = 1, \, 2,\, 3, ...$) gauge theories in $D= 2p$ dimensions of spacetime and established that such 
{\it massive} and {\it massless} gauge theories are the tractable field-theoretic examples 
of Hodge theory where the discrete and continuous symmetry transformations  (and corresponding conserved Noether charges)
provide the {\it physical} realizations of the de Rham cohomological operators (see, e.g. [9-12]) of differential geometry.
One of the highlights of such studies has been the observation that a set of fields, with the  {\it negative} kinetic terms, appear  
in the theory on the ground of symmetry considerations  {\it alone}. These ``exotic" fields are important in the context of 
the cyclic, bouncing and self-accelerated cosmological models of the Universe where they have been christened as the ``ghost" 
and/or ``phantom" fields (see, e.g. [13-15] and references therein). One of the possible candidates of the dark matter/dark energy is 
{\it also} intimately related with the existence of the  fields/particles with 
{\it negative} kinetic terms with {\it well-defined} mass and {\it without} mass (see, e.g. [16, 17] for details).

In a very recent work [18], we have exploited the theoretical beauty  of the augmented version of
superfield approach (AVSA) to BRST formalism to derive $(i)$ the complete set of (anti-)BRST symmetry transformations, 
and $(ii)$ the (anti-)BRST invariant Curci-Ferrari (CF) type restrictions [19] for the  
St$\ddot u$ckelberg-modified  {\it massive} Abelian 3-form gauge theory in any arbitrary D-dimension of spacetime.
The existence of the {\it latter} (i.e., CF-type restrictions) is one of the key signatures of a BRST-quantized theory as it is
connected with the idea of the geometrical objects called gerbes (see, e.g., [20, 21]). 
We would like to lay emphasis on the fact that for the BRST-quantized theory, the 
existence of the Curci-Ferrari type restrictions is as {\it fundamental} as the existence of the first-class constraints
in the definition of a {\it classical} gauge theory. 
The central theme of our present endeavor 
is to perform a thorough constraint analysis of the D-dimensional 
St$\ddot u$ckelberg-modified Abelian 3-form {\it massive}  gauge theory (cf. Sec. 3) and show that its first-class constraints generate the continuous 
and infinitesimal {\it classical} gauge symmetry transformations that can be easily exploited for the {\it quantum}
versions of the off-shell nilpotent (anti-)BRST symmetry transformations. 
It is pertinent to point out, at this stage, that
in our earlier works (see, e.g. [6-8,18]), the constraint analysis of the {\it modified} massive Abelian 3-form theory
has {\it not} been performed.
In our present investigation, 
we derive the conserved Noether currents and corresponding conserved charges for our  
system and show that the underlying nilpotent (anti-)BRST symmetry transformations 
and the  infinitesimal ghost-scale symmetry transformations lead to the derivation of the 
conserved (anti-)BRST and ghost charges.  
It turns out that the standard Noether conserved (anti-)BRST charges [$Q_{(a)b}$] are {\it not} nilpotent of 
order two (i.e. $[Q_{(a)b}]^2 \neq 0$). A systematic method has been proposed   in [22] to obtain the 
off-shell nilpotent versions of the (anti-)BRST charges. These {\it latter} nilpotent charges and ghost charge 
obey the  {\it standard} BRST algebra (cf. Appendix C).
The physicality criteria w.r.t. the above nilpotent versions of the
(anti-)BRST charges are crucial as far as the appearance of the first-class constraints at the quantum level is concerned.
One of the highlights of our present investigation is the observation that the 
proof of the absolute anticommutativity property of the nilpotent (anti-)BRST charges (cf. Appendix B) leads to the 
derivation of the (anti-)BRST invariant CF-type restrictions. The {\it latter} are also derived by the requirement of the 
straightforward {\it equality} of the {\it coupled} (anti-)BRST invariant Lagrangian densities (cf. Appendix A). It is worthwhile to point out that, in our present endeavor,  we have confined ourselves to the 
theoretical techniques of the BRST formalism because our system is simple and endowed with the first-class constraint {\it only}.
Mathematically more sophisticated and more general BV formalism has {\it not} been adopted in our present endeavor
because of the simplicity of our system.

Even though, our main results have been mentioned in a somewhat haphazard manner throughout our earlier paragraphs, we would like to assimilate systematically 
our key results in a single paragraph for the readers' convenience. We have been able to  show the existence of the first-class constraints at the 
{\it classical} level in the 
Lagrangian formulation (cf. Sec. 3) as well as at the {\it quantum} level within the framework of BRST formalism (cf. Sec. 6). In our present endeavor,
we have computed all the conserved charges corresponding to the continuous symmetry transformations of our theory and explicitly
shown that the  {\it Noether} conserved (anti-)BRST charges are {\it not} nilpotent of order two (cf. Sec. 5). The
off-shell nilpotent versions  of these charges have been derived 
and  their usefulness have been shown in the context of the physicality criteria (cf. Sec. 6). Two of the
sacrosanct properties of the BRST formalism are the nilpotency property and absolute anticommutativity property that are 
associated with the
(anti-)BRST symmetries. We have devoted an Appendix (cf. Appendix E) of our present endeavor to establish the connections between {\it these} properties
and the CF-type restrictions at the level of the
(anti-)BRST transformed fields (that are present in our theory)
. Two appendices (i.e. Appendices A and B) deal with the derivations of these CF-type restrictions
from different theoretical angles. These derivations are completely different from such derivations in our previous work [18]. We have
derived the standard BRST algebra amongst the {\it appropriate} conserved charges of our theory.

Our present endeavor is essential and important on the following counts. First, we have already derived the coupled (but equivalent)
Lagrangian densities for the St$\ddot u$ckelberg-modified  {\it massive} Abelian 3-form theory that respect the  
off-shell nilpotent and absolutely anticommuting (anti-)BRST symmetry transformations. 
The {\it latter} and associated CF-type restrictions  have been deduced by exploiting the theoretical strength of the AVSA to BRST formalism (see, e.g. [18]).
However, we have {\it not} derived the expressions for the conserved charges corresponding to the underlying continuous 
symmetry transformations of our theory. In our present investigation, we accomplish this goal and deduce the explicit expressions for the 
(anti-)BRST and ghost charges. Second, we have been able to show the existence of the CF-type restrictions
(on our theory) by symmetry considerations {\it alone} in our earlier work [18]. In our present endeavor, we show the presence of the above 
CF-type restrictions by proving the absolute anticommutativity of the nilpotent (anti-)BRST charges
and straightforward equality of the {\it coupled} Lagrangian densities. 
Third, we have established that the operator forms of the 
first-class constraints annihilate the physical quantum states of our theory through the physicality criteria w.r.t. the
nilpotent versions of the (anti-)BRST charges (cf. Sec. 6). This observation is one of the key results of our present investigation.
Finally, our present investigation is a modest step in the direction to prove that the 6D {\it massive} Abelian 3-form
theory is a field-theoretic example of Hodge theory where the fields with {\it negative} kinetic terms are expected to appear  on the 
basis of symmetry considerations  {\it alone}. Such ``exotic" fields are one of the possible candidates for 
dark matter/dark energy [16, 17] and these fields
(popularly christened as  the ``ghost" and/or ``phantom" fields) 
are found to play a crucial role in the
cyclic, bouncing and self-accelerated cosmological models of the Universe [13-15].

The theoretical contents of our present investigation are organized as follows. First of all,
in Sec. 2, we discuss the infinitesimal and local gauge symmetry transformations for the St$\ddot u$ckelberg-modified Lagrangian density
and (anti-)BRST symmetries for the coupled (but equivalent) Lagrangian density. In Sec.  3, we write down the
standard generator for the infinitesimal and local {\it classical} gauge symmetry transformations in terms of the first-class constraints
and establish a connection between {\it this} generator and the Noether conserved charge corresponding to the 
infinitesimal local gauge symmetry transformations (at the {\it classical} level). Our Sec. 4 deals with
 the derivation of the Euler-Lagrange (EL) equations of motion (EoM) from the coupled (but equivalent) 
Lagrangian densities where we comment on the existence of the CF-type restrictions [cf. Eq. (29) below]. 
The subject matter of  our Sec. 5 is connected with the derivations of the 
Noether conserved currents, corresponding conserved charges and a few comments on the 
non-nilpotency of the standard Noether (anti-) BRST charges. Our Sec. 6 is devoted to the discussion on the physicality criteria 
w.r.t. the off-shell nilpotent versions of the (anti-)BRST charges where we show the existence of the first-class
constraints of the theory at the {\it quantum} level in their operator 
form. Finally, in Sec. 7, we summarize 
our key results and point out a few future directions of further investigation(s) in the light of our present investigation.

In our Appendices A and B, we demonstrate the existence of a set of (anti-)BRST invariant CF-type restrictions by (i) requiring the 
straightforward equality  of the coupled Lagrangian densities, and (ii) demanding the absolute anticommutativity of the
off-shell nilpotent versions of the conserved (anti-)BRST charges. Our Appendix C is devoted to a brief description of the
standard BRST algebra which exists amongst the off-shell nilpotent versions of the (anti-)BRST charges and the 
conserved ghost charge of our BRST-quantized theory.
In Appendix D, we provide a glossary of the key properties and nature of {\it all} the tower of fields
that are present in our BRST-quantized theory (cf. Sec. 2). Our Appendix E deals with the proof that the 
sanctity and usefulness of the (anti-)BRST invariant CF-type restrictions (that are present on our theory) 
are {\it true} even at the level of (anti-)BRST transformed fields. \\

\noindent
{\it {General Convention and Notations}}:
We adopt the convention of the {\it left} derivative w.r.t. all the {\it fermionic} fields of our theory in the 
computations of the canonical conjugate momenta, equations of motion,  Noether's conserved currents/charges, etc. The background {\it flat}
D-dimensional Minkowskian spacetime manifold is endowed with a metric tensor  ${\eta}_{\mu \nu } =$ diag (+1, -1, -1,...) 
so that the dot product between two non-null vectors $P_\mu$ and $Q_\mu$ is defined as: 
$P \cdot Q  = {\eta}_{\mu \nu } \, P^\mu\, Q^\nu \equiv P_0\, Q_0 - P_i\, Q_i$ 
where the Latin indices $i, j, k, ... = 1, 2,..., D-1$ stand for the space directions {\it only}
and the Greek indices $\mu,\, \nu, \, \lambda, ... = 0, 1, 2, ..., D-1$
correspond to the time and space directions {\it together}. Einstein's summation convention has been taken 
into account in the whole body of the text. We denote the (anti-)BRST symmetry transformations 
by the symbols $s_{(a)b}$ and corresponding conserved charges have been represented by the notations $Q_{(a)b}$. 
Due to their fermionic nature, the (anti-)BRST symmetry transformation operators $s_{(a)b}$ commute with the bosonic
fields of our theory and they anticommute with their counterparts fermionic fields.


\section{Preliminaries: Local and Infinitesimal Gauge and Off-shell Nilpotent (Anti-)BRST Symmetries}

Within the framework of Lagrangian formulation for the D-dimensional {\it massive}
Abelian 3-form theory (see e.g. [18] for details), we discuss the infinitesimal, continuous and local
gauge symmetry transformations for the following St$\ddot u$ckelberg-modified Lagrangian density [${\cal L}_S^{(A)}$]
\begin{eqnarray}
{\cal L}_{S}^{(A)} = \frac {1}{24}\,H^{\mu \nu \lambda \zeta }\,H_{\mu \nu \lambda \zeta} - \frac {m^2}{6} 
\, A^{\mu\nu\lambda }\,A_{\mu\nu\lambda } \pm \frac{m}{3}\,A^{\mu\nu\lambda} \, \Sigma_{\mu \nu \lambda }
- \frac{1}{6}\, \Sigma^{\mu \nu \lambda }\, \Sigma_{\mu \nu \lambda },
\end{eqnarray}
where the totally antisymmetric field-strength (curvature) tensor $H_{\mu \nu \lambda \zeta}$ has been derived from the 4-form
$H^{(4)} = d\, A^{(3)}$ [18] and superscript  $(A)$ denotes the {\it modified} Lagrangian density  [${\cal L}_S^{(A)}$] for the {\it massive} Abelian gauge
field $A_{\mu\nu\lambda}$.
The Abelian 3-form  $A^{(3)} = \frac{1}{3 !}\,A_{\mu\nu\lambda}\, (d\,x^\mu \wedge d\,x^\nu \wedge d\, x^\lambda)$
defines the {\it totally} antisymmetric tensor gauge field $A_{\mu\nu\lambda}$ and $d=  \partial_\mu \, d\, x^\mu$ 
(with $d^2 =  \frac{1}{2 !}\, (\partial_\mu\, \partial_\nu - \partial_\nu\, \partial_\mu ) \, (d\,x^\mu \wedge d\,x^\nu) = 0)$
is the exterior derivative of differential geometry [9-12]. The explicit form of $H_{\mu \nu \lambda \zeta}$, in terms 
of the 3-form Abelian basic  gauge field $A_{\mu\nu\lambda}$, is as follows:
\begin{eqnarray}
H_{\mu \nu \lambda \zeta} = \partial_\mu\, A_{\nu\lambda\zeta} - \partial_\nu\,
 A_{\lambda \zeta \mu } + \partial_\lambda\, A_{\zeta \mu \nu }  - \partial_\zeta \,A_{\mu \nu \lambda }.
\end{eqnarray}
In the above equation (1), the symbol $m$ is the rest mass of the 3-form field  $A_{\mu\nu\lambda}$ and the notation $\Sigma_{\mu\nu\lambda}$
stands for the totally antisymmetric field-strength tensor
\begin{eqnarray}
\Sigma_{\mu \nu \lambda } = \partial_\mu\, \Phi_{\nu\lambda} + \partial_\nu\, \Phi_{\lambda \mu } + 
\partial_\lambda\, \Phi_{\mu\nu},
\end{eqnarray}
for the St$\ddot u$ckelberg 2-form field $\Phi_{\mu\nu}$ defined through the 2-form: 
$\Phi^{(2)} = \frac{1}{2 !}\, \Phi_{\mu\nu}\, (d\,x^\mu \wedge d\,x^\nu) $. It is self-evident that the 3-form 
$\Sigma^{(3)} = d\, \Phi^{(2)}$ defines the field-strength tensor (3) with the help of the exterior derivative
$d$ and the St$\ddot u$ckelberg 2-form field $\Phi_{\mu\nu}$ 
because $\Sigma^{(3)} = \frac{1}{3 !}\, \Sigma_{\mu\nu\lambda}\,  (d\,x^\mu \wedge d\,x^\nu \wedge d\, x^\lambda)$.
The infinitesimal, continuous and local gauge symmetry transformations ($\delta_g$) for the fields $A_{\mu\nu\lambda}$, $\Phi_{\mu\nu}$,  
$\Sigma_{\mu \nu \lambda }$ and $H_{\mu \nu \lambda \zeta}$ are as follows [18]
\begin{eqnarray}
&&\delta_g\, A_{\mu\nu\lambda} = \partial_\mu\, \Lambda_{\nu\lambda} + \partial_\nu\, \Lambda_{\lambda\mu} 
+ \partial_{\lambda} \Lambda_{\mu\nu}, \nonumber\\
&&\delta_g\, \Sigma_{\mu\nu\lambda} = \pm \, m\,(\partial_\mu\, \Lambda_{\nu\lambda} + \partial_\nu\, \Lambda_{\lambda\mu} 
+ \partial_{\lambda} \Lambda_{\mu\nu}),\nonumber\\
&&\delta_g\, \Phi_{\mu\nu} = \pm \, m\, \Lambda_{\mu\nu} - (\partial_\mu\,\Lambda_\nu - \partial_\nu\,\Lambda_\mu), \qquad
\delta_g\, H_{\mu \nu \lambda \zeta } = 0,
\end{eqnarray}
where the antisymmetric $[\Lambda_{\mu\nu} (x) = -\, \Lambda_{\nu\mu} (x)]$ tensor and Lorentz vector $\Lambda_\mu (x)$
are the local gauge symmetry transformation parameters. 
It is straightforward to note that $\delta_g\, {\cal L}_S^{(A)} = 0$. The above transformations are {\it generated} 
by the first-class constraints that exist for the Lagrangian density 
${\cal L}_S^{(A)}$ (cf. Sec. 3 below for details).

The St$\ddot u$ckelberg-modified Lagrangian density (1) has been generalized to the coupled (but equivalent)
(anti-)BRST invariant Lagrangian densities in our earlier work [18]. The {\it perfectly} BRST invariant Lagrangian 
density $({\cal L}_B)$, incorporating the gauge-fixing as well as the Faddeev-Popov (FP) ghost terms,  is as follows 
\begin{eqnarray}
{\cal L}_{B} &=&  {\cal L}_{S}^{(A)} + (\partial_\mu A^{\mu\nu\lambda}) B_{\nu\lambda}  - \frac{1}{2} B_{\mu\nu}\, B^{\mu\nu}
+ \frac{1}{2}\, B^{\mu\nu}\, \Big[ \partial_\mu \, \phi_\nu - \partial_\nu\, \phi_\mu \mp \, m\, \Phi_{\mu\nu}  \Big] \nonumber\\
&-&\, (\partial_\mu\, \Phi^{\mu\nu})\, B_\nu - \frac{1}{2}\, B^\mu\, B_\mu + \frac{1}{2}\, B^{\mu}\, \Big[\pm \, m\phi_\mu - \partial_\mu\, \phi\Big] 
 + \frac{m^2}{2}\, \bar C_{\mu\nu}\, C^{\mu\nu} \nonumber\\
 &+& (\partial_\mu \bar C_{\nu\lambda} + \partial_\nu \bar C_{\lambda\mu} + 
\partial_\lambda \bar C_{\mu\nu}) (\partial^\mu C^{\nu\lambda}) 
\pm m\, (\partial_\mu\, \bar C^{\mu\nu})\, C_\nu \pm \, m\, \bar C^\nu\, (\partial^\mu\, C_{\mu\nu}) \nonumber\\
&+& (\partial_\mu\, \bar C_\nu - \partial_\nu\, \bar C_\mu)\, (\partial^\mu \ C^\nu)  
- \frac{1}{2}\, \Big[\pm m\, \bar \beta^\mu 
- \partial^\mu\, \bar\beta \Big]\, \Big[\pm m\,  \beta_\mu - \partial_\mu\, \beta \Big] \nonumber\\
&-&\,  (\partial_\mu \, \bar\beta_\nu - \partial_\nu\, \bar\beta_\mu)\, (\partial^\mu\,\beta^\nu) - \partial_\mu \bar C_2 \partial^\mu  C_2
 - \, m^2\, \bar C_2\, C_2 + [(\partial \cdot \bar\beta) \mp m\, \bar \beta]\, B \nonumber\\
 &-& [(\partial \cdot \phi) \mp m\, \phi]\, B_1 - [(\partial \cdot \beta) \mp m\, \beta]\, B_2
+ \Big[\partial_\nu \bar C^{\nu\mu}  + \partial^\mu\, \bar C_1 \mp \frac {m}{2}\, \bar C^\mu\Big]\, f_\mu \nonumber\\
&-&  2\, F^\mu f_\mu - 2  F\, f - \Big[\partial_\nu  C^{\nu\mu}  + \partial^\mu  C_1 \mp \frac {m}{2}  C^\mu\Big] F_\mu
 +  \Big[\frac{1}{2} (\partial \cdot C) \mp m C_1\Big]\, F \nonumber\\
  &-& \Big[\frac{1}{2}\, (\partial \cdot \bar C) \mp m\, \bar C_1\Big]\, f-\, B\, B_2 -\, \frac{1}{2}\, B_1^2,
\end{eqnarray}
where the antisymmetric $(B_{\mu\nu} = -\, B_{\nu\mu})$ tensor $(B_{\mu\nu})$, bosonic and fermionic sets of the  Lorentz vector $(B_\mu, \,  F_\mu,  \, f_\mu)$ and scalar $(B, \, B_1,\, B_2,\, F, \,  f)$ {\it auxiliary} fields, respectively,  
appear in the above Lagrangian density along with the {\it fermionic} (anti-)ghost fields $(\bar C_2) C_2, \, 
(\bar C_{\mu\nu}) C_{\mu\nu}, \, 
(\bar C_{\mu}) C_{\mu}, (\bar C_1) C_1 $ as well as the {\it bosonic} (anti-)ghost fields
$(\bar\beta_\mu)\, \beta_\mu$ and $ (\bar \beta)\, \beta$. 
In addition to the antisymmetric ($\Phi_{\mu\nu} = -\, \Phi_{\nu\mu}$) St$\ddot u$ckelberg {\it bosonic} field $\Phi_{\mu\nu}$, 
we also have the vector {\it bosonic} field $\phi_\mu$ and scalar {\it bosonic} field $\phi$.
We have a set of special kinds of auxiliary fields $(B_1, \, B, \, B_2)$ which carry the ghost numbers ($0, \, +2, \, -2$), respectively. 
Similarly, we have {\it perfectly} anti-BRST invariant Lagrangian density $({\cal L}_{\bar B})$ as
\begin{eqnarray}
{\cal L}_{\bar B} &=& {\cal L}_{S}^{(A)} - (\partial_\mu A^{\mu\nu\lambda}) \bar B_{\nu\lambda} 
 - \frac{1}{2} \bar B_{\mu\nu}\, \bar B^{\mu\nu}
+ \frac{1}{2}\, \bar B^{\mu\nu}\, \Big[ \partial_\mu \, \phi_\nu - \partial_\mu\, \phi_\nu \pm \, m\, \Phi_{\mu\nu}  \Big] \nonumber\\
&+&(\partial_\mu\, \Phi^{\mu\nu})\, \bar B_\nu - \frac{1}{2}\,  \bar B^\mu\, \bar B_\mu +
  \frac{1}{2}\,  \bar B^{\mu}\, \Big[\pm \, m\phi_\mu - \partial_\mu\, \phi \Big] 
+ \frac{m^2}{2}\, \bar C_{\mu\nu}\, C^{\mu\nu} \nonumber\\
&+& (\partial_\mu \bar C_{\nu\lambda} + \partial_\nu \bar C_{\lambda\mu} + 
\partial_\lambda \bar C_{\mu\nu}) (\partial^\mu C^{\nu\lambda}) 
\pm m\, (\partial_\mu\, \bar C^{\mu\nu})\, C_\nu
\pm \, m\, \bar C^\nu\, (\partial^\mu\, C_{\mu\nu}) \nonumber\\
&+& (\partial_\mu\, \bar C_\nu - \partial_\nu\, \bar C_\mu)\, (\partial^\mu \ C^\nu) 
 - \frac{1}{2}\, \Big[\pm m\, \bar \beta^\mu 
- \partial^\mu\, \bar\beta \Big]\, \Big[\pm m\,  \beta_\mu - \partial_\mu\, \beta \Big] \nonumber\\
&-&\, (\partial_\mu \, \bar\beta_\nu - \partial_\nu\, \bar\beta_\mu)\, (\partial^\mu\,\beta^\nu) - \partial_\mu \bar C_2 \partial^\mu  C_2
 - \, m^2\, \bar C_2\, C_2 + [(\partial \cdot \bar\beta) \mp m\, \bar \beta]\, B \nonumber\\
 &-& [(\partial \cdot \phi) \mp m\, \phi]\, B_1 -  [(\partial \cdot \beta) \mp m\, \beta]\, B_2
+ \Big[\partial_\nu C^{\nu\mu}  - \partial^\mu\,  C_1 \mp \frac {m}{2}\,  C^\mu\Big]\, \bar f_\mu \nonumber\\
&+&  \, 2 \bar F^\mu \bar f_\mu + 2  \bar F\, \bar f - \Big[ \partial_\nu  \bar C^{\nu\mu}  - \partial^\mu  
\bar C_1 \mp \frac {m}{2}  \bar C^\mu\Big] \bar F_\mu
 + \Big[ \frac{1}{2} (\partial \cdot \bar C) \pm m\bar C_1 \Big] \bar F \nonumber\\
 &-& \Big[\frac{1}{2}\, (\partial \cdot C) \pm m\, C_1 \Big]\, \bar f  -\, B\, B_2 -\, \frac{1}{2}\, B_1^2,
\end{eqnarray}
which {\it also} incorporates the gauge-fixing and FP-ghost terms. The Lagrangian density ${\cal L}_{\bar B}$ is 
characterized by the two {\it bosonic} as well as the four {\it fermionic} auxiliary fields 
$(\bar B_{\mu\nu}, \, \bar B_\mu, \, \bar F_\mu, \, \bar f_\mu, \, \bar F, \, \bar f)$, respectively, 
and the {\it rest} of the symbols are {\it same} as in the Lagrangian density (5).
For the sake of readers’ convenience, we have provided a glossary 
of all the fields that are present in (5) and (6), various aspects that describe
their true nature, their ghost numbers, etc., in a tabulated form in our Appendix
D.

The generalizations of the {\it classical} gauge symmetry transformations (4) at the {\it quantum}
level has been obtained in our earlier work [18] where the 
AVSA to BRST formalism has been exploited in the derivations of the {\it proper} (anti-)BRST transformations.
It is a bit involved but straightforward to note that the following off-shell nilpotent $(s_b^2 = 0)$
BRST transformations ($s_b$)
\begin{eqnarray}
&&s_b A_{\mu\nu\lambda} = \partial_\mu C_{\nu\lambda} + \partial_\nu C_{\lambda\mu} + \partial_\lambda C_{\mu\nu}, \quad
s_b C_{\mu\nu} = \partial_\mu \beta_\nu - \partial_\nu \beta_\mu, \nonumber\\
&&s_b \bar B_{\mu\nu} =  (\partial_\mu\, f_\nu - \partial_\nu\, f_\mu), 
 \quad s_b \bar C_{\mu\nu} = B_{\mu\nu},\nonumber\\
&&s_b\, \Phi_{\mu\nu} = \pm\, m\, C_{\mu\nu} - \, (\partial_\mu\, C_\nu - \partial_\nu\, C_\mu), 
\qquad \quad s_b \bar F_\mu = - \partial_\mu  B,\nonumber\\
&&s_b\, C_\mu = \pm\, m\, \beta_\mu - \partial_\mu\, \beta, \qquad  s_b \phi_\mu = f_\mu, 
\qquad  s_b \beta_\mu = \partial_\mu C_2, \nonumber\\
&&s_b\, \bar B_\mu = \pm\, m\, f_\mu - \partial_\mu f,  
\qquad s_b \bar f_\mu = -\, \partial_\mu B_1,  \quad s_b \bar \beta_\mu = F_\mu,\nonumber\\
&&s_b\, \bar C_\mu = B_\mu, \qquad  s_b \bar C_2 = B_2, \qquad  s_b \bar C_1 = -\, B_1,\nonumber\\
&&s_b\, \phi = f, \qquad   s_b\, \beta = \pm \, m\, C_2, \qquad s_b\, \bar\beta = F,\nonumber\\
&&s_b\, \bar F = \mp\, m\, B, \qquad s_b\, \bar f = \mp\, m\, B_1, \qquad  s_b \, C_1 = - B, \nonumber\\
&&s_b\, [H_{\mu\nu\lambda\zeta}, \, B_{\mu\nu}, \, B_\mu, \, f_\mu, \, F_\mu, \, F, \, f,\,   B, \, B_1,\, B_2, \, C_2] = 0,
\end{eqnarray}
are the {\it symmetry} transformations for the action integral $S_1 = \int d^D x \,{\cal L}_{ B} $
because the Lagrangian density (${\cal L}_{B} $) transforms, under $s_b$, to the {\it total} spacetime derivative as:
\begin{eqnarray}
s_b\, {\cal L}_{B} &=& \partial_\mu\, \Big[ (\partial^\mu\, C^{\nu\lambda} + \partial^\nu\, C^{\lambda\mu}
+ \partial^\lambda\, C^{\mu\nu}) \, B_{\nu\lambda}  + B^{\mu\nu}\, f_\nu - B_2\, \partial^\mu\,C_2  \nonumber\\
&-& B_1\, f^\mu + B\, F^\mu - (\partial^\mu\, \beta^\nu - \partial^\nu\, \beta^\mu)\, F_\nu 
+ \frac{1}{2}\, (\pm\, m\, \beta^\mu - \partial^\mu\, \beta)\, F \nonumber\\
 &+& (\partial^\mu\,  C^\nu - \partial^\nu\,  C^\mu)\, B_\nu \pm \, m\, B^{\mu\nu}\, C_\nu  
-\, \frac{1}{2} \,  B^\mu\,  f \nonumber\\
&\mp& m\,\bar C^{\mu\nu}\, \big(\pm m\,  \beta_\nu - \partial_\nu\, \beta \big)
\mp\, m\, (\partial^\mu\, \beta^\nu - \partial^\nu\,\beta^\mu)\,\bar C_\nu  \Big].
\end{eqnarray}
The above observation establishes the {\it invariance} of the action integral when we exploit the theoretical 
strength of Gauss's divergence theorem. We have also the generalization of the classical gauge 
symmetry transformations (4) to their {\it quantum} counterparts that are nothing but the off-shell nilpotent $(s_{ab}^2 = 0)$
anti-BRST symmetry transformations ($s_{ab}$) as
\begin{eqnarray}
&&s_{ab} A_{\mu\nu\lambda} = \partial_\mu \bar C_{\nu\lambda} + \partial_\nu \bar C_{\lambda\mu} + \partial_\lambda \bar C_{\mu\nu}, \quad
s_{ab} \bar C_{\mu\nu} = \partial_\mu \bar \beta_\nu - \partial_\nu \bar\beta_\mu,\nonumber\\
&&s_{ab} B_{\mu\nu} = (\partial_\mu  \bar f_\nu - \partial_\nu  \bar f_\mu),
 \quad s_{ab}  C_{\mu\nu} = \bar B_{\mu\nu},\nonumber\\
&&s_{ab}\, \Phi_{\mu\nu} = \pm\, m\, \bar C_{\mu\nu} - \, (\partial_\mu\, \bar C_\nu - \partial_\nu\, \bar C_\mu), \quad s_{ab}\,  F_\mu = 
-\, \partial_\mu\, B_2,\nonumber\\
&&s_{ab}\, \bar C_\mu = \pm\, m\, \bar\beta_\mu
 - \partial_\mu\, \bar\beta, \qquad s_{ab} \phi_\mu = \bar f_\mu, \quad  s_{ab}  \bar \beta_\mu = \partial_\mu\, \bar C_2, \nonumber\\
&&s_{ab}\, B_\mu = \pm \, m\, \bar f_\mu - \, \partial_\mu \, \bar f, \quad  
 \quad s_{ab}  \beta_\mu = \bar F_\mu,  \quad s_{ab}  f_\mu =  \partial_\mu B_1,\nonumber\\
&&s_{ab}\, C_\mu = \bar B_\mu, \qquad  s_{ab}\, f = \pm\, m\, B_1,  \qquad s_{ab}\,\phi = \bar f,\nonumber\\
&&s_{ab}\,\bar\beta = \pm\, m\, \bar C_2, \qquad 
s_{ab}\,\beta = \bar F, \qquad s_{ab}\, F = \mp\, m\, B_2 \nonumber\\
&&s_{ab} \bar C_1 = - B_2, \qquad  s_{ab} C_1 =  B_1, \qquad s_{ab} C_2 =  B,\nonumber\\
&&s_{ab}\, [H_{\mu\nu\lambda\zeta}, \, \bar B_{\mu\nu}, \, \bar B_\mu, \;  \bar F_\mu,\,\bar f_\mu, \, 
 \bar F,\,\bar f, \,  B, \, B_1, \, B_2, \, \bar C_2] = 0,
\end{eqnarray}
which transform the {\it perfectly} anti-BRST invariant Lagrangian density 
$({\cal L}_{\bar B})$ to the following {\it total} spacetime derivative, namely; 
\begin{eqnarray}
s_{ab}\, {\cal L}_{\bar B} &=& \partial_\mu\, \Big[\bar B^{\mu\nu}\, \bar f_\nu  - (\partial^\mu\, \bar C^{\nu\lambda} 
+ \partial^\nu\, \bar C^{\lambda\mu}+ \partial^\lambda\, \bar C^{\mu\nu}) \, \bar B_{\nu\lambda}  + B \, \partial^\mu\, \bar C_2 \nonumber\\
&-& B_2 \, \bar F^\mu - B_1\, \bar f^\mu - (\partial^\mu\, \bar \beta^\nu - \partial^\nu\, \bar \beta^\mu)\, \bar F_\nu
+ \frac{1}{2}\, (\pm\, m\, \bar \beta^\mu - \partial^\mu\, \bar\beta)\, \bar F   \nonumber\\
 &-& (\partial^\mu\, \bar C^\nu -\partial^\nu\, \bar C^\mu)\, \bar B_\nu \mp \, m\, \bar B^{\mu\nu}\, \bar C_\nu  
-\, \frac{1}{2} \, \bar B^\mu\, \bar f \nonumber\\ 
&\pm& m\, C^{\mu\nu}\, (\pm m\, \bar \beta_\nu - \partial_\nu\, \bar \beta)
\pm\, m\, (\partial^\mu\, \bar\beta^\nu - \partial^\nu\,\bar\beta^\mu)\,C_\nu\Big].
\end{eqnarray}
This observation, once again, establishes the fact that the action integral  $S_2 = \int d^D x \,{\cal L}_{\bar B} $ remains invariant 
under the anti-BRST symmetry transformations $(s_{ab})$ which have been listed in (9). This happens due to the
application of Gauss's divergence theorem which implies that all the {\it physical} fields vanish off as $x \longrightarrow \pm \infty$.

We end this section with a few {\it final} remarks. First, the (anti-)BRST invariant coupled
(but equivalent) Lagrangian densities (6) and (5) respect {\it both} the BRST and anti-BRST symmetry transformations 
{\it together} on the submanifold of the quantum Hilbert space of fields where the Curci-Ferrari (CF) type restrictions 
($B_{\mu\nu} + \bar B_{\mu\nu} = \partial_\mu \phi_\nu - \partial_\nu \phi_\mu,\,
 B_\mu + \bar B_\mu = \pm m\, \phi_\mu - \partial_\mu \, \phi, \, f_\mu +  \bar F_\mu = \partial_\mu C_1,\, 
\bar f_\mu +  F_\mu = \partial_\mu \bar C_1,\, f + \bar F = \pm m\, C_1$ and $\bar f + F = \pm \, m\, \bar C_1$)
of our St$\ddot u$ckelberg-modified theory are satisfied (see, e.g. [18] for details). Second, it is quite straightforward to note that the
{\it original} St$\ddot u$ckelberg-modified Lagrangian density ${\cal L}_S^{(A)}$ [cf. Eq. (1)] remains invariant 
$[s_{(a)b}\, {\cal L}_S^{(A)} = 0$] under the infinitesimal, off-shell nilpotent and continuous 
(anti-)BRST symmetry transformations $s_{(a)b}$ [cf. Eqs (9), (7)]. Finally, 
it is worthwhile to mention here that the absolute anticommutativity property $\{s_b, \, s_{ab}\} = 0$ is  {\it also} satisfied when we 
invoke the validity of the above CF-type restrictions. 
For, instance, we note the following explicit computations, namely; 
\begin{eqnarray}
&&\{s_b, \, s_{ab} \}\, A_{\mu\nu\lambda} = \partial_\mu \, (B_{\nu\lambda} + \bar B_{\nu\lambda}) 
+ \partial_\nu\, (B_{\lambda\mu} + \bar B_{\lambda\mu})   + \partial_\lambda \, (B_{\mu\nu} + \bar B_{\mu\nu}), \nonumber\\
&&\{s_b, \, s_{ab} \}\, C_{\mu\nu} = \partial_\mu\, (f_\nu + \bar F_\nu) - \partial_\nu\, (f_\mu + \bar F_\mu), \nonumber\\
&&\{s_b, \, s_{ab} \}\, \bar C_{\mu\nu} = \partial_\mu\, (\bar f_\nu +  F_\nu) - \partial_\nu\, (\bar f_\mu +  F_\mu), \nonumber\\
 &&\{s_b, \, s_{ab} \}\, \phi_{\mu\nu} =  \pm m\, (B_{\mu\nu} + \bar B_{\mu\nu}) 
- \big\{ \partial_\mu\, (B_\nu + \bar B_\nu) - \partial_\nu \, (B_\mu + \bar B_\mu) \big\}, \nonumber\\
&&\{s_b, \, s_{ab} \}\, C_{\mu} = \pm m \, (f_\mu + \bar F_\mu) - \partial_\mu \, (f + \bar F),\nonumber\\
&&\{s_b, \, s_{ab} \}\, \bar C_{\mu} =  \pm m \, (\bar f_\mu + F_\mu) - \partial_\mu \, (\bar f +  F), 
\end{eqnarray}
which demonstrate that $\{s_b, \, s_{ab} \} = 0$ for the above fields provided we invoke the validity of: 
$B_{\mu\nu} + \bar B_{\mu\nu} = \partial_\mu \phi_\nu - \partial_\nu \phi_\mu,\,
B_\mu + \bar B_\mu = \pm m\, \phi_\mu - \partial_\mu \, \phi, \, f_\mu +  \bar F_\mu = \partial_\mu C_1,\, 
\bar f_\mu +  F_\mu = \partial_\mu \bar C_1,\, f + \bar F = \pm m\, C_1$ and $\bar f + F = \pm \, m\, \bar C_1$. 
It is straightforward to check  that the absolute anticommutativity property 
(i.e. $\{s_b, \, s_{ab}  \} = 0$) is automatically satisfied for the rest of {\it all} the fields of our theory 
which is described by the Lagrangian densities ${\cal L}_B$ and ${\cal L}_{\bar B}$. We {\it also} point out that only a {\it single} CF-type restriction
is required in the proof of the absolute anticommutativity property of the (anti-)BRST symmetry transformations for the 
fields $A_{\mu\nu\lambda}, \, C_{\mu\nu}, \, \bar C_{\mu\nu}$. On the other hand, for the fields $\Phi_{\mu\nu}, \, C_\mu, \, \bar C_{\mu}$,
the absolute anticommutativity property of the (anti-)BRST symmetry transformations is satisfied when  a set of {\it two} CF-type restrictions are invoked {\it together}. Finally, we establish that the sanctity of the CF-type restrictions is maintained even at the level of the transformed fields under the
(anti-)BRST symmetry transformations where we find that the off-shell nilpotency, absolute anticommutativity and (anti-)BRST invariant CF-type restrictions are all intertwined
together in an elegant manner (cf. Appendix E for details).\\


\section{Generator for the Gauge Symmetry Transformations: First-Class Constraints and Conserved Charge
 of our theory}

In this section, we perform the constraint analysis for the St$\ddot u$ckelberg-modified Lagrangian density
(${\cal L}_S^{(A)}$) and establish that it is endowed with the {\it first-class} constraints in the terminology of Dirac's 
prescription for the classification of constraints [23-26]. These first-class constraints are present in the generator for 
the infinitesimal, continuous and local {\it classical} gauge symmetry transformations (4). In this context, we note that 
we have the following canonical conjugate momenta for the 
St$\ddot u$ckelberg-modified Lagrangian density (1), namely; 
\begin{eqnarray}
&&\Pi^{\mu\nu\lambda}_{(A)} = \frac{\partial\, {\cal L}_S}{\partial\, (\partial_0\, A_{\mu\nu\lambda})} \equiv \frac{1}{3}\, H^{0 \mu\nu\lambda}, \nonumber\\
&&\Pi^{\mu\nu}_{(\phi)} = \frac{\partial\, {\cal L}_S}{\partial\, (\partial_0\, \Phi_{\mu\nu})} \equiv 
-\, \Sigma^{0\mu\nu} \pm m\, A^{0\mu\nu},
\end{eqnarray}
w.r.t. the {\it basic} 3-form field $A_{\mu\nu\lambda}$ and St$\ddot u$ckelberg 2-form field $\Phi_{\mu\nu}$. 
From (12), it is straightforward to check that we have the following {\it primary} constraints, respectively, namely; 
\begin{eqnarray}
&&\Pi^{0 i j}_{(A)} = \frac{1}{3}\, H^{0 0 i j} \approx 0,  \qquad 
\Pi^{0 i}_{(\phi)} = -\, \Sigma^{0 0 i} \pm m\, A^{0 0 i} \approx  0.
\end{eqnarray}
These constraints are {\it weakly} zero in the terminology of Dirac. As a consequence, the first-order time derivative on them can be performed to derive  the 
{\it secondary} constraints on the theory. The Hamiltonian formalism is the {\it best} approach to obtain the {\it successive} constraints. 
However, for our simple case, the Lagrangian formalism (see, e.g. [27]) is simple, beautiful and straightforward. 
The Euler-Lagrange (EL) equation of motion (EoM) w.r.t. the {\it basic} fields  $A_{\mu \nu \lambda }$ and $\Phi_{\mu \nu }$, 
 obtained from ${\cal L}_S^{(A)}$ [cf. Eq. (1)],  are as follows:  
\begin{eqnarray}
&&\partial_\rho\, H^{\rho\mu\nu\lambda} + m^2\, A^{\mu\nu\lambda} \mp m\, \Sigma^{\mu\nu\lambda} = 0, \qquad
\partial_\rho\, \Sigma^{\rho\mu\nu} \mp m\, \partial_\rho\, A^{\rho\mu\nu} = 0.
\end{eqnarray}
It is clear that, for the choices: $\mu = 0, \nu = j, \,  \lambda = k$, we have the following:
\begin{eqnarray}
&&\partial_0\, H^{00jk} + \partial_i\, H^{i0jk}  + m^2\, A^{0jk} \mp m\, \Sigma^{0jk} = 0, \nonumber\\
&&\partial_0\, \Sigma^{00j} + \partial_i\, \Sigma^{i0j} \mp m\, \partial_0\, A^{00j} \mp m\,  \partial_i\, A^{i0j}  = 0.
\end{eqnarray}
The above equations lead to the derivations of the {\it secondary} constraints on our theory as these are nothing but the outcome of 
setting the first-order ``time" derivative of the primary constraints (13) equal to zero. 
In other words, we have the following {\it secondary} constraints emerging out from the time-evolution invariance [27] of the {\it primary} constraints:
\begin{eqnarray}
&&3\, \frac{\partial\, \Pi_{(A)}^{0ij}}{\partial\, t}\,  =  3\, \partial_k\, \Pi^{kij}_{(A)} \mp m\, \Pi^{ij}_{(\phi)} \approx 0
\; \Longrightarrow \;  \frac{\partial\, \Pi_{(A)}^{0ij}}{\partial\, t}\, 
 =  \partial_k\, \Pi^{kij}_{(A)} \mp \frac{m}{3}\, \Pi^{ij}_{(\phi)} \approx 0, \nonumber\\
&&\frac{\partial\, \Pi_{(\phi)}^{0i}}{\partial\, t}\,  = \partial_j\, \Pi^{ji}_{(\phi)} \approx 0.
\end{eqnarray}
In the above, we have used, primarily,  the definitions of the space components 
of the canonical conjugate momenta w.r.t. $A_{\mu\nu\lambda}$and $\Phi_{\mu\nu}$ [cf. Eq. (12)], namely; 
\begin{eqnarray}
&&\Pi^{ i j k}_{(A)} = \frac{1}{3}\, H^{0  i j k}, \qquad \qquad 
\Pi^{ i j}_{(\phi)} = -\, \Sigma^{0  i j} \pm m\, A^{0  i j}, 
\end{eqnarray}
which are  deduced from the {\it original} definitions of the canonical conjugate momenta in (12). There are {\it no} further 
constraints on the theory. Since the {\it primary} as well as {\it secondary} 
constraints are expressed in terms of the components of the canonical conjugate momenta {\it only}, 
it is pretty obvious that their commutator(s) will be zero. Hence, our theory is endowed with the {\it first-class} constraints
in the terminology of Dirac's prescription for the classification scheme of the constraints [23-26].
In other words, our theory is an example of
 a {\it massive} gauge theory where the mass and gauge invariance co-exist {\it together}.  This is, primarily, due to the existence of the first-class constraints on our theory.

Before we write down the {\it final} expression for the generator of the gauge transformations (4) in terms of the constraints (13) and (16), 
we define the {\it non-trivial} equal-time canonical commutators  for the St$\ddot u$ckelberg-modified Lagrangian density (${\cal L}_S^{(A)}$) as:
\begin{eqnarray}
\big[A_{0ij} \, (\vec x, t), \, \Pi^{0kl}_{(A)}\, (\vec y, t)\big] &=& \frac{i}{2!}\, \big( \delta^k_i\, \delta^l_j -  \delta^l_i\, \delta^k_j \big)\, 
\delta^{(D-1)}\, (\vec x - \vec y), \nonumber\\
\big[A_{ijk} \, (\vec x, t), \, \Pi^{lmn}_{(A)}\, (\vec y, t)\big] &=& \frac{i}{3!}\, 
\big[\delta_i^l\, \big( \delta^m_j\, \delta^n_k -  \delta^n_j\, \delta^m_k \big)
+ \delta_i^m\, \big( \delta^n_j\, \delta^l_k -  \delta^l_j\, \delta^n_k \big) \nonumber\\
 &+& \delta_i^n\, \big( \delta^l_j\, \delta^m_k -  \delta^m_j\, \delta^l_k \big)
\big] \delta^{(D-1)}\, (\vec x - \vec y), \nonumber\\
\big[\Phi_{0i} \, (\vec x, t), \, \Pi^{0j}_{(\phi)}\, (\vec y, t)\big] &=& i\, \delta^j_i\, \delta^{(D-1)}\, (\vec x - \vec y), \nonumber\\
\big[\Phi_{ij} \, (\vec x, t), \, \Pi^{kl}_{(\phi)}\, (\vec y, t)\big] &=&    \frac{i}{2!}\, \big( \delta^k_i\, \delta^l_j - 
 \delta^l_i\, \delta^k_j \big)\,    \delta^{(D-1)}\, (\vec x - \vec y).
\end{eqnarray}
The rest of {\it all} the  brackets are  trivially  zero. 
In terms of all the four first-class constraints,
the generator $G^{(sm)}$,  for our {\it modified} massive Abelian 3-form theory,  is [28,29]
\begin{eqnarray}
G^{(sm)} &=& \int d^{D-1} x \, \Big[ \Pi ^{0ij} \, (\partial_0\, \Lambda_{ij})
 + \Pi ^{ij0} \,(\partial_i\, \Lambda_{j0})  + \Pi ^{j0i} \,(\partial_j\, \Lambda_{0i}) \nonumber\\
&-& \big[\partial_i\, \Pi ^{ijk} \mp \frac{m}{3}\, \Pi^{jk}_{(\phi)}\big]\, \Lambda_{jk} 
- \big[\partial_j\, \Pi ^{jki} \mp \frac{m}{3}\, \Pi^{ki}_{(\phi)}\big]\, \Lambda_{ki} \nonumber\\
&-& \big[\partial_k\, \Pi ^{kij} \mp \frac{m}{3}\, \Pi^{ij}_{(\phi)}\big]\, \Lambda_{ij}
+ \Pi_{(\phi)}^{0i} \big[ \pm m\, \Lambda_{0i} 
- (\partial_0\, \Lambda_{i} - \partial_i\, \Lambda_{0})\big] \nonumber\\
&+& \partial_i\, \Pi_{(\phi)}^{ij}\, \Lambda_{j} + \partial_j\, \Pi_{(\phi)}^{ji}\, \Lambda_{i}\Big],
\end{eqnarray}  
where the superscript $(sm)$ on the generator denotes that it is explicitly written 
for the St$\ddot u$ckelberg-modified {\it massive}  Abelian 3-form theory. 
In equation (19), the 
totally antisymmetric nature of $\Pi^{0ij}, \, \Pi^{ijk}, \, \Lambda_{ij}, \, \Lambda_{0i}$, etc.,  has been taken into account.  
Using the Gauss divergence theorem (where the gauge transformation parameters and fields have been treated as the physically well-defined 
objects which vanish off as $x \longrightarrow \pm\, \infty$), the above gauge symmetry generator $G^{(sm)}$ can be 
re-expressed, in a more transparent, beautiful  and useful fashion, as follows 
\begin{eqnarray}
G^{(sm)} &=& \int d^{D-1} x \, \Big[\Pi ^{0ij}\,(\partial_0\, \Lambda_{ij} + \partial_i\, \Lambda_{j0} + \partial_j\, \Lambda_{0i})\nonumber\\
&+& \Pi^{ijk}\, (\partial_i\, \Lambda_{jk} + \partial_j\, \Lambda_{ki} + \partial_k\, \Lambda_{ij}) \nonumber\\
&+& \,\Pi_{(\phi)}^{0i}  \, \big\{\pm m\, \Lambda_{0i} - (\partial_0\, \Lambda_{i} - \partial_i\, \Lambda_{0} )\big\} \nonumber\\
&+& \,\Pi_{(\phi)}^{ij}\, \big\{\pm m\, \Lambda_{ij} - (\partial_i\, \Lambda_{j} - \partial_j\, \Lambda_{i} )\big\} \Big], 
\end{eqnarray}
where we have exploited the potential of a simple (but cute) algebraic trick: 
$m\, \Pi^{ij}\, \Lambda_{ij} = (m/3)\,  \Pi^{ij}\, \Lambda_{ij} + (m/3)\,  \Pi^{jk}\, \Lambda_{jk} + (m/3)\,  \Pi^{ki}\, \Lambda_{ki} $
where {\it three} Latin indices have been used.  
It is straightforward, using the {\it non-trivial}  existing canonical brackets of Eq. (18),  to check that we have the following
expressions for the infinitesimal  gauge symmetry transformations for 
the {\it basic}  fields $(A_{\mu \nu \lambda }, \Phi_{\mu\nu})$ in their independent component form, namely;  
\begin{eqnarray}
&&\delta_g\, A_{0ij}\,(\vec x, t) = -\, i\, \big[A_{0ij}\,(\vec x, t), G  \big] \equiv 
(\partial_0\, \Lambda_{ij} + \partial_i\, \Lambda_{j0} + \partial_j\, \Lambda_{0i}),   \nonumber\\
&&\delta_g\, A_{ijk}\,(\vec x, t) = -\, i\, \big[A_{ijk}\,(\vec x, t), G  \big] \equiv 
(\partial_i\, \Lambda_{jk} + \partial_j\, \Lambda_{ki} + \partial_k\, \Lambda_{ij}), \nonumber\\
&&\delta_g\, \Phi_{0i}\,(\vec x, t) = -\, i\, \big[\Phi_{0i}\,(\vec x, t), G  \big] \equiv 
 \pm\, m\, \Lambda_{0i} - (\partial_0\, \Lambda_{i} - \partial_i\, \Lambda_{0} ), \nonumber\\
&&\delta_g\, \Phi_{ij}\,(\vec x, t) = -\, i\, \big[\Phi_{ij}\,(\vec x, t), G  \big] \equiv
 \pm \, m\, \Lambda_{ij} - (\partial_i\, \Lambda_{j} - \partial_j\, \Lambda_{i}),
\end{eqnarray}
which establish that the  first-class constraints are {\it indeed} the generator for the {\it classical} local, continuous 
and infinitesimal gauge symmetry transformations (4). A close look at (21) shows that we have already obtained: 
$\delta_g\, A_{\mu\nu\lambda} = \partial_\mu\, \Lambda_{\nu\lambda} + \partial_\nu\, \Lambda_{\lambda\mu} 
+ \partial_{\lambda} \Lambda_{\mu\nu}$ and  $\delta_g\, \Phi_{\mu\nu} = \pm \, m\, \Lambda_{\mu\nu} - (\partial_\mu\,\Lambda_\nu - \partial_\nu\,\Lambda_\mu)$.
These results automatically imply that we {\it also} have: $\delta_g\, H_{\mu \nu \lambda \zeta } = 0$ and 
$\delta_g\, \Sigma_{\mu\nu\lambda} = \pm \, m\,(\partial_\mu\, \Lambda_{\nu\lambda} + \partial_\nu\, \Lambda_{\lambda\mu} 
+ \partial_{\lambda} \Lambda_{\mu\nu})$ due to the local and infinitesimal
gauge symmetry transformations  (i.e. $\delta_g A_{\mu\nu\lambda}, \; \delta_g \Phi_{\mu\nu})$ 
on the basic fields $A_{\mu\nu\lambda}$ and $\Phi_{\mu\nu}$. Thus, we have derived 
{\it all} the infinitesimal gauge transformations that are listed in (4).

We establish now a deep connection between the Noether conserved current and charge 
(corresponding to the infinitesimal gauge transformations)
and the first-class constraints of the theory under consideration. In this context, 
we can compute the Noether conserved current for the infinitesimal gauge symmetry transformations (4) for the 
St$\ddot u$ckelberg-modified {\it massive} Abelian 3-form theory. This current $[J^\mu _{(sm)}]$ 
turns out to be the following
\begin{eqnarray}
J^\mu _{(sm)} & = & \frac {1}{3}\, H^{\mu\nu\lambda\zeta}\, (\partial_\nu\Lambda_{\lambda\zeta}
 + \partial_\lambda\Lambda_{\zeta\nu} \partial_\zeta\Lambda_{\nu\lambda})\nonumber\\
& + &   (\pm\, m\, A^{\mu\nu\lambda} - \Sigma^{\mu\nu\lambda})\,\big[\pm\,  
m\,\Lambda_{\nu\lambda} - (\partial_\nu\Lambda_\lambda - \partial_\lambda\Lambda_\nu)\big],
\end{eqnarray}
which can be shown to be conserved $(\partial_\mu \, J^\mu _{(sm)} = 0)$ due to the EoMs (15) that have already been derived from the 
St$\ddot u$ckelberg-modified Lagrangian density $({\cal L}_S^{(A)})$ w.r.t. the basic fields $A_{\mu\nu\lambda}$ and $\Phi_{\mu\nu}$ of our theory.  
Here the subscript ($sm$) on the Noether current denotes that it is written for the St$\ddot u$ckelberg-modified {\it massive} Abelian 
3-form theory. This Noether conserved current leads to the definition of the conserved charge $Q_{(sm)}$ as 
\begin{eqnarray}
Q_{(sm)} = \int d^{D - 1} x \, J^0_{(sm)} \equiv  \int d^{D - 1} x \Big[ \frac {1}{3}\, H^{0 \nu \lambda \zeta}\, \big(\partial_\nu \Lambda_{\lambda \zeta } 
+ \partial_\lambda \Lambda_{\zeta  \nu } +  \partial_\zeta \Lambda_{\nu \lambda }\big) \nonumber\\ 
+ \, \big (\pm m \, A^{0 \nu \lambda } - \Sigma^{0 \nu \lambda }\big)\, \big[\pm m\, \Lambda_{\nu\lambda}
 - (\partial_\nu\, \Lambda_\lambda - \partial_\lambda \, \Lambda_\nu)\big]\Big]
\end{eqnarray}
In view of the fact that our theory is endowed with a set of {\it four} first-class constraints, we have to expand the 
r.h.s. of the above equation carefully. In other words, we do {\it not} have to set the constraints strongly equal to zero
which appear in the expression for the zeroth component  of the conserved Noether current (i.e. $J^0_{(sm)}$).
To be precise, we have the following expression for the conserved charge, namely;  
\begin{eqnarray}
Q_{(sm)} &=&  \int d^{D - 1} x \, \Big[\frac {1}{3}\, H^{00ij}\, (\partial_0\Lambda_{ij} 
+ \partial_i\Lambda_{j0} +  \partial_j\Lambda_{0i})  \nonumber\\
&+& \frac {1}{3}\, H^{0ijk}\, (\partial_i\Lambda_{jk} 
+ \partial_j\Lambda_{ki} +  \partial_k\Lambda_{ij}) \nonumber\\
&+& \, \big(\pm m \, A^{0 0 i } - \Sigma^{0 0 i }\big)\, \big [\pm m\, \Lambda_{0 i}
 - \big(\partial_0\, \Lambda_i - \partial_i \, \Lambda \big)\big] \nonumber\\
&+& \, \big(\pm m \, A^{0 i j } - \Sigma^{0 i j }\big)\, \big[\pm m\, \Lambda_{i j} 
 - (\partial_i \, \Lambda_j  - \partial_j  \, \Lambda_i)\big]
\Big].
\end{eqnarray}
Taking into account, the definition of the constraints in the equations (13) and (16), we have the following
charge in terms of the first-class constraints, namely; 
\begin{eqnarray}
Q_{(sm)} & =& \int d^{D-1} x \, \Big[\Pi ^{0ij}\,(\partial_0\, \Lambda_{ij} + \partial_i\, \Lambda_{j0} + \partial_j\, \Lambda_{0i}) \nonumber\\
&+& \Pi^{ijk}\, (\partial_i\, \Lambda_{jk} + \partial_j\, \Lambda_{ki} + \partial_k\, \Lambda_{ij}) \nonumber\\
&+& \,\Pi_{(\phi)}^{0i}  \, \big\{\pm m\, \Lambda_{0i} - (\partial_0\, \Lambda_{i} - \partial_i\, \Lambda_{0} )\big\}  \nonumber\\
&+& \,\Pi_{(\phi)}^{ij}\, \big\{\pm m\, \Lambda_{ij} - (\partial_i\, \Lambda_{j} - \partial_j\, \Lambda_{i} )\big\} \Big].
\end{eqnarray}
which is nothing but the {\it final} form of the generator $G^{(sm)}$ that is quoted in (20). Taking into account the 
Gauss divergence theorem and the following  {\it simple}  algebraic trick 
\begin{eqnarray}
 \pm m\, \, \Pi_{(\phi)}^{ij}\, \Lambda_{ij}  =  \pm \frac{m}{3}\, \Pi_{(\phi)}^{ij}\, \Lambda_{ij} 
 \pm \frac{m}{3}\, \Pi_{(\phi)}^{jk}\, \Lambda_{jk}  \pm \frac{m}{3}\, \Pi_{(\phi)}^{ki}\, \Lambda_{ki},  
\end{eqnarray}
which involves {\it three} Latin indices, 
we can express the above conserved charge as  follows 
\begin{eqnarray}
Q_{(sm)} &=& \int d^{D-1} x \, \Big[\Pi ^{0ij}\,(\partial_0\, \Lambda_{ij} + \partial_i\, \Lambda_{j0} + \partial_j\, \Lambda_{0i}) \nonumber\\
&+& \,\Pi_{(\phi)}^{0i}  \, \big\{\pm m\, \Lambda_{0i} - (\partial_0\, \Lambda_{i} - \partial_i\, \Lambda_{0} )\big\} \nonumber\\ 
&-& \big(\partial_i \, \Pi^{ijk}\,  \mp  \frac{m}{3}\, \Pi_{(\phi)}^{jk}\big)\,  \Lambda_{jk}  
- \big(\partial_j \, \Pi^{jki}\,  \mp  \frac{m}{3}\, \Pi_{(\phi)}^{ki}\big)\,  \Lambda_{ki}  \nonumber\\
&-& \big(\partial_k \, \Pi^{kij}\,  \mp  \frac{m}{3}\, \Pi_{(\phi)}^{ij}\big)\,  \Lambda_{ij}  
+ \partial_i \, \Pi_{(\phi)}^{ij}\, \Lambda_{j} + \partial_j \, \Pi_{(\phi)}^{ji}\, \Lambda_{i} \Big], 
\end{eqnarray}
which is nothing but the expression for the generator $G^{(sm)}$ that has been written in the equation (19) 
in terms of the primary and secondary constraints. Hence, there is a deep connection between the classical 
generator and the  Noether conserved charge for the D-dimensional St$\ddot u$ckelberg-modified massive Abelian 3-form gauge theory.

We conclude this section with the following crucial and clinching remarks. First, the St$\ddot u$ckelberg-technique 
of compensating fields(s) is responsible for the conversion of the second-class constraints of the  original 
{\it massive} Abelian 3-form  theory into their counterparts first-class constraints. 
Second, the resulting first-class constraints generate the 
{\it classical} local, continuous and infinitesimal gauge symmetry transformations for the St$\ddot u$ckelberg-modified Lagrangian
density (${\cal L}_S^{(A)}$). Third, appropriate power(s) of the rest-mass $(m)$ 
has to be taken into account so that, in the gauge symmetry transformations, 
the different gauge symmetry transformation parameters appear with the
appropriate mass dimension (in the natural units: $\hbar = c = 1$) as is  the case for $\delta_g\, \Phi_{\mu\nu}$ [cf. Eqs.  (4), (21)].
Finally, if the expressions for the first-class
constraints appear in the zeroth component of the Noether conserved current [e.g. $J^0_{(sm)}$], 
they should {\it not} be set  strongly equal 
to zero in the derivation of the final expression for the Noether conserved charge.


\section{Equations of Motion: Coupled (but Equivalent) (Anti-)BRST Invariant Lagrangian Densities}

It is very interesting to pinpoint the fact that the Euler-Lagrange (EL) equations of motion (EoM) from the
coupled Lagrangian densities ${\cal L}_{B}$ and ${\cal L}_{\bar B}$ [cf. Eqs. (5), (6)] with respect 
to the auxiliary fields ($B_{\mu\nu}, \, \bar B_{\mu\nu}, \, B_\mu,\, \bar B_\mu, \, f_\mu,\, F_\mu, 
\bar f_\mu, \, \bar F_\mu, \, f,\, F,\, \bar f,\, \bar F$) are:
\begin{eqnarray} 
&&  B_{\mu\nu} = (\partial^\rho A_{\rho\mu\nu}) 
+ \frac {1}{2}\,(\partial_\mu \phi_\nu - \partial_\nu\phi_\mu \mp \, m\, \Phi_{\mu\nu}),\nonumber\\
 &&\bar B_{\mu\nu} = -\,( \partial^\rho A_{\rho\mu\nu}) 
+ \frac {1}{2}\,(\partial_\mu \phi_\nu - \partial_\nu\phi_\mu \pm \, m\, \Phi_{\mu\nu}), \nonumber\\
&& B_\mu = -\, (\partial^\rho \, \Phi_{\rho \mu}) + \frac {1}{2}\,(\pm\, m\, \phi_\mu - \partial_\mu \, \phi),  \; \;
\bar B_\mu =  (\partial^\rho \, \Phi_{\rho \mu}) + \frac {1}{2}\,(\pm\, m\, \phi_\mu - \partial_\mu \, \phi), \nonumber\\
&& F_\mu = \frac {1}{2}\, (\partial^\rho\, \bar C_{\rho\mu}) +\frac {1}{2}\, \partial_\mu\, \bar C_1 \mp \frac{m}{4}\, \bar C_\mu,   \; \;
\bar f_\mu = -\,\frac {1}{2}\, (\partial^\rho\,  \bar C_{\rho\mu}) +\frac {1}{2}\, \partial_\mu\, \bar C_1 \pm \frac{m}{4}\,  \bar C_\mu, \nonumber\\
&&f_\mu = \frac {1}{2}\,(\partial^\rho\,  C_{\rho\mu}) +\frac {1}{2}\, \partial_\mu\,  C_1 \mp \frac{m}{4}\,  C_\mu, \;\;
 \bar F_\mu = -\, \frac {1}{2}\,(\partial^\rho\,  C_{\rho\mu}) +\frac {1}{2}\, \partial_\mu\,  C_1 \pm \frac{m}{4}\,  C_\mu,  \nonumber\\
&& F = -\, \frac{1}{4}\, (\partial \cdot \bar C) \pm  \frac {m}{2}\, \, \bar C_1, \qquad \qquad \qquad  \quad 
 \bar f = \, \frac{1}{4}\, (\partial \cdot  \bar C) \pm \frac {m}{2}\,\,  \bar C_1, \nonumber\\
&& f =  -\, \frac{1}{4}\, (\partial \cdot  C) \pm \frac {m}{2}\, \,  C_1,  \qquad \quad  \quad \qquad  \quad 
  \bar F = \frac{1}{4}\, (\partial \cdot  C) \pm \frac {m}{2}\, \,  C_1.
\end{eqnarray}
It is evident that we obtain the following relationships using the straightforward algebra
\begin{eqnarray}
&&B_{\mu\nu} + \bar B_{\mu\nu} = \partial_\mu \phi_\nu - \partial_\nu \phi_\mu, \qquad
B_\mu + \bar B_\mu = \pm m\, \phi_\mu - \partial_\mu \, \phi, \nonumber\\ 
&& \bar f_\mu +  F_\mu = \partial_\mu \bar C_1, \qquad\qquad \qquad 
f_\mu +  \bar F_\mu = \partial_\mu C_1, \nonumber\\ 
&&  f + \bar F = \pm m\, C_1,\qquad\qquad \qquad 
\bar f + F = \pm \, m\, \bar C_1,
\end{eqnarray}
which are nothing but the (anti-)BRST invariant CF-type restrictions that are responsible for the 
absolute anticommutativity (i.e. $\{s_b, \, s_{ab}  \} = 0$) of the (anti-)BRST symmetry transformations
$[s_{(a)b}]$. Furthermore, we obtain the following {\it common} EL-EoMs from the coupled (but equivalent)  Lagrangian densities
${\cal L}_{B}$ and ${\cal L}_{\bar B}$, namely;
\begin{eqnarray}
&&B_1   = -\, (\partial \cdot \phi \mp m\, \phi), \qquad  B_2  = (\partial \cdot \bar \beta \mp m\, \bar \beta), \qquad
B  = -\, (\partial \cdot  \beta \mp m\,  \beta), \nonumber\\
&&\Box \beta \mp m\, (\partial \cdot \beta) \mp 2\, m \, B = 0, \qquad 
\Box \bar\beta \mp m\, (\partial \cdot \bar\beta) \pm 2\, m \, B_2 = 0, \nonumber\\
&& \Box\, \bar\beta_\mu - \partial_\mu\, (\partial\cdot \bar \beta) + \partial_\mu\, B_2
 - \frac{m^2}{2}\, \bar\beta_\mu \pm \frac{m}{2}\,\partial_\mu\, \bar\beta =0, \nonumber\\
&& \Box\, \beta_\mu - \partial_\mu\, (\partial\cdot  \beta) - \partial_\mu\, B
 - \frac{m^2}{2}\, \beta_\mu \pm \frac{m}{2}\,\partial_\mu\, \beta =0, \nonumber\\
&& (\Box - m^2)\, C_2 = 0, \qquad  (\Box - m^2)\, \bar C_2 = 0,
\end{eqnarray}
w.r.t. the auxiliary fields $(B_1,\, B,\, B_2)$ and the {\it bosonic} (anti-)ghost vector and scalar fields
$(\beta_\mu,\, \bar\beta_\mu,\, \beta,\, \bar\beta)$ as well as the {\it fermionic} (anti-)ghost fields $(\bar C_2)C_2$. 
It is clear, from the above equations, that the auxiliary fields 
$(B_1,\,B, \,  B_2)$ carry the ghost numbers ($0, \, +2,\, -2$), respectively, because the basic fields $(\phi, \, \beta, \, \bar\beta)$
{\it also} have the {\it same} ghost numbers.  
Before we proceed further, it is worthwhile to mention that the
(anti-)BRST invariant CF-type restrictions (29) are {\it not} the EL-EoMs as they are 
not derived from a single Lagrangian density and/or the minimization of the corresponding action integral.

We now focus on the EL-EoM, derived from the Lagrangian densities ${\cal L}_{B}$ and ${\cal L}_{\bar B}$,
that are {\it different} in appearance but {\it equivalent} on the submanifold of the Hilbert space of quantum fields where 
the CF-type restrictions (29) are satisfied/valid. At this stage, first of all, we derive explicitly the EL-EoMs from the Lagrangian density 
${\cal L}_{B}$ w.r.t. the fields: $A_{\mu \nu \lambda },\, \Phi_{\mu\nu}, C_{\mu\nu}, \, \bar C_{\mu\nu}, \, \phi_\mu,\,
C_\mu, \, \bar C_\mu, \phi,\,  C_1, \,\bar C_1$. These are as follows: 
\begin{eqnarray}
&&\partial^\rho\, H_{\rho\mu\nu\lambda} + m^2\, A_{\mu\nu\lambda} \mp\, m\, (\partial_\mu\, \Phi_{\nu\lambda} 
 + \partial_\nu\, \Phi_{\lambda\mu} + \partial_\lambda\, \Phi_{\mu\nu})  \nonumber\\
&&+ (\partial_\mu\, B_{\nu\lambda} + \partial_\nu\, B_{\lambda\mu} + \partial_\lambda\, B_{\mu\nu}) = 0,\nonumber\\
&&  \Box\, \Phi_{\mu\nu}\, -\, \partial_\mu\, (\partial^{\rho}\, \Phi_{\rho\nu}) + \partial_\nu\, (\partial^\rho\, \Phi_{\rho\mu}) 
\mp\, m\, (\partial^\rho\, A_{\rho\mu\nu}) \nonumber\\
&& + \frac{1}{2}\, (\partial_\mu\, B_\nu - \partial_\nu\, B_\mu) \mp \frac{m}{2}\, B_{\mu\nu}= 0, \nonumber\\
&&  \Box\, \bar C_{\mu\nu} - \partial_\mu\, (\partial^\rho\, \bar C_{\rho\nu}) + \partial_\nu\, (\partial^\rho\, \bar C_{\rho\mu})
\pm\, \frac{m}{2}\, (\partial_\mu \, \bar C_\nu - \partial_\nu\,\bar   C_\mu) \nonumber\\
&& + \frac{1}{2}\,  (\partial_\mu \, F_\nu - \partial_\nu\, F_\mu)  -\,\frac{m^2}{2}\,\bar C_{\mu\nu}  = 0, \nonumber\\
&&  \Box\, C_{\mu\nu} - \partial_\mu\, (\partial^\rho\, C_{\rho\nu}) + \partial_\nu\, (\partial^\rho\, C_{\rho\mu})
\pm\, \frac{m}{2}\, (\partial_\mu \, C_\nu - \partial_\nu\, C_\mu) \nonumber\\
&&+ \frac{1}{2}\,  (\partial_\mu \, f_\nu - \partial_\nu\, f_\mu)  -\,\frac{m^2}{2}\,C_{\mu\nu} =0,\nonumber\\
&&  \partial_\mu\, B^{\mu\nu} - \partial^\nu \, B_1\, \mp\, \frac{m}{2}\,  B^\nu =0, 
\qquad \frac{1}{2}\, (\partial\cdot B) \pm \, m\, B_1 = 0, \nonumber\\
 && \Box \bar C_\mu - \partial_\mu\, (\partial \cdot \bar C) - \frac{1}{2}\, \partial_\mu\, F 
\mp\, m\, (\partial^\rho\, \bar C_{\rho\mu}) \pm \frac{m}{2}\, F_\mu= 0, \nonumber\\
&& \Box C_\mu - \partial_\mu\, (\partial \cdot C) - \frac{1}{2}\, \partial_\mu\, f 
\mp\, m\, (\partial^\rho\,C_{\rho\mu}) \pm \frac{m}{2}\, f_\mu= 0, \nonumber\\
&& (\partial \cdot F) \mp m\, F = 0, \qquad (\partial \cdot f) \mp m\, f = 0.
\end{eqnarray}
The above kinds of EL-EoMs can be derived from the Lagrangian density 
${\cal L}_{\bar B}$, too. These equations of motion are: 
\begin{eqnarray}
&&\partial^\rho\, H_{\rho\mu\nu\lambda} + m^2\, A_{\mu\nu\lambda} \mp\, m\, (\partial_\mu\, \Phi_{\nu\lambda}
 + \partial_\nu\, \Phi_{\lambda\mu} + \partial_\lambda\, \Phi_{\mu\nu}) \nonumber\\
&& -\, (\partial_\mu\, \bar B_{\nu\lambda} + \partial_\nu\, \bar B_{\lambda\mu} + \partial_\lambda\, \bar B_{\mu\nu}) = 0,\nonumber\\
&&  \Box\, \Phi_{\mu\nu}\, -\, \partial_\mu\, (\partial^{\rho}\, \Phi_{\rho\nu}) + \partial_\nu\, (\partial^\rho\, \Phi_{\rho\mu}) 
\mp\, m\, (\partial^\rho\, A_{\rho\mu\nu})  \nonumber\\
&&- \frac{1}{2}\, (\partial_\mu\, \bar B_\nu - \partial_\nu\, \bar B_\mu) 
\pm \frac{m}{2}\, \bar B_{\mu\nu}= 0, \nonumber\\
&&  \Box\, \bar C_{\mu\nu} - \partial_\mu\, (\partial^\rho\, \bar C_{\rho\nu}) + \partial_\nu\, (\partial^\rho\, \bar C_{\rho\mu})
\pm\, \frac{m}{2}\, (\partial_\mu \, \bar C_\nu - \partial_\nu\,\bar   C_\mu) \nonumber\\
&&- \frac{1}{2}\,  (\partial_\mu \, \bar f_\nu - \partial_\nu\, \bar f_\mu) -\, \frac{m^2}{2}\, \bar C_{\mu\nu} = 0, \nonumber\\
&&  \Box\, C_{\mu\nu} - \partial_\mu\, (\partial^\rho\, C_{\rho\nu}) + \partial_\nu\, (\partial^\rho\, C_{\rho\mu})
\pm\, \frac{m}{2}\, (\partial_\mu \, C_\nu - \partial_\nu\, C_\mu) \nonumber\\
&& -\, \frac{1}{2}\,  (\partial_\mu \, \bar F_\nu - \partial_\nu\, \bar F_\mu) -\,\frac{m^2}{2}\,C_{\mu\nu}   = 0, \nonumber\\
&&  \partial_\mu\, \bar B^{\mu\nu} - \partial^\nu \, B_1\, \mp\, \frac{m}{2}\, \bar B^\nu =0, 
\qquad \frac{1}{2}\, (\partial \cdot \bar B) \pm m\, B_1 = 0, \nonumber\\
&& \Box \bar C_\mu - \partial_\mu\, (\partial \cdot \bar C) + \frac{1}{2}\, \partial_\mu\, \bar f 
\mp\, m\, (\partial^\rho\, \bar C_{\rho\mu}) \mp \frac{m}{2}\, \bar f_\mu= 0, \nonumber\\
&& \Box C_\mu - \partial_\mu\, (\partial \cdot C) + \frac{1}{2}\, \partial_\mu\, \bar F 
\mp\, m\, (\partial^\rho\,C_{\rho\mu}) \mp \frac{m}{2}\, \bar F_\mu= 0, \nonumber\\
&& (\partial \cdot \bar F) \mp m\, \bar F = 0, \qquad (\partial \cdot \bar f) \mp m\, \bar f = 0.
\end{eqnarray}
It is quite straightforward to note that the above EL-EoMs [cf. Eqs (31) and (32)] are {\it equivalent} on the 
submanifold of the Hilbert space of the quantum fields where the CF-type conditions (29) are satisfied. 
Let us take an example to corroborate this claim. The equation of motion w.r.t. $\Phi_{\mu\nu}$ fields from 
${\cal L}_{B}$ and ${\cal L}_{\bar B}$ can be subtracted from each-other to yield the following explicit expression:
\begin{eqnarray}
\frac{1}{2}\, \Big[\partial_\mu\, (B_\nu + \bar B_\nu) - \partial_\nu (B_\mu + \bar B_\mu) \Big] 
\mp \frac{m}{2}\, (B_{\mu\nu} + \bar B_{\mu\nu}) = 0. 
\end{eqnarray}
Now it is elementary to check that the substitution of the CF-type restrictions: 
$B_{\mu\nu} + \bar B_{\mu\nu} = \partial_\mu \phi_\nu - \partial_\nu \phi_\mu, \,
B_\mu + \bar B_\mu = \pm m\, \phi_\mu - \partial_\mu \, \phi$ in the above equation establishes that the equations of motion,
derived from ${\cal L}_{B}$ and ${\cal L}_{\bar B}$, are {\it equivalent} on the submanifold
in the Hilbert space of quantum fields which is defined by the CF-type restrictions. Similarly, 
it can be seen that all the equations of motion (31) and (32) are {\it equivalent}.

We end this section with the following {\it final} remarks. First of all, one can deduce some very beautiful equations from the 
{\it ones} that have been obtained as the EL-EoMs from ${\cal L}_{ B}$ and ${\cal L}_{\bar B}$. For instance, we observe that the following 
\begin{eqnarray}
&&(\Box - m^2)\, C_1 = 0, \qquad (\Box - m^2)\, \bar C_1 = 0, \qquad(\Box - m^2)\, (\partial \cdot C) = 0,\nonumber\\
&&(\Box - m^2)\, (\partial \cdot \bar C) = 0, \qquad (\Box - m^2)\, B_1 = 0, \qquad (\Box - m^2)\, B_2 = 0,\nonumber\\
&& (\Box - m^2)\, B = 0, \qquad \Box\, B_\mu  - \partial_\mu\, (\partial \cdot B)\mp\, m\, (\partial^\rho\,B_{\rho\mu}) = 0,\nonumber\\
&&\pm m\, [\Box\, \bar C_\mu - \partial_\mu\, (\partial \cdot \bar C)] 
+  [\Box\, F_\mu - \partial_\mu\, (\partial \cdot F)] -\, {m^2} (\partial^\rho\, \bar C_{\rho\mu})= 0,\nonumber\\
&&\pm\,  m [\Box\,  C_\mu - \partial_\mu\, (\partial \cdot  C)] 
+  [\Box\, f_\mu - \partial_\mu\, (\partial \cdot f)]-\,{m^2} (\partial^\rho\, C_{\rho\mu}) =  0, 
\end{eqnarray}
can be deduced from the EL-EoM [cf. Eq. (31)] that have been derived from the 
{\it perfectly} BRST invariant Lagrangian density ${\cal L}_{B}$. 
In exactly similar fashion, the concise and beautiful equations of motion can be derived from the 
Lagrangian density  ${\cal L}_{\bar B}$, too. Second, it can be seen, once again, that the above  equations
(that have been derived from ${\cal L}_{B}$) will be {\it equivalent} to the {\it ones} that are obtained from 
${\cal L}_{\bar B}$ on the submanifold in the quantum Hilbert space of the fields that are defined by the 
CF-type restrictions (29). Third, it is very interesting to observe that the CF-type restrictions can be 
derived from the EL-EoMs from the coupled (but equivalent) Lagrangian densities ${\cal L}_{B}$ 
and ${\cal L}_{\bar B}$, too, besides their derivations from the requirement of the absolute anticommutativity 
property of the (anti-)BRST symmetry transformations  [cf. Eq. (11)]. Finally, it is worthwhile to 
point out that the CF-type restrictions (29)  have been precisely  derived from the augmented superfield 
approach to BRST formalism in our earlier work [18]. It turns out that they are {\it physical} restrictions
on our theory because they are found to be (anti-)BRST invariant quantities. In other words, it can be checked that the following are true, namely; 
\begin{eqnarray}
&&s_{(a)b}\, [B_{\mu\nu} + \bar B_{\mu\nu} - (\partial_\mu \phi_\nu - \partial_\nu \phi_\mu)] = 0,\qquad 
s_{(a)b}\, [\bar f +  F \mp m\, \bar C_1] = 0,
 \nonumber\\
 &&s_{(a)b}\, [\bar f_\mu +  F_\mu - \partial_\mu \, \bar C_1] = 0, \qquad \qquad 
 s_{(a)b}\, [B_{\mu} + \bar B_{\mu}  \mp m\, \phi_\mu + \partial_\mu\, \phi] = 0, 
 \nonumber\\
&&s_{(a)b}\, [f + \bar F \mp m\, C_1] = 0, \qquad \qquad  \quad 
 s_{(a)b}\, [f_\mu + \bar F_\mu - \partial_\mu \, C_1] = 0,
\end{eqnarray}
where $s_{(a)b}$ are the off-shell nilpotent (anti-)BRST symmetry transformations 
(9) and (7), respectively. We shall see later that the (anti-)BRST invariant CF-type restrictions will be also derived from (i) the equality  of the coupled Lagrangian densities ${\cal L}_{B}$ and ${\cal L}_{\bar B}$ in our 
Appendix A, and (ii) the requirement of the absolute anticommutativity of the 
off-shell nilpotent versions of the (anti-)BRST charges $[Q_{(a)b}^{(1)}]$ in our  Appendix B. 
It is worthwhile to mention, in passing, that the CF-type restrictions have {\it also} been derived from the 
(anti-) BRST invariance of ${\cal L}_{B}$ and ${\cal L}_{\bar B}$ (see, e.g. [18] for details).\\


\section{Continuous Symmetries: Conservation Laws}

\noindent
This section is divided into two parts. In Subsec. A, we derive the Noether conserved currents and provide the
proof of their conservation laws. Our Subsec. B is devoted to the derivation of the Noether conserved
charges where their salient features have been pointed out in a quite elaborate manner.

\subsection{Conserved  Currents: Noether Theorem}

Whenever the action integral (or the Lagrangian density) remains invariant under the 
{\it continuous} symmetry transformation, according to Noether's theorem, 
there is always existence of the  conserved current (as well as corresponding conserved charge). 
In this context, first of all, we focus on the following ghost-scale
 {\it continuous} symmetry transformations with a global (i.e. spacetime independent)
{\it bosonic} scale transformation parameter $\Omega$ as
\begin{eqnarray}
&&C_2 \longrightarrow e^{3\,\Omega}\, C_2, \;\; \bar C_2 \longrightarrow e^{-\, 3\,\Omega}\, \bar C_2, \;\;
\beta_\mu \longrightarrow e^{2\,\Omega}\, \beta_\mu, \;\; \bar \beta_\mu \longrightarrow e^{-\, 2\,\Omega}\, \bar \beta_\mu, \nonumber\\
&&\beta \longrightarrow e^{2\,\Omega}\, \beta, \quad  \bar \beta \longrightarrow e^{-\, 2\,\Omega}\, \bar \beta,  \quad
B \longrightarrow e^{2\,\Omega}\, B, \quad  B_2 \longrightarrow e^{-\, 2\,\Omega}\, B_2, \nonumber\\
&&C_1 \longrightarrow e^{\Omega}\, C_1, \quad \bar C_1 \longrightarrow e^{-\, \Omega}\, \bar C_1,
\quad C_{\mu\nu} \longrightarrow e^{\Omega}\, C_{\mu\nu}, \quad \bar C_{\mu\nu} \longrightarrow e^{-\, \Omega}\, \bar C_{\mu\nu}, \nonumber\\
&& f \longrightarrow e^{\Omega}\, f, \quad \bar f \longrightarrow e^{-\, \Omega}\, \bar f, \quad
F \longrightarrow e^{-\, \Omega}\, F, \; \bar F \longrightarrow e^{ \Omega}\, \bar F, \quad f_\mu \longrightarrow e^{\Omega}\, f_\mu, \nonumber\\
&& \bar f_\mu \longrightarrow e^{-\, \Omega}\, \bar f_\mu, \quad
F_\mu \longrightarrow e^{-\, \Omega}\, F_\mu, \quad \bar F_\mu \longrightarrow e^{ \Omega}\, \bar F_\mu, \quad 
 \Sigma \longrightarrow e^0\,  \Sigma,
\end{eqnarray}
where the numerical factors, in the exponents, denote the ghost numbers of the specific fields. The generic field
$\Sigma$ stands for {\it all} the fields: $A_{\mu\nu \lambda }, \; \Phi_{\mu\nu}, \; \phi_\mu,  \phi, \; B_1, \; B_{\mu\nu}, \; 
\bar B_{\mu\nu}, \; B_\mu, \; \bar B_\mu$  which carry the ghost number equal to {\it zero}.
It can be readily checked that the coupled (but equivalent) Lagrangian densities ${\cal L}_B$ and ${\cal L}_{\bar B}$
[cf. Eqs. (5), (6)] remain invariant under the ghost-scale continuous symmetry transformations (36).

The infinitesimal version $(s_g)$ of the above ghost-scale symmetry transformations (36), with the 
very {\it simple} choice $\Omega = 1$ for the sake of bravity, can be expressed as: 
\begin{eqnarray}
&&s_g\, C_2 = 3\, C_2, \qquad s_g\, \bar C_2 = -\,  3\, \bar C_2, \qquad
s_g\, \beta_\mu = 2\, \beta_\mu, \quad s_g\, \bar \beta_\mu = -\, 2\, \bar \beta_\mu, \nonumber\\
&&s_g\, \beta = 2\, \beta, \qquad \quad s_g\, \bar \beta = -\, 2\, \bar \beta, \quad
\qquad s_g\, B = 2 \, B, \qquad s_g\, B_2 = -\,  2\, B_2, \nonumber\\
&& s_g\, C_1 = C_1, \qquad \; \; s_g\, \bar C_1 = -\,  \bar C_1, \qquad
s_g\, C_{\mu\nu} = C_{\mu\nu} , \qquad s_g\, \bar C_{\mu\nu}  = -\,  \bar C_{\mu\nu}, \nonumber\\
&&  s_g\, f = f, \qquad s_g\, \bar f = -\,  \bar f, \qquad
 s_g\, F = -\, F, \quad s_g\, \bar F =  \bar F, \qquad  s_g\, f_\mu = f_\mu, \nonumber\\
 &&s_g\, \bar f_\mu = -\,  \bar f_\mu, \quad
 s_g\, F_\mu = -\, F_\mu, \qquad s_g\, \bar F_\mu =  \bar F_\mu, \qquad s_g\, \Sigma = 0.
\end{eqnarray}
According to the basic concept behind the Noether theorem, the expression for the ghost conserved current 
$[J_{(g)}^\mu]$, derived from the Lagrangian density ${\cal L}_B$, is as follows
\begin{eqnarray}
&&J_{(g)}^\mu = (s_g\, C_{2})\, \Big[ \frac{\partial\, {\cal L}_B}{\partial \, (\partial_\mu \,  C_{2})}\Big]
+ (s_g\, \bar C_{2})\, \Big[ \frac{\partial\, {\cal L}_B}{\partial \, (\partial_\mu \, \bar C_{2})}\Big]
+  (s_g\,  C_{1})\, \Big[ \frac{\partial\, {\cal L}_B}{\partial \, (\partial_\mu \, C_{1})}\Big] \nonumber\\
 &&+ (s_g\,  \bar C_{1})\, \Big[ \frac{\partial\, {\cal L}_B}{\partial \, (\partial_\mu \, \bar C_{1})}\Big] 
+ (s_g\, \beta)\, \Big[ \frac{\partial\, {\cal L}_B}{\partial \, (\partial_\mu \, \beta)}\Big]
+  (s_g\, \bar\beta)\, \Big[ \frac{\partial\, {\cal L}_B}{\partial \, (\partial_\mu \, \bar\beta)}\Big] \nonumber\\ 
&&+  (s_b\,  C_{\nu })\, \Big[ \frac{\partial\, {\cal L}_B}{\partial \, (\partial_\mu \, C_{\nu})}\Big]
+  (s_b\,  \bar C_{\nu })\, \Big[ \frac{\partial\, {\cal L}_B}{\partial \, (\partial_\mu \, \bar C_{\nu})}\Big] 
+  (s_b\,  \beta_{\nu })\, \Big[ \frac{\partial\, {\cal L}_B}{\partial \, (\partial_\mu \, \beta_{\nu})}\Big]\nonumber\\
&&+  (s_b  \bar \beta_{\nu }) \Big[ \frac{\partial {\cal L}_B}{\partial  (\partial_\mu \, \bar \beta_{\nu})}\Big]
+  (s_b\,  C_{\nu\lambda  })\, \Big[ \frac{\partial\, {\cal L}_B}{\partial \, (\partial_\mu \, C_{\nu\lambda })}\Big]
+  (s_b\,  \bar C_{\nu \lambda })\, \Big[ \frac{\partial\, {\cal L}_B}{\partial \, (\partial_\mu \, \bar C_{\nu\lambda })}\Big],
\end{eqnarray}
where the convention of the left-derivative w.r.t. the fermionic (anti-)ghost fields has been adopted. 
The explicit computation of the above equation, with the help of (37) and the Lagrangian density ${\cal L}_B$,
 leads to the following  expression for the ghost current:
\begin{eqnarray}
J_{(g)}^\mu  &=& (\partial^\mu\, \bar C^{\nu\lambda} + \partial^\nu\,\bar C^{\lambda\mu} + \partial^\lambda\,\bar C^{\mu\nu})\, C_{\nu \lambda }
\mp m\, \bar C^{\mu\nu}\, C_\nu - \bar C^{\mu\nu}\, f_\nu \nonumber\\
&+& (\partial^\mu\, C^{\nu\lambda} + \partial^\nu\, C^{\lambda\mu} + \partial^\lambda\, C^{\mu\nu})\, \bar C_{\nu \lambda }
\mp m\,  C^{\mu\nu}\, \bar C_\nu - C^{\mu\nu}\, F_\nu \nonumber\\
&+& (\partial^\mu\, \bar C^{\nu} - \partial^\nu\, \bar C^{\mu})\, C_\nu + (\partial^\mu\,  C^{\nu} - \partial^\nu\,  C^{\mu})\, \bar C_\nu 
+ \frac{1}{2}\, (\bar C^\mu\, f + C^\mu \, F) \nonumber\\
&-&\, C_1\, F^\mu - \bar C_1\, f^\mu + 2\, (\partial^\mu\, \beta^\nu - \partial^\nu\, \beta^\mu)\, \bar\beta_\nu
-\,  2\, (\partial^\mu\, \bar \beta^\nu - \partial^\nu\, \bar \beta^\mu)\, \beta_\nu \nonumber\\
&+& (\pm m\, \bar\beta^\mu - \partial^\mu\, \bar\beta)\, \beta - (\pm m\, \beta^\mu - \partial^\mu\, \beta)\, \bar \beta
-\, 2\, \beta^\mu\, B_2 - \, 2\, \bar\beta^\mu\,B \nonumber\\
&+& 3\, \bar C_2\, \partial^\mu\, C_2 + 3\,  C_2\, \partial^\mu\,\bar C_2.
\end{eqnarray}
Using the EL-EoMs (that have been derived in Sec. 4) from the Lagrangian density ${\cal L}_B$, it is {\it not} very difficult 
to check that $\partial_\mu \, J_{(g)}^\mu  = 0 $ which proves the conservation law for the ghost Noether current. It will be 
worthwhile to mention here that the ghost Noether current $[\bar J_{(g)}^\mu]$ (that
has been computed from ${\cal L}_{\bar B}$ using the {\it same} formula as (38) with the replacement:
 ${\cal L}_B \longrightarrow {\cal L}_{\bar B}$) is:
\begin{eqnarray}
\bar J_{(g)}^\mu  &=& (\partial^\mu\, \bar C^{\nu\lambda} + \partial^\nu\,\bar C^{\lambda\mu} + \partial^\lambda\,\bar C^{\mu\nu})\, C_{\nu \lambda }
\mp m\, \bar C^{\mu\nu}\, C_\nu + \bar C^{\mu\nu}\, \bar F_\nu \nonumber\\
&+& (\partial^\mu\, C^{\nu\lambda} + \partial^\nu\, C^{\lambda\mu} + \partial^\lambda\, C^{\mu\nu})\, \bar C_{\nu \lambda }
\mp m\,  C^{\mu\nu}\, \bar C_\nu + C^{\mu\nu}\, \bar f_\nu \nonumber\\
&+& (\partial^\mu\, \bar C^{\nu} - \partial^\nu\, \bar C^{\mu})\, C_\nu + (\partial^\mu\,  C^{\nu} - \partial^\nu\,  C^{\mu})\, \bar C_\nu 
- \frac{1}{2}\, ( C^\mu\, \bar f + \bar C^\mu \, \bar F) \nonumber\\
&+& (\pm m\, \bar\beta^\mu - \partial^\mu\, \bar\beta)\, \beta - (\pm m\, \beta^\mu - \partial^\mu\, \beta)\, \bar \beta
-\, 2\, \beta^\mu\, B_2 - \, 2\, \bar\beta^\mu\,B \nonumber\\
&-&\, C_1\, \bar f^\mu - \bar C_1\, \bar F^\mu + 2\, (\partial^\mu\, \beta^\nu - \partial^\nu\, \beta^\mu)\, \bar\beta_\nu
-\,  2\, (\partial^\mu\, \bar \beta^\nu - \partial^\nu\, \bar \beta^\mu)\, \beta_\nu \nonumber\\
&+&  3\,  C_2\, \partial^\mu\,\bar C_2 + 3\, \bar C_2\, \partial^\mu\, C_2.
\end{eqnarray}
Using the appropriate equations of motion from Sec. 4 (that have been derived from ${\cal L}_{\bar B}$), it is {\it not} very hard to 
prove  that $\partial_\mu \, \bar J_{(g)}^\mu  = 0 $ which establishes the validity of the conservation law.

Some of the key points to be noted, at this juncture,  are as follows. First, we observe that a part of the ghost currents in (39) and (40),
depending on the transformations of the fields $\beta, \; \bar\beta, \beta_\mu$ and $\bar\beta_\mu$, 
remains conserved on its own. Second, in exactly similar fashion, a part of the ghost currents, generated by the 
transformations: $s_g\, C_2 = 3\, C_2$ and $s_g\, \bar C_2 = -\, 3\, \bar C_2$, also remains conserved on 
its own. Third, we have to use the expressions for the auxiliary fields of (28) and $B = -\, (\partial \cdot \beta \mp m\, \beta)$
as well as $B_2 = (\partial \cdot \bar \beta \mp m\, \bar\beta)$ at appropriate places for the proof of the conservation laws.
Finally, the part of the ghost currents, derived from the transformations for the fields 
$C_1, \;\bar C_1, \; C_\mu, \; \bar C_\mu, \; C_{\mu\nu}, \; \bar C_{\mu\nu}$, etc., are conserved due to the equations of motion
of {\it all} these fields (that have been  derived from the Lagrangian densities ${\cal L}_{ B}$ and ${\cal L}_{\bar B}$ in Sec. 4).
To sum up, we observe that the part of the ghost currents, corresponding to the infinitesimal ghost-scale transformations for the 
(anti-)ghost fields with ghost numbers: $(-\, 3, + 3), \, (-\, 2, + 2)$ and $(-\, 1, +1)$, are separately 
and independently {\it conserved} on their own due to the EL-EoMs.

We are in the position now to discuss the continuous and infinitesimal BRST symmetry transformations (7) and corresponding 
Noether current.
Due to our observation in (8), it is clear that we shall
have conserved BRST current (and corresponding charge) for the Lagrangian density ${\cal L}_B$. 
In exactly similar fashion, our observation in (10) would lead to the derivation of the anti-BRST conserved current 
(and corresponding conserved anti-BRST charge). First of all, we concentrate on the derivation of Noether's current 
corresponding to the BRST symmetry transformations. 
The expression for the Noether current [$J^{\mu}_{(b)}$] is 
\begin{eqnarray}
J^{\mu}_{(b)} &=& (s_b A_{\alpha \beta \gamma })\Big[ \frac{\partial{\cal L}_B}{\partial  (\partial_\mu  A_{\alpha \beta \gamma })}\Big]
+ (s_b \Phi_{\alpha \beta }) \Big[ \frac{\partial {\cal L}_B}{\partial  (\partial_\mu  \Phi_{\alpha \beta})}\Big]
+  (s_b \bar C_{\alpha \beta }) \Big[ \frac{\partial {\cal L}_B}{\partial  (\partial_\mu  \bar C_{\alpha \beta})}\Big] \nonumber\\
&+&  (s_b\,  C_{\alpha \beta })\, \Big[ \frac{\partial\, {\cal L}_B}{\partial \, (\partial_\mu \, C_{\alpha \beta})}\Big]
+  (s_b\,  \phi_{\alpha })\, \Big[ \frac{\partial\, {\cal L}_B}{\partial \, (\partial_\mu \, \phi_{\alpha})}\Big]
+  (s_b\,  C_{\alpha })\, \Big[ \frac{\partial\, {\cal L}_B}{\partial \, (\partial_\mu \, C_{\alpha})}\Big] \nonumber\\
&+&  (s_b\,  \bar C_{\alpha })\, \Big[ \frac{\partial\, {\cal L}_B}{\partial \, (\partial_\mu \, \bar C_{\alpha})}\Big]
+  (s_b\,  \beta_{\alpha })\, \Big[ \frac{\partial\, {\cal L}_B}{\partial \, (\partial_\mu \, \beta_{\alpha})}\Big]
+  (s_b\,  \bar \beta_{\alpha })\, \Big[ \frac{\partial\, {\cal L}_B}{\partial \, (\partial_\mu \, \bar \beta_{\alpha})}\Big] \nonumber\\
&+&  (s_b\, \bar C_{2})\, \Big[ \frac{\partial\, {\cal L}_B}{\partial \, (\partial_\mu \, \bar C_{2})}\Big]
 +  (s_b\,  C_{1})\, \Big[ \frac{\partial\, {\cal L}_B}{\partial \, (\partial_\mu \, C_{1})}\Big]
 + (s_b\,  \bar C_{1})\, \Big[ \frac{\partial\, {\cal L}_B}{\partial \, (\partial_\mu \, \bar C_{1})}\Big] \nonumber\\
&+& (s_b\,  \phi)\, \Big[ \frac{\partial\, {\cal L}_B}{\partial \, (\partial_\mu \,\phi)}\Big] 
+ (s_b\,  \beta)\, \Big[ \frac{\partial\, {\cal L}_B}{\partial \, (\partial_\mu \,\beta)}\Big]
+ (s_b\, \bar \beta)\, \Big[ \frac{\partial\, {\cal L}_B}{\partial \, (\partial_\mu \,\bar \beta)}\Big]- X^\mu,
\end{eqnarray}
where ${\cal L}_B$ is the {\it perfectly} BRST invariant
Lagrangian density (5) and $X^{\mu}$ is expression in the square bracket of Eq. (8).
In other words, we have:
\begin{eqnarray}
 X^\mu &=&(\partial^\mu\, C^{\nu\lambda} + \partial^\nu\, C^{\lambda\mu}
+ \partial^\lambda\, C^{\mu\nu}) \, B_{\nu\lambda}  + B^{\mu\nu}\, f_\nu - B_2\, \partial^\mu\,C_2  \nonumber\\
&-& B_1\, f^\mu + B\, F^\mu - (\partial^\mu\, \beta^\nu - \partial^\nu\, \beta^\mu)\, F_\nu 
+ \frac{1}{2}\, (\pm\, m\, \beta^\mu - \partial^\mu\, \beta)\, F \nonumber\\
 &+& (\partial^\mu\,  C^\nu -\partial^\nu\,  C^\mu)\, B_\nu \pm \, m\, B^{\mu\nu}\, C_\nu  
-\, \frac{1}{2} \,  B^\mu\,  f \nonumber\\
&\mp& m\,\bar C^{\mu\nu}\, \big(\mp m\,  \beta_\nu - \partial_\nu\, \beta \big)
\mp\, m\, (\partial^\mu\, \beta^\nu - \partial^\nu\,\beta^\mu)\,\bar C_\nu.
\end{eqnarray}
We 
call a Lagrangian density as the {\it perfectly} symmetry invariant Lagrangian density where {\it no}
EL-EoMs and/or CF-type restrictions are invoked for its symmetry invariance (from outside). For instance, the Lagrangian density 
${\cal L}_B $ is a perfectly symmetry invariant Lagrangian density w.r.t. the BRST symmetry transformations ($s_b$).
It will be noted that, in (41), we have {\it not} taken the BRST symmetry transformations on the 
{\it auxiliary} fields because they do {\it not} have their derivatives in the Lagrangian density ${\cal L}_B$. 
Furthermore, due to our observation that $s_b\, C_2 = 0$, we have {\it not} taken {\it its} contribution to (41)
even though the derivative on $C_2$ field exists. The substitutions of the BRST symmetry transformations (7) and 
the appropriate expressions for the derivatives w.r.t. the suitable fields of the Lagrangian density ${\cal L}_B$ lead to the following: 
\begin{eqnarray}
J^{\mu}_{(b)} &=&  H^{\mu \nu \lambda \zeta } \, (\partial_\nu\, C_{\lambda\zeta })
\pm \frac{m}{2}\,[\pm m\, \bar\beta^\mu -  \partial^\mu\, \bar\beta ]\, C_2  \pm m\, \bar C^{\mu\nu}\, (\pm m\, \beta_\nu - \partial_\nu\, \beta) \nonumber\\
&+& (\partial^\mu\,  C^{\nu\lambda} 
 + \partial^\nu\,  C^{\lambda\mu} + \partial^\lambda\,  C^{\mu\nu})\, B_{\nu\lambda} 
- (\partial^\mu\, \bar C^{\nu\lambda} + \partial^\nu\, \bar C^{\lambda\mu}  \nonumber\\
&+& \partial^\lambda\, \bar C^{\mu\nu})
\, (\partial_\nu\, \beta_\lambda - \partial_\lambda\, \beta_\nu)
\mp m\, C^{\mu\nu}\, B_\nu -\, B_1\, f^\mu + B\,  F^\mu 
- \, \frac{1}{2}\, B^\mu\, f  \nonumber\\
 & +& [\pm m\, A^{\mu\nu\lambda} -\, \Sigma^{\mu\nu\lambda} ]\, [\pm m\, C_{\nu\lambda} - (\partial_\nu\, C_\lambda - \partial_\lambda\, C_\nu)]\, 
 - B_2 \, \partial^\mu \, C_2 + B^{\mu\nu}\, f_\nu  \nonumber\\
&-& \, (\partial^\mu\, \bar\beta^\nu - \partial^\nu\, \bar\beta^\mu)\, (\partial_\nu\, C_2)
-\, (\partial^\mu\, \beta^\nu - \partial^\nu\, \beta^\mu)\, F_\nu + (\partial^\mu\, C^\nu - \partial^\nu\, C^\mu)\, B_\nu  \nonumber\\
 &  +& \, \frac{1}{2}\, \big(\pm m\, \beta^\mu - \partial^\mu\, \beta \big)\, F - \, 
(\partial^\mu\, \bar C^\nu - \partial^\nu\, \bar C^\mu)\, (\pm m\, \beta_\nu - \partial_\nu\, \beta).
\end{eqnarray}
According to the basic tenets of Noether's theorem, we know that the conservation law $\partial_\mu\, J^\mu_{(b)} = 0$
can be proven by using the appropriate EL-EoMs that have been derived in our Sec. 4 where some of the 
common EL-EoMs [cf. Eq. (30)] have been listed (that emerge out from both the Lagrangian densities
${\cal L}_{B}$ and ${\cal L}_{\bar B}$). In the proof of $\partial_\mu\, J^{\mu}_{(b)} = 0$, we have utilized the EL-EoMs (31)
that have been derived from  ${\cal L}_{B}$. Furthermore, we have invoked the explicit expressions (28) for the 
auxiliary fields $(B_{\mu\nu}, \, B_\mu, \, F_\mu, \, f_\mu, \, F, \, f)$ in the above proof.
We lay emphasis on the fact that the expressions for these auxiliary fields in Eq. (28) are nothing but the 
EL-EoMs derived from  ${\cal L}_{B}$ w.r.t. these auxiliary fields themselves. Ultimately, we note that the conservation law 
[$\partial_\mu\, J^\mu_{(b)} = 0$] is {\it true} which proves the sanctity of Noether's theorem.

We now concentrate on the proof of the conservation of the Noether current $[J^{\mu}_{( ab)}]$ corresponding to the 
infinitesimal, continuous and off-shell nilpotent $(s_{ab}^2 = 0)$ anti-BRST symmetry transformations (9). 
The explicit expression for $J^{\mu}_{( ab)}$ can be computed in {\it exactly} similar fashion as 
$J^{\mu}_{(b)}$ [cf. Eq. (41)] where we have to replace: $s_b \rightarrow  s_{ab}$ and $ {\cal L}_B \rightarrow {\cal L}_{\bar B}$.
In this context, we note that the anti-BRST symmetry transformations ($s_{ab}$) are listed in (9) and the explicit expression for ${\cal L}_{\bar B}$
is given in (6). In addition, we here to replace $X^\mu$ of equation (41) by $Y^\mu$ which is as follows:
\begin{eqnarray}
 Y^\mu &=& \bar B^{\mu\nu}\, \bar f_\nu  - (\partial^\mu\, \bar C^{\nu\lambda} 
+ \partial^\nu\, \bar C^{\lambda\mu}+ \partial^\lambda\, \bar C^{\mu\nu}) \, \bar B_{\nu\lambda}  + B \, \partial^\mu\, \bar C_2 \nonumber\\
&-& B_2 \, \bar F^\mu - B_1\, \bar f^\mu - (\partial^\mu\, \bar \beta^\nu - \partial^\nu\, \bar \beta^\mu)\, \bar F_\nu
+ \frac{1}{2}\, (\pm\, m\, \bar \beta^\mu - \partial^\mu\, \bar\beta)\, \bar F   \nonumber\\
 &-& (\partial^\mu\, \bar C^\nu -\partial^\nu\, \bar C^\mu)\, \bar B_\nu \mp \, m\, \bar B^{\mu\nu}\, \bar C_\nu  
-\, \frac{1}{2} \, \bar B^\mu\, \bar f \nonumber\\ 
&\pm& m\, C^{\mu\nu}\, (\pm m\, \bar \beta_\nu - \partial_\nu\, \bar \beta)
\pm\, m\, (\partial^\mu\, \bar\beta^\nu - \partial^\nu\,\bar\beta^\mu)\,C_\nu.
\end{eqnarray}
The above expression in (44) is nothing but the terms that are present in the square bracket of equation (10) on the r.h.s. 
The exact mathematical form of the anti-BRST Noether current is as follows\\
\begin{eqnarray}
J^{\mu}_{( ab)} &=& H^{\mu \nu \lambda \zeta } \, (\partial_\nu\, \bar C_{\lambda\zeta })
 \pm \frac{m}{2}\,[\pm m\, \beta^\mu -  \partial^\mu\,\beta ] \,  \bar C_2 \mp m\, C^{\mu\nu}\, 
(\pm m\, \bar \beta_\nu - \partial_\nu\, \bar\beta)  \nonumber\\
&-& (\partial^\mu\, \bar C^{\nu\lambda}
+ \partial^\nu\,  \bar C^{\lambda\mu} + \partial^\lambda\,  \bar C^{\mu\nu})\, \bar B_{\nu\lambda} 
+ \, (\partial^\mu\,  C^{\nu\lambda}
 + \partial^\nu\, C^{\lambda\mu}  \nonumber\\
 &+& \partial^\lambda\, C^{\mu\nu})\,  (\partial_\nu\, \bar \beta_\lambda - \partial_\lambda\, \bar \beta_\nu)
 \pm m\, \bar C^{\mu\nu}\, \bar B_\nu 
 - \bar f^\mu \,  B_1  - B_2\, \bar F^\mu - \, \frac{1}{2}\,\bar  B^\mu\, \bar f \nonumber\\
 &+&
[ \pm m\, A^{\mu\nu\lambda} -\, \Sigma^{\mu\nu\lambda} ] \, [\pm m\, \bar C_{\nu\lambda} - (\partial_\nu\, \bar C_\lambda - \partial_\lambda\, \bar C_\nu)]
+ B\, \partial^\mu \,\bar C_2 +  \bar B^{\mu\nu}\, \bar f_\nu \nonumber\\
&-& \, (\partial^\mu\, \beta^\nu - \partial^\nu\, \beta^\mu)\, (\partial_\nu\, \bar C_2)
-\, (\partial^\mu\, \bar \beta^\nu - \partial^\nu\, \bar \beta^\mu)\, \bar F_\nu
 - \, (\partial^\mu\, \bar C^\nu - \partial^\nu\, \bar C^\mu)\,\bar B_\nu  \nonumber\\
  & +& \frac{1}{2}\, \big(\pm m\, \bar \beta^\mu - \partial^\mu\, \bar \beta \big)\, \bar F 
+ (\partial^\mu\,  C^\nu - \partial^\nu\, C^\mu)\, (\pm m\, \bar \beta_\nu - \partial_\nu\, \bar \beta),
\end{eqnarray}
where it should be noted that $s_b\, \bar C_2$ will be replaced by $s_{ab}\, C_2$ because we note that 
$s_{ab}\, \bar C_2 = 0$ but $s_{ab}\, C_2 = B$.
The conservation law (i.e. $\partial_\mu\, J^{\mu}_{( ab)} = 0$) can be proven by using the appropriate equations of motion
of Sec. 4. We would like to lay emphasis on the fact that 
the following points are pertinent in our present discussion. First, we have {\it not} used the CF-type restrictions (29) anywhere
in the proof of $\partial_\mu\, J^{\mu}_{(b)} = 0$ and $\partial_\mu\, J^{\mu}_{( ab)} = 0$
as the (anti-)BRST invariant CF-type restrictions, as pointed out earlier,  are {\it not} the EL-EoMs. Second, the explicit 
expressions for the auxiliary fields in (28) have been used in the proof of the conservation laws because the entries of equation (28)
are primarily the EL-EoMs from the Lagrangian densities ${\cal L}_{B}$ and ${\cal L}_{\bar B}$.
Finally, we have used quite frequently the {\it common} equations of motion (30) from ${\cal L}_{B}$ and ${\cal L}_{\bar B}$.
On top of it, in the proof of 
$\partial_\mu\, J^{\mu}_{(b)} = 0$ and $\partial_\mu\, J^{\mu}_{( ab)} = 0$,
we have invoked the validity of EL-EoMS (31) and (32), respectively.

\subsection{Conserved Charges from Conserved Currents}

The central purpose of our present subsection is to derive the conserved Noether charges $Q_{r} \equiv \int d^{D-1} x \,
 J^0_{(r)} \; (r = a, \, ab, \, g)$ from the conserved Noether currents $J^{\mu}_{(r)} \; (r = a, \, ab, \, g)$ 
of our previous subsection and comment on some salient features. For instance, 
we show that the standard Noether conserved charges [$Q_{(a)b}$] are {\it not} 
off-shell nilpotent (i.e. $Q^2_{(a)b} \ne 0$) of order two (where we have taken the help of
the celebrated relationship between the continuous symmetries and their generators as the conserved Noether charges). 
 In this context, first of all, we concentrate on the anti-BRST
charge $Q_{ab}$ which is explicitly written, in any arbitrary D-dimension of spacetime, as follows 
\begin{eqnarray}
Q_{ab} &=&  \int  d^{D-1} x \, J^{0}_{( ab)} \equiv  \int  d^{D-1} x \,\Big [ H^{0 i j k } \, (\partial_i\, \bar C_{jk })
 \pm \frac{m}{2}\,[\pm m\, \beta^0 -  \partial^0\,\beta ] \,  \bar C_2 \nonumber\\ 
&\mp& m\, C^{0i}\, (\pm m\, \bar \beta_i 
- \partial_i\, \bar\beta) 
- (\partial^0\, \bar C^{ij} + \partial^i\,  \bar C^{j0} + \partial^j\,  \bar C^{0i})\, \bar B_{ij}  - B_2\, \bar F^0\nonumber\\ 
&+& \, (\partial^0\,  C^{ij}
 + \partial^i\, C^{j0} + \partial^j\, C^{0i})\,  (\partial_i\, \bar \beta_j - \partial_j\, \bar \beta_i) + B\, \partial^0 \,\bar C_2  
+  \bar B^{0i}\, \bar f_i - \, \frac{1}{2}\,\bar  B^0\, \bar f \nonumber\\
 &\pm&  m\, \bar C^{0i}\, \bar B_i 
 - \bar f^0 \,  B_1  +
[ \pm m\, A^{0ij} -\, \Sigma^{0ij} ] \, [\pm m\, \bar C_{ij} - (\partial_i\, \bar C_j - \partial_j\, \bar C_i)]
\nonumber\\
& -& \, (\partial^0\, \beta^i - \partial^i\, \beta^0)\, (\partial_i\, \bar C_2)
-\, (\partial^0\, \bar \beta^i - \partial^i\, \bar \beta^0)\, \bar F_i
 - \, (\partial^0\, \bar C^i - \partial^i\, \bar C^0)\,\bar B_i  \nonumber\\
  & +& \frac{1}{2}\, \big(\pm m\, \bar \beta^0 - \partial^0\, \bar \beta \big)\, \bar F 
+ (\partial^0\,  C^i - \partial^i\, C^0)\, (\pm m\, \bar \beta_i - \partial_i\, \bar \beta) \Big],
\end{eqnarray}
where the expression for $J^0_{(ab)}$ has been derived from the Noether conserved anti-BRST current  
$J^\mu_{(ab)}$ [cf. Eq. (45)] in a  straightforward fashion. In exactly similar manner, we write down the BRST charge 
$Q_{b} \equiv \int d^{D-1} x \, J^0_{(b)}$ from the expression for conserved current [$J^\mu_{(b)}$] that has been 
written  in (43). To be precise, we have the following expression for $Q_b$  in any arbitrary D-dimensions of spacetime
 for our present system, namely; 
\begin{eqnarray}
Q_{b} &=&  \int  d^{D-1} x \, J^{0}_{( b)} \equiv  \int  d^{D-1} x \,\Big [H^{0ijk} \, (\partial_i\, C_{jk })
\pm \frac{m}{2}\,[\pm m\, \bar\beta^0 -  \partial^0\, \bar\beta ]\, C_2  \nonumber\\ 
&\pm& m\, \bar C^{0i}\, (\pm m\, \beta_i  
- \partial_i\, \beta)
+ (\partial^0\,  C^{ij}  + \partial^i\,  C^{j0} + \partial^j\,  C^{0i})\, B_{ij}  \nonumber\\
&- &(\partial^0\, \bar C^{ij} + \partial^i\, \bar C^{j0} + \partial^j\, \bar C^{0i}) 
\, (\partial_i\, \beta_j - \partial_j\, \beta_i)  - B_2 \, \partial^0 \, C_2
+ B\,  F^0   + B^{0i}\, f_i \nonumber\\
 &-& B_1\, f^0  \mp m\, C^{0i}\, B_i + [\pm m\, A^{0ij} -\, \Sigma^{0ij} ]\, [\pm m\, C_{ij} - (\partial_i\, C_j - \partial_j\, C_i)]\, 
 \nonumber\\
&-& \, (\partial^0\, \bar\beta^i - \partial^i\, \bar\beta^0)\, (\partial_i\, C_2)
-\, (\partial^0\, \beta^i - \partial^i\, \beta^0)\, F_i + (\partial^0\, C^i - \partial^i\, C^0)\, B_i  \nonumber\\
 &-&  \frac{1}{2} B^0  f   +  \frac{1}{2} \big(\pm m \beta^0 - \partial^0\beta \big)\, F -  
(\partial^0 \bar C^i  - \partial^i \bar C^0)\, (\pm m\, \beta_i - \partial_i\, \beta)\Big],
\end{eqnarray}
where $J^0_{(b)}$ is the zeroth component of $J^\mu_{(b)}$ that has been quoted in (43). Finally, we write
down the expression for the 
ghost charge $Q_{g} = \int d^{D-1} x \, J^0_{(g)}$ as follows
\begin{eqnarray}
Q_{g} &=&  \int  d^{D-1} x \, J^{0}_{(g)} \equiv  \int  d^{D-1} x \,\Big [(\partial^0\, \bar C^{ij} 
+ \partial^i\,\bar C^{j0} + \partial^j\,\bar C^{0i})\, C_{ij } \mp m\, \bar C^{0i}\, C_i \nonumber\\ 
&-& \bar C^{0i}\, f_i 
+ (\partial^0\, C^{ij} + \partial^i\, C^{j0} + \partial^j\, C^{0i})\, \bar C_{ij }
\mp m\,  C^{0i}\, \bar C_i - C^{0i}\, F_i -\, C_1\, F^0  \nonumber\\
&+& (\partial^0\, \bar C^{i} - \partial^i\, \bar C^{0})\, C_i + (\partial^0\,  C^{i} - \partial^i\,  C^{0})\, \bar C_i 
+ \frac{1}{2}\, (\bar C^0\, f + C^0 \, F) - \bar C_1\, f^0  \nonumber\\
&+& 2\, (\partial^0\, \beta^i - \partial^i\, \beta^0)\, \bar\beta_i
-\,  2\, (\partial^0\, \bar \beta^i - \partial^i\, \bar \beta^0)\, \beta_i  + (\pm m\, \bar\beta^0 - \partial^0\, \bar\beta)\, \beta\nonumber\\
 &-& (\pm m\, \beta^0 - \partial^0\, \beta)\, \bar \beta
-\, 2\, \beta^0\, B_2 - \, 2\, \bar\beta^0\,B 
+ 3\, \bar C_2\, \partial^0\, C_2 + 3\,  C_2\, \partial^0\,\bar C_2 \Big],
\end{eqnarray}
where the mathematical form of  of the conserved ghost current $[J^\mu_{(g)}]$ has been quoted in Eq. (39) 
for our system. A similar kind of ghost charge $\bar Q_{g} = \int d^{D-1} x \, \bar J^0_{(g)}$
can be computed from (40). However, we shall {\it not} explicitly  express it here and we shall only concentrate on 
$Q_{g} = \int d^{D-1} x \, J^0_{(g)}$ [cf. the above Eq. (48)].

We conclude this subsection with the following crucial remarks. First of all, we note that the standard expressions for the
Noether conserved charges $ Q_r$ (with $r = b, ab, g $) lead the derivation of the infinitesimal continuous symmetry transformations
(7), (9) and (37) which are fermionic (i.e. off-shell nilpotent) as well as bosonic in nature, respectively. In other words, these standard
Noether charges are the generators for the above infinitesimal continuous symmetry transformations. Mathematically, this statement
can be captured in the following (anti)commutators between the generic 
fermionic/bosonic field $\Phi$ of our theory (described by 
$ {\cal L}_B$ and ${\cal L}_{\bar B}$) and the 
Noether conserved charges $ Q_r$ (with $r = b, ab, g $),
namely;
\begin{eqnarray}
s_r\, \Phi &=& - i\, \big [ \Phi, \; Q_r \big ]_{(\pm)}, \qquad r = b, ab,  \nonumber\\
s_g \, \Phi &=& - i\, \big [ \Phi, \; Q_g \big ],
\end{eqnarray}
where the subscripts $(\pm)$ on the square bracket, in the first entry of the above equation, stand for the (anti)commutator for the 
generic field $\Phi$ being fermionic/bosonic in nature. Since the infinitesimal ghost-scale symmetry transformations (37) are
bosonic in nature, we have a commutator in the second entry of the above equation between the generic 
fermionic/bosonic field $\Phi$ and the conserved
Noether ghost charge $Q_g$. Second, the relationship between the continuous symmetry transformation and its generator (as the
standard conserved Noether charge) is {\it general} in nature. Thus, it can be readily checked that we have 
\begin{eqnarray}
s_b\, Q_b &=& - i\, \big \{ Q_b, Q_b \big \} = -\,2\,i\, Q_b^2,  \nonumber\\
s_{ab}\, Q_{ab} &=& - i\, \big \{ Q_{ab}, Q_{ab} \big \} = -\,2\,i\, Q_{ab}^2,
\end{eqnarray}
where the l.h.s. can be computed by the direct applications of the (anti-) BRST symmetry transformations [cf. Eqs. (9), (7)]
on the precise expressions for the Noether conserved 
(anti-)BRST charges [cf. Eqs. (46), (47)]. Ultimately, we observe that the following 
are true, namely;
\begin{eqnarray}
s_b\, Q_b  & = & \int d^{D - 1} x\, \Big [\pm \frac {m}{2}\, \Big(\pm\, m\,F^0 - \partial^0 F\Big)\, C_2 
 - (\partial^0 F^i - \partial^i F^0)\, (\partial_i C_2)\nonumber\\
 & -&  (\partial^0 \, B^{ij} + \partial^i \, B^{j0}  + \partial^j \, B^{0i} )  \, (\partial_i \beta_j - \partial_j \beta_i)\nonumber\\ 
 & + & \Big[\pm\,m\, B^{0i} - (\partial^0 B^i - \partial^i B^0)]\, (\pm \, m\, \beta_i - \partial_i \beta)\Big] \neq 0,\nonumber\\ 
 s_{ab}\, Q_{ab} & = & \int d^{D - 1} x\, \Big [\pm \frac {m}{2}\, \Big(\pm\, m\,\bar F^0 
- \partial^0 \bar F\Big)\, \bar C_2  - (\partial^0 \bar F^i - \partial^i \bar F^0)\, (\partial_i \bar C_2)\nonumber\\
 &  + &  (\partial^0 \, \bar B^{ij} + \partial^i \, \bar B^{j0}  + \partial^j \, \bar B^{0i} )  \, (\partial_i \bar \beta_j - \partial_j \bar \beta_i)\nonumber\\ 
 &  -  & \Big[\pm\, m\, \bar B^{0i} - (\partial^0 \bar B^i - \partial^i \bar B^0)]\, (\pm \, m\, \bar \beta_i - \partial_i \bar \beta)\Big] \neq 0, 
\end{eqnarray}
which demonstrate that the standard Noether conserved (anti-)BRST charges are {\it not} nilpotent of order two
(i.e. $Q_{(a)b}^2 \neq 0 $). As a consequence, one can {\it not} obtain the standard BRST algebra with the
Noether conserved charges $Q_{(a)b}$ and $Q_g$. Furthermore, the BRST cohomology w.r.t. the Noether conserved 
(anti-)BRST charges  can {\it not} be performed.

\section{Physicality Criteria: Nilpotent Charges}

Our starting {\it classical} theory is described by the Lagrangian density (1) for the {\it modified} massive Abelain 3-form theory
which is endowed with a set of {\it four} first-class constraints [cf. Eqs. (13), (16)] in the terminology of Dirac's prescription
for the classification scheme of constraints (see, e.g. [23-27] for details).  According to the celebrated Dirac quantization scheme,
for the quantization of a {\it classical} system endowed with any kind of constraints,
the Dirac quantization conditions  require that the operator form of {\it these}
constraints must annihilate the physical states (i.e. $|phys> $) existing in  the total quantum  Hilbert space of states. 
In other words, the Dirac quantization conditions are as follows [23,24]:
\begin{eqnarray}
&&\Pi^{0ij}_{(A)}\; |phys> = 0, \qquad  \qquad \qquad \qquad \; \Pi^{0i}_{(\phi)} \; |phys> = 0, \nonumber\\
&& \Big [\partial_i\, \Pi^{ijk}_{(A)} \mp \frac{m}{3}\, \Pi^{jk}_{(\phi)} \Big] \;|phys> = 0, \qquad \big[\partial_i \Pi^{ij}_{(\phi)} \big]\;|phys> = 0.
\end{eqnarray}
Within the framework of BRST-quantization scheme, the conditions (52) must be realized through the physicality criteria w.r.t. the 
conserved and nilpotent (anti-)BRST charges where we demand that the physical states (i.e. $|phys> $), in the total quantum
Hilbert space of states, are {\it those} that are annihilated by the {\it above} conserved and nilpotent charges [25-27, 30, 31]. However, we have 
already noted (at the fag end of our previous subsection) that the standard Noether conserved (anti-)BRST charges are {\it not} 
nilpotent (i.e. $Q_{(a)b}^2 \neq 0 $)
of order two. In other words, they are conserved {\it but} non-nilpotent.

It has been shown in our earlier work (see, e.g. [22] for details) that the Noether theorem does {\it not}
lead to the derivations of the {\it nilpotent} (anti-)BRST charges for the BRST-quantized theories where the
CF-type restrictions are {\it non-trivial}. A systematic theoretical method has been developed  in [22] where we have 
discussed the precise theoretical methodology to obtain the nilpotent (i.e. $[Q^{(1)}_b]^2 = 0$) version of the BRST charge $Q^{(1)}_b$ 
from the standard Noether conserved charge $Q_b$  which is found to be non-nilpotent (i.e. $Q_b^2 \neq 0)$.
It is obvious here 
that we have chosen the notation for the non-nilpotent Noether conserved charge  as $Q_b$ (which is derived by using the 
standard Noether's theorem).
The nilpotent version  of the conserved BRST charge $Q_b^{(1)}$, for our {\it modified} D-dimensional massive Abelian 3-form theory, is as follows [22]:    
\begin{eqnarray}
Q^{(1)}_b  & = & \int d^{D - 1} x \Big[(\partial^0 C^{ij} + \partial^i  C^{j0} + \partial^j C^{0i}) B_{ij}
 -  [\pm m C^{0i} - (\partial^0 C^i - \partial^i C^0)]B_i \nonumber\\
 & - &   (\partial^0 B^{ij} + \partial^i  B^{j0} + \partial^j B^{0i})\, C_{ij} + [\,\pm\,m\,B^{0i} - (\partial^0 B^i - \partial^i B^0)]\, C_i \nonumber\\
& - &  [\,\pm\, m\,\bar C^{0i} - (\partial^0 \bar C^i - \partial^i \bar C^0)\,(\pm\,m\,\beta_i - \partial_i \beta)] 
  + 2\,(\partial^0 \bar\beta ^i - \partial^i \bar \beta ^0)\, \partial_i C_2\nonumber\\
& \mp & \,m\,(\pm\,m\,\bar\beta^0 - \partial^0\bar\beta)\, C_2 + (\partial^0 \bar C^{ij} + \partial^i \bar C^{j0} + \partial^j \bar C^{0i})\, (\partial_i \beta_j 
- \partial_j \beta_i)   \nonumber\\
& + &2\, (\partial^0 F^i - \partial^i F^0)\,\beta_i - (\pm\,m\,F^0 - \partial^0 F)\, \beta + 3\,\dot B_2\,C_2  - B_2\,\dot C_2 +  B\, F^0\nonumber\\
& - & B_1 f^0 - \frac {1}{2} B^0 f + B^{0i} f_i + \frac {1}{2}  (\pm m \beta^0 - \partial^0 \beta) F 
- (\partial^0 \beta ^i - \partial^i \beta ^0)\,F_i \Big].
\end{eqnarray} 
At this juncture, it is pertinent to point out that in our earlier work [22], we have exploited the interplay of (i) the appropriate 
El-EoMs from the Lagrangian density ${\cal L}_B$, (ii) the Gauss divergence theorem, and (iii) the off-shell nilpotent BRST symmetry
transformations $s_b$ {\it together} to obtain the above off-shell nilpotent version of the BRST charge $Q_b^{(1)}$. 
It is interesting to point out that, out of
the whole tower of the EL-EoMs (cf. Sec. 4 for details), our proposal has utilized 
{\it only} the following {\it three} relevent and useful EL-EoMs (see, e.g. [22])
\begin{eqnarray}
&& \partial_\mu\, H^{\mu\nu\lambda\xi} + m^2 \, A^{\nu\lambda\xi} \mp\, m \, \Sigma^{\nu\lambda\xi}
+ \bigl (\partial^\nu B^{\lambda\xi}  + \partial^\lambda B^{\xi\nu}  + \partial^\xi B^{\nu\lambda} \bigr ) = 0, \nonumber\\
&& \partial_\mu \, (\partial^\mu \bar C^\nu - \partial^\nu \bar C^\mu) \mp m \, (\partial_\mu \bar C^{\mu\nu}) \pm \frac{m}{2}\, F^\nu
- \frac{1}{2}\, \partial^\nu F = 0, \nonumber\\ 
&& \partial_\mu \, (\partial^\mu \bar \beta^\nu - \partial^\nu \bar \beta^\mu) + \partial^\nu \, B_2 
\pm \frac{m}{2}\, \partial^\nu \bar \beta
- \frac{1}{2}\,m^2\, \bar \beta^\nu = 0,
\end{eqnarray} 
to obtain the off-shell nilpotent version of the BRST charge $Q^{(1)}_b$ from the non-nilpotent standard conserved  Noether BRST charge $Q_b$ (which has been derived using the Noether theorem).

The  coupled (but equivalent) (anti-)BRST invariant Lagrangian densities ${\cal L}_B $ and ${\cal L}_{\bar B}$ do {\it not} support
any kind of constraints unlike the starting  {\it classical} 
Lagrangian density ${\cal L}_S^{(A)}$ [cf. Eq. (1)] for the {\it modified} massive
Abelian 3-form theory where we have a set of {\it four} first-class constraints [cf. Eqs. (13), (16)] in the terminology of 
Dirac's prescription for the classification scheme of constraints.  However, these constraints have been traded with the 
Nakanishi-Lautrup auxiliary fields and the specific combination of derivatives on them for the Lagrangian densities ${\cal L}_B $ and ${\cal L}_{\bar B}$ at the {\it quantum} level.
To corroborate this statement, first of all, 
 let us focus on the following expressions for the canonical conjugate momenta w.r.t. the basic fields
$A_{\mu\nu\lambda} $ and $ \Phi_{\mu\nu}$ for the BRST invariant Lagrangian density ${\cal L}_B$, namely; 
\begin{eqnarray}
\Pi^{\mu \nu \lambda}_{(A)} & = & \frac{\partial\, {\cal L}_B}{\partial (\partial_0\, A_{\mu \nu \lambda })} =   \frac {1}{3}\, H^{0\mu\nu\lambda}
 + \frac {1}{3}\,(\eta^{\mu 0}\,B^{\nu\lambda} + \eta^{\nu 0}\, B^{\lambda\mu} + \eta^{\lambda 0}\,B^{\mu\nu}),\nonumber\\
\Pi^{\mu \nu }_{(\phi)} & = & \frac{\partial\, {\cal L}_B}{\partial (\partial_0\, \Phi_{\mu \nu})} =  -\, \Sigma^{0\mu\nu} \pm m\,A^{0\mu\nu}
- \frac {1}{2}\, (\eta^{\mu0}\,B^{\nu} - \eta^{\nu0}\,B^\mu).
\end{eqnarray} 
It is clear, from the above expressions, that we have the following explicit components 
\begin{eqnarray}
\Pi^{0ij}_{(A)} & = &   \frac {1}{3}\, H^{00ij}
 + \frac {1}{3}\, B^{ij} \equiv \frac {1}{3}\, B_{ij} \qquad \Longrightarrow \qquad B_{ij} = 3\, \Pi^{0ij}_{(A)}, \nonumber\\
\Pi^{0i}_{(\phi)} & = &   -\, \Sigma^{00i} \pm m\,A^{00i}
- \frac {1}{2}\,B^{i} \equiv \frac {1}{2}\, B_i \qquad \Longrightarrow \qquad B_{i} = 2\, \Pi^{0i}_{(\phi)}, \nonumber\\
\Pi^{ijk}_{(A)} & = &   \frac {1}{3}\, H^{0ijk} \qquad \Longrightarrow \qquad  H^{0ijk} = 3\, \Pi^{ijk}_{(A)}, \nonumber\\
\Pi^{ij}_{(\phi)} & = &   -\, \Sigma^{0ij} \pm m\,A^{0ij},
\end{eqnarray}
which demonstrate that the original primary constraints (i.e. $\Pi^{0ij}_{(A)} \approx 0, \; \Pi^{0i}_{(\phi)} \approx 0)$
have been traded with the Nakanishi-Lautrup auxiliary fields $B_{ij}$ and $B_{i}$ of the BRST-invariant theory (which are 
present in  the Lagrangian density ${\cal L}_{B}$). Now, we concentrate on the top equation of (54) {\it and} the 
following EL-EoM derived from ${\cal L}_{B}$ w.r.t. $\Phi_{\mu\nu}$: 
\begin{eqnarray}
\partial_\mu\, \Sigma^{\mu\nu\lambda}  \mp m\, (\partial_\mu\, A^{\mu\nu\lambda}) + \frac{1}{2}\, (\partial^\nu\, B^\lambda - \partial^\lambda\, B^\nu)
\mp \, \frac{m}{2}\, B^{\nu\lambda}  = 0. 
\end{eqnarray}
Making the choices: $\nu = 0, \, \lambda = j $ and $\xi = k$, we obtain the following   from the top equation of (54) and (57), namely; 
\begin{eqnarray}
&& (\partial^0 B^{ij} + \partial^i  B^{j0} + \partial^j B^{0i})  = 3\, \Big [\partial_k\, \Pi^{kij}_{(A)} \mp   \frac{m}{3}\, \Pi^{ij}_{(\phi)}  \Big], \nonumber\\
 &&(\partial^0\, B^i - \partial^i\, B^0) \mp m \, B^{0i}  = 2\, [\partial_j\, \Pi^{ji}_{(\phi)} ], 
\end{eqnarray}
where we have used the notations from (56) for the purely space components of the canonical conjugate momenta. 
Thus, we have observed here that the {\it secondary} constraints [cf. Eq. (16)] of our {\it classical} St$\ddot u$ckelberg-modified 
massive Abelian 3-form theory have been traded with the specific combinations  of derivatives (as well as the specific components 
 of) on the Nakanishi-Lautrup auxiliary fields $(B_{\mu\nu}, \, B_\mu)$
 of the BRST-invariant Lagrangian density ${\cal L}_B$ at the {\it quantum} level [cf. Eq. (58)].

In view of the definitions (56), it is straightforward to note that the  physicality criteria [25-27, 30, 31] w.r.t. the conserved 
standard Noether charge (47) leads to the following 
\begin{eqnarray}
Q_b\, | phys > \,  = \,  0   \qquad \Longrightarrow \qquad B_{ij} \, | phys > \,  =\,  0, \qquad B_i\, | phys > \, =\,  0, 
\end{eqnarray}
where we have taken into account the fact that the physical states are {\it those} that are annihilated by the 
conserved BRST charge ($Q_b$). Thus, we note that {\it only} the 
 operator form of the primary constraints (i.e. $\Pi^{0ij}_{(A)} \approx 0, \; \Pi^{0i}_{(\phi)} \approx 0)$ of the 
original {\it classical} theory (which have been traded with the Nakanishi-Lautrup auxiliary fields: 
$B_{ij} = 3\, \Pi^{0ij}_{(A)}, \;  B_{i} = 2\, \Pi^{0i}_{(\phi)}$) are able to annihilate the 
physical state (i.e. $| phys > $). Hence, the quantization conditions (59) do {\it not} say anything about the 
secondary constraints [cf. Eq. (16)] of the original classical {\it modified} Abelian 3-form theory.
We, ultimately, infer that the Noether conserved (but non-nilpotent)
BRST charge $Q_b$ [cf. Eq. (47)] fails in satisfying the Dirac quantization conditions for the precise  quantization of our 
classical theory [cf. Eq. (1)] that is endowed with a set of {\it four} first-class constraints. We draw the conclusion  that the standard 
conserved (but non-nilpotent) Noether charge is {\it not} suitable 
for the physicality criterion at the quantum level (for our present {\it classical} system that is endowed 
with {\it four} first-class constraints).

Against the backdrop of the above paragraph, we concentrate on the physicality criteria 
(i.e. $Q_b^{(1)} \, | phys >  = 0$) w.r.t. the nilpotent version of the BRST charge [cf. Eq. (53)] which leads to the following quantization
conditions [25-27, 30, 31] on the physical states (i.e.  $| phys >  = 0$) existing in the {\it total} quantum Hilbert space of states, namely;  
\begin{eqnarray}
B_{ij} \, | phys >  = 0  & \Longrightarrow &  \Pi^{0ij}_{(A)} \, | phys >  = 0,  \nonumber\\
B_{i} \, | phys >  = 0 & \Longrightarrow & \Pi^{0i}_{(\phi)} \, | phys >  = 0,  \nonumber\\
 (\partial^0 B^{ij} + \partial^i  B^{j0} + \partial^j B^{0i}) \, | phys >  = 0  &\Longrightarrow& 
\Big (\partial_k\, \Pi^{kij}_{(A)} \mp   \frac{m}{3}\, \Pi^{ij}_{(\phi)}  \Big)  \, | phys >  = 0, \nonumber\\
\big[(\partial^0 B^i - \partial^i B^0) \mp m\, B^{0i} \big] \, | phys > = 0  & \Longrightarrow & 
(\partial_i\, \Pi^{ij}_{(\phi)})\, |phys > = 0, 
\end{eqnarray}
where we have taken into consideration the 
theoretical strength of the EL-EoMs (58). Thus, we have obtained the conditions 
on the physical states from the coefficients of the {\it basic} ghost fields
(i.e. $C_{\mu\nu},\,  C_\mu$) in (the expression for $Q_b^{(1)}$) where the first {\it four} 
terms have played the crucial roles. In other words, the physical states are annihilated by only the field operators which are 
endowed with the ghost numbers equal to zero. To be precise, the physical states 
(i.e. $| phys > $), existing in the total quantum Hilbert space of states, are 
unaffected by (i) the presence of the basic ghost fields, (ii) the auxiliary fields with specific non-zero ghost numbers, and (iii) the ordinary derivatives acting on {\it both} of them. This is why, the first four terms of  $Q_b^{(1)}$ have contributed to 
the physicality criterion w.r.t. it in the equation (60).
It will be worthwhile to point out that we have {\it not} taken 
$B_1\, | phys > = 0, \, B^{0i}\, | phys > = 0$ and $B^0\, | phys > = 0$ because these bosonic auxiliary fields are associated with the 
components of the {\it auxiliary} ghost fields $f_\mu$ and $f$ which are {\it not} independent 
as they have to obey: $f_\mu + \bar F_\mu = \partial_\mu\, C_1, \; f + \bar F = \pm m\, C_1$.

For the completion of our arguments, we now say a few words about the anti-BRST charge. As discussed in the context of the conserved 
BRST charge, in exactly similar fashion, we find that the 
{\it standard} Noether conserved anti-BRST charge (46) leads to the following conditions [25-27, 30, 31] on the physical states, namely; 
\begin{eqnarray}
Q_{ab}\, | phys > = 0   \qquad \Longrightarrow \qquad \bar B_{ij} \, | phys >  = 0, \qquad \bar B_i\, | phys > = 0, 
\end{eqnarray}
which is {\it not} different from the conditions (59) that have been derived from the conserved Noether 
charge $Q_b$ (except the change  of notations for the Nakanishi-Lautrup fields). We also observe that the 
anti-BRST charge (46) does {\it not} say anything about the secondary constraints [cf. Eq. (16)] of the original {\it classical} 
modified massive Abelian 3-form theory [described by the Lagrangian density (1)]. Thus, we focus on the 
physicality criteria (i.e. $Q_{ab}^{(1)}\, | phys > = 0$) w.r.t. the nilpotent version of the anti-BRST charge 
$Q_{ab}^{(1)}$ which has been explicitly derived in our earlier work [22] as follows 
\begin{eqnarray}
Q_{ab}^{(1)} & = & \int d^{D - 1}  x  \Big[(\partial^0 \bar B^{ij} + \partial^i  \bar B ^{j0} + \partial^j \bar B ^{0i})  \bar C_{ij}
 - \, [\pm m  \bar B^{0i} - (\partial^0 \bar B^i - \partial^i \bar B^0)] \bar C_i \nonumber\\
 & - &   (\partial^0 \bar C^{ij} + \partial^i  \bar C ^{j0} + \partial^j  \bar C^{0i})\, \bar B _{ij} + [\,\pm\,m\,\bar C ^{0i} - (\partial^0 \bar C ^i - \partial^i \bar C ^0)]\,\bar  B_i \nonumber\\
& + &  [\,\pm\,m\,  C^{0i} - (\partial^0 C^i - \partial^i  C^0)]\,(\pm\,m\,\bar\beta_i - \partial_i \bar\beta) 
  + 2\,(\partial^0 \beta ^i - \partial^i  \beta ^0)\, \partial_i \bar C_2\nonumber\\
& \mp &m\, (\pm\,m\, \beta^0 - \partial^0\beta)\, \bar C_2 - (\partial^0  C^{ij} + \partial^i  C^{j0} + \partial^j  C^{0i})\, (\partial_i \bar \beta_j 
- \partial_j \bar \beta_i)   \nonumber\\
& + & 2\, (\partial^0 \bar F^i - \partial^i \bar F^0)\,\bar \beta_i - (\pm\,m\,\bar F^0 - \partial^0 \bar F)\, \bar \beta -  3\,\dot B\,\bar C_2   + B\,\dot { \bar C}_2 -   B_2\, \bar F^0\nonumber\\
& - & B_1 \bar f ^0 - \frac {1}{2} \bar B^0 \bar f + \bar B^{0i} \bar f_i+ \frac {1}{2}  (\pm\,m\,\bar \beta^0 - \partial^0 \bar \beta)\, \bar F 
- (\partial^0 \bar \beta ^i - \partial^i \bar \beta ^0)\bar F_i \Big], 
\end{eqnarray}
where the above form of the nilpotent anti-BRST charge has been derived using the interplay of 
(i) the appropriate EL-EoMs derived from ${\cal L}_{\bar B}$, (ii) the Gauss divergence theorem, and 
(iii) the off-shell nilpotent anti-BRST symmetry transformations $s_{ab}$ [cf. Eq. (9)]
at appropriate places.

We observe that the physicality criteria 
(i.e. $Q_{ab}^{(1)} | phys > = 0$) w.r.t. $Q_{ab}^{(1)}$, leads to the following conditions  on the physical state (i.e.  $| phys >  = 0$)
with the help of the
first {\it four} terms, namely; 
\begin{eqnarray*}
\bar B_{ij} \, | phys >  = 0  &\Longrightarrow &  \Pi^{0ij}_{(A)} \, | phys >  = 0,  \nonumber\\
\bar B_{i} \, | phys >  = 0 &\Longrightarrow & \Pi^{0i}_{(\phi)} \, | phys >  = 0,  \nonumber\\
\end{eqnarray*}
\begin{eqnarray}
 (\partial^0 \bar B^{ij} + \partial^i \bar B^{j0} + \partial^j \bar B^{0i}) \, | phys >  = 0  &\Longrightarrow & 
\Big (\partial_k  \Pi^{kij}_{(A)} \mp   \frac{m}{3}  \Pi^{ij}_{(\phi)}  \Big)  \, | phys >  = 0, \nonumber\\
\big[\pm m\, \bar B^{0i} - (\partial^0 \bar B^i - \partial^i \bar B^0) \big] \, | phys > = 0  &\Longrightarrow & 
\bigl (\partial_i\, \Pi^{ij}_{(\phi)} \big )\, |phys > = 0, 
\end{eqnarray}
where we have used the following definitions of the canonical conjugate momenta w.r.t. 
the basic fields $A_{\mu\nu\lambda}$ and $ \Phi_{\mu\nu}$ for our theory described by the Lagrangian density  ${\cal L}_{\bar B}$, namely; 
\begin{eqnarray}
\Pi^{\mu \nu \lambda}_{(A)} & = & \frac{\partial\, {\cal L}_{\bar B}}{\partial (\partial_0\, A_{\mu \nu \lambda })} =  
 \frac {1}{3}\, H^{0\mu\nu\lambda}
 - \frac {1}{3}\,(\eta^{\mu 0}\,\bar B^{\nu\lambda} + \eta^{\nu 0}\, \bar B^{\lambda\mu} + \eta^{\lambda 0}\, \bar B^{\mu\nu}),\nonumber\\
\Pi^{\mu \nu }_{(\phi)} & = & \frac{\partial\, {\cal L}_{\bar B}}{\partial (\partial_0\, \Phi_{\mu \nu})} =  -\, \Sigma^{0\mu\nu} \pm m\,A^{0\mu\nu}
+ \frac {1}{2}\, (\eta^{\mu0}\,\bar B^{\nu} - \eta^{\nu0}\, \bar B^\mu).
\end{eqnarray} 
From the above, it is clear that we have the following explicit independent components of the 
canonical conjugate momenta (derived from the Lagrangian density ${\cal L}_{\bar B} $):
\begin{eqnarray}
&&\Pi^{0ij}_{(A)}  =  \frac {1}{3}\, \bar B_{ij}, \qquad  \qquad \; 
\Pi^{0i}_{(\phi)}  =    -\, \frac {1}{2}\, \bar B_i, \nonumber\\
&&\Pi^{ijk}_{(A)}  =    \frac {1}{3}\, H^{0ijk} \qquad \qquad 
\Pi^{ij}_{(\phi)}  =    -\, \Sigma^{0ij} \pm m\,A^{0ij}. 
\end{eqnarray}
In obtaining the result (63), we have also used the following EL-EoMs 
(derived from the Lagrangian density ${\cal L}_{\bar B}$),  namely;
\begin{eqnarray}
&&\partial_\mu\, H^{\mu\nu\lambda\xi} - (\partial^\nu\, \bar B^{\lambda\xi} 
+ \partial^\lambda \, \bar B^{\xi\nu} + \partial^\xi\, \bar B^{\nu\lambda}) = 0, 
\nonumber\\
&&\partial_\mu\, \Sigma^{\mu\nu\lambda}  \mp m\, (\partial_\mu\, A^{\mu\nu\lambda}) - \frac{1}{2}\, (\partial^\nu\,\bar  B^\lambda - \partial^\lambda\, \bar B^\nu)
\pm  \, \frac{m}{2}\, \bar B^{\nu\lambda}  = 0, 
\end{eqnarray}
with the specific choices: $\nu = 0, \lambda = j$ and $\xi = k$ which lead to the relationships between  the specific combinations of the 
derivatives on the Nakanishi-Lautrup type auxiliary fields and the secondary constraints [cf. Eq. (63)]. In fact, the last two entries of (63)
are due to the EL-EoMs (66) that are derived from the Lagrangian density ${\cal L}_{\bar B} $. 
We would like to add here that the physical states, in the physicality criterion (63) w.r.t. the nilpotent version 
of the anti-BRST charge (i.e. $Q_{ab}^{(1)} | phys > = 0$), remains unaffected by the presence of
(i) the basic anti-ghost fields, (ii) the auxiliary fields with specific non-zero ghost numbers, 
(iii) the ordinary derivatives acting on {\it both} of them. This happens because all such terms, 
present in the expression for the nilpotent anti-BRST charge $Q_{ab}^{(1)}$, do not act on the physical states. 
To be precise, the field operator with ghost number equal to {\it zero} act on the 
physical states. Such field operators turn out to be the coefficient of the {\it basic} 
(anti-)ghost fields and/or the derivatives on them. The field operators with ghost numbers 
equal to zero do {\it not} contribute to the physicality criteria if they are associated 
with the {\it auxiliary} (anti-)ghost fields. For instance,  the terms 
($\bar B^{0i}\, \bar f_i, \, -\, B_i\, \bar f^0, \, -\, \frac{1}{2}\, \bar B^0\, \bar f$)
in (62) do {\it not} contribute. This is why, we do not have: 
$\bar B^{0i} |phys> = 0, \, B_1|phys> = 0, \,$ and $\bar B^0|phys> = 0$
in (63).

We end our present section with a couple of crucial and clenching remarks. First of all, we have {\it not} 
taken the conditions:  ($B^{0i} | phys > = 0, \, B_1\, | phys > = 0, \, 
B^0\, | phys > = 0$) and  ($\bar B^{0i} | phys > = 0, \, B_1\, | phys > = 0, \, B^0\, | phys > = 0$) from the physicality 
criteria w.r.t. the nilpotent and conserved BRST charge (i.e. $Q_b^{(1)} |phys > = 0$
and nilpotent anti-BRST charge $Q_{ab}^{(1)} |phys > = 0$) because ($B^{0i}, \, B_1, \, B^0$) and ($\bar B^{0i}, \, \bar B_1, \, \bar B^0$)
are {\it not} associated with the {\it basic} (anti-)ghost fields (e.g. $C_{\mu\nu},\,  C_\mu, \, 
\bar C_{\mu\nu},\,  \bar C_\mu, \, \beta_\mu, \, \bar\beta_\mu$, etc.). Rather, we observe that the terms 
($B^{0i} f_i, \, B_1\, f^0, \, B^0\, f$) and ($\bar B^{0i}\, \bar f_i, \, B_1\, \bar f^0, \, B^0 \bar f$) are present in the expression for $Q_b^{(1)}$ and $Q_{ab}^{(1)}$, 
respectively. However, the fermionic (anti-)ghost  auxiliary fields $(\bar f_\mu)f_\mu$  and $(\bar f)f$ are {\it not}
the basic (anti-)ghost fields of our theory and they are also {\it not} independent in the sense that they obey the CF-type restrictions: 
$f + \bar F = m\, C_1, \; \bar f + F = m\, \bar C_1$.  
Furthermore, we observe, from the expressions for the  
following canonical conjugate momenta, derived from the Lagrangian densities 
${\cal L}_{ B} $ and ${\cal L}_{\bar B}$ w.r.t. the fields $\phi_{\mu}$ and $\phi$, namely; 
\begin{eqnarray}
&&\Pi^\mu_{(\phi)} = \frac{\partial\, {\cal L}_{B}}{\partial (\partial_0\, \phi_{\mu})} \; \; \equiv  \; \; B^{0\mu} - \eta^{0\mu}\, B_1, \nonumber\\
&&\Pi^\mu_{(\phi)} = \frac{\partial\, {\cal L}_{\bar B}}{\partial (\partial_0\, \phi_{\mu})} \; \; \equiv  \; \; \bar B^{0\mu} - \eta^{0\mu}\,  B_1, \nonumber\\
\nonumber\\
&&\Pi_{(\phi)} = \frac{\partial\, {\cal L}_{B}}{\partial (\partial_0\, \phi)} 
\; \; \equiv  \;\; 
-\, \frac{1}{2}\, B^0,  \nonumber\\
&&\Pi_{(\phi)} = \frac{\partial\, {\cal L}_{\bar B}}{\partial (\partial_0\, \phi)} \; \; \equiv  \; \;
-\, \frac{1}{2}\, \bar B^0, 
\end{eqnarray} 
that $\Pi^i_{(\phi)} = B^{0i}, \; \Pi^0_{(\phi)} = -\, B_1, \;
\Pi_{(\phi)} = -\, \frac{1}{2}\, B^0$ for the Lagrangian density ${\cal L}_B$
and  $\Pi^i_{(\phi)} = \bar B^{0i}, \, \Pi^0_{(\phi)} = -\, B_1, \;
\Pi_{(\phi)} = -\, \frac{1}{2}\,\bar B^0$ for the Lagrangian density ${\cal L}_{\bar B}$.
It is pertinent to point out that the subscript $(\phi)$ has been taken in the expressions
for the canonical conjugate momenta w.r.t. both the fields $\Phi_{\mu}$  and $\phi$. However, 
the contravariant indices differentiate them clearly.
It is pretty obvious that these components of momenta [$\Pi^\mu_{(\phi)}$], and $\Pi_{(\phi)}$ are {\it not} any 
type of constraints on our theory. Hence, we can {\it not} take $B^{0i} | phys > = 0, \, B_1 |phys> = 0, \, B^0\, |phys> = 0$,
w.r.t. the  nilpotent charge $Q_b^{(1)}$ and, similarly, the physicality criterion w.r.t. the 
anti-BRST charge $Q_{ab}^{(1)} | phys > = 0$ can {\it not} produce $\bar B^{0i} | phys > = 0, \, B_1 |phys> = 0, \, \bar B^0\, |phys> = 0$,
as one of the quantization conditions on the physical states (i.e. $|phys>$) existing in the total quantum Hilbert 
space of states of our {\it modified} massive Abelian 3-form theory in any arbitrary D-dimension of spacetime.
The second very important point is the observation that the Dirac quantization conditions, for our
D-dimensional {\it modified} massive Abelian 3-form theory, are precisely satisfied through the physicality
criteria w.r.t. the conserved and off-shell nilpotent (anti-)BRST charges (i.e. $Q_{(a)b}^{(1)} | phys > = 0$)
because we observe that the quantization conditions on the physical states are the {\it same}  [cf. Eqs. (63), (60)]
in terms of the primary and secondary first-class constraints (of our present theory which is endowed with a set of {\it four}
first-class constraints).

\section{Conclusions}

In our present investigation, we have taken a D-dimensional St$\ddot u$ckelberg-modified {\it massive} Abelian 3-form theory and 
proven it to be an example of a massive model of a gauge theory because of the presence of the first-class constraints on it
(cf. Sec. 3). The {\it latter} generate the gauge symmetry transformations [cf. Eqs. (4), (21)] for the St$\ddot u$ckelberg-modified
Lagrangian density (1) where the mass and gauge symmetry co-exist.
 The {\it classical} infinitesimal, local and continuous gauge symmetry transformations have been elevated to 
their {\it quantum} counterparts as the (anti-)BRST symmetry transformations [cf. Eqs. (9), (7)] which are respected by a set of coupled 
(but equivalent)  Lagrangian densities. These {\it latter} set of Lagrangian densities are nothing but the generalizations
of the classical St$\ddot u$ckelberg-modified Lagrangian density (1). 
The existence of the coupled (but equivalent) (anti-)BRST invariant Lagrangian densities 
[cf. Eqs. (5), (6)] owe their origin to the existence of the (anti-)BRST invariant Curci-Ferrari restrictions [cf. Eq. (29)]
which are {\it also} responsible for the absolute anticommutativity property of the nilpotent (anti-)BRST transformations 
and corresponding conserved and nilpotent (anti-)BRST charges (cf. Appendix B).

The central objective of our present endeavor has been to derive the Noether conserved currents 
and corresponding charges which are the generators for the infinitesimal and continuous BRST, 
anti-BRST and ghost-scale symmetry transformations. We have derived the {\it standard} BRST algebra (cf. Appendix C)  
from the ideas and concepts behind the continuous symmetry transformations and their generators (according to 
Noether's theorem in the context of quantum field theory).
The BRST algebra physically implies $(i)$ the off-shell nilpotency of the (anti-)BRST conserved charges which establishes their {\it fermionic} nature, 
$(ii)$ the absolute anticommutativity of the (anti-)BRST conserved charges (and, hence, their independent identities), and 
$(iii)$ the ghost number of a field increases by {\it one} by the BRST symmetry transformations generated by the BRST
charge. On the contrary, the ghost number of a field decreases by {\it one} by the anti-BRST symmetry transformations that 
are generated by the conserved anti-BRST charge. These claims are {\it true} for any arbitrary (anti-)BRST invariant theory in 
any arbitrary dimension of spacetime [cf. Eq. (C.2)].

One of the highlights of our present investigations is the derivation of the (anti-)BRST invariant CF-type restrictions
from the proof of the absolute anticommutativity of the conserved and nilpotent 
(anti-)BRST charges [$Q_{(a)b}^{(1)}$]. These restrictions automatically appear when we demand 
the absolute anticommutativity [cf. Eq. (11)] of the (anti-)BRST symmetry transformations that are listed in (9) and (7), respectively. 
It is worthwhile to point out that the CF-type restrictions (29) {\it also} appear when we demand the equivalence of the 
coupled Lagrangian densities (5) and (6) w.r.t. the (anti-)BRST symmetry transformations. This proof has been 
established in our earlier recent work [18]. In our Appendix A, we have shown the direct {\it equality} of the total Lagrangian densities
${\cal L}_{B}$ and ${\cal L}_{\bar B}$ in a straightforward manner without any consideration of symmetry transformations and/or principles
where, once again, the CF-type restrictions (29) appear in a beautiful fashion. 
In addition, the (anti-)BRST invariant CF-type restrictions (29) have {\it also} been derived 
from the EL-EoMs (w.r.t. the fermionic/bosonic auxiliary fields) from the coupled (but equivalent) 
Lagrangian densities ${\cal L}_{B}$ and ${\cal L}_{\bar B}$, 
respectively (cf. Sec. 4 for details). It is worthwhile to mention here that the {\it massless} limit of 
the CF-type restrictions [cf. Eq. (29)] has been found in our earlier work [32] in the context of the
BRST approach to the D-dimensional Abelian 3-form {\it gauge} theory. 
The existence of the (non-)trivial CF-type restrictions is one of the hallmarks of the BRST-quantized gauge/diffeomorphism invariant
theories. The D-dimensional Abelian 1-form theory is endowed with a {\it trivial} CF-type 
restriction. However, the {\it latter} turns out to be the limiting case of the non-trivial celebrated CF-condition [19] 
that exists for the BRST-quantized D-dimensional non-Abelian 1-form theory [30, 31].

It is gratifying to state that we have established a deep connection between the first-class constraints 
and the Noether conserved charge at the {\it classical} level (cf. Sec. 3 for details) for the 
St$\ddot u$ckelberg-modified massive Abelian 3-form theory which is described by the Lagrangian density (1).
On the contrary, at the {\it quantum} level, we find that the standard Noether conserved (anti-)BRST charges 
$(Q_{(a)b})$, derived from the infinitesimal, continuous and nilpotent (anti-)BRST 
symmetry transformations [cf. Eq. (9), (7)],  are found to be {\it non-nilpotent}.
Furthermore, the physicality criteria ($Q_{(a)b} |phys > = 0$) w.r.t. these standard Noether (anti-)BRST 
charges produce specific conditions on the physical states (i.e. $|phys> $) that are {\it not} complete and consistent 
with Dirac's quantization conditions [cf. Eqs. (59), (61)]. Thus, we have derived the off-shell nilpotent versions of the 
(anti-) BRST charges [cf. Eqs. (62), (53)] by exploiting the theoretical proposal in our earlier work [22]. 
The physicality criteria (i.e. $Q_{(a)b}^{(1)} |phys > = 0$) 
w.r.t. these nilpotent  (i.e. $[Q_{(a)b}^{(1)}]^2  = 0)$  (anti-BRST) charges $[Q_{(a)b}^{(1)}]$
produce the quantization conditions [cf. Eqs. (63), (60)]
which are found to be consistent with Dirac's quantization conditions [cf. Eq. (52)] where the 
operator forms of the first-class constraints annihilate the physical states (i.e. $|phys> $) of the quantum version of our theory.

It will be an interesting future project to capture the off-shell nilpotency and absolute anticommutativity of the 
conserved (anti)-BRST charges within the framework of superfield approach to BRST formalism. The central goal of our
present investigation (and earlier one on the modified {\it massive} Abelian 3-form theory [18]) is to show that the  
St$\ddot u$ckelberg-modified massive 6D Abelian 3-form theory is a {\it massive} model of Hodge theory where 
$(i)$ the standard St$\ddot u$ckelberg technique gets modified, $(ii)$ the off-shell nilpotent (anti-)co-BRST
symmetries exist, too, in addition to the off-shell nilpotent (anti-)BRST symmetry transformations, and $(iii)$
there are new ``exotic" fields that are introduced in the theory which possess {\it negative} kinetic terms. 
These ``exotic" fields, as mentioned earlier, are now-a-days very popular in the context of the cyclic, bouncing and self-accelerated cosmological 
models of the Universe where they have been christened as the ``ghost" and/or ``phantom" particles/fields [13-15]. 
These ``exotic" fields also provide a set of possible candidates of dark matter [16, 17] as they obey standard Klein-Gordon equation
with a well-defined rest mass.
It is obvious that the {\it massless} limit of the ``exotic" fields would represent the idea of dark energy where there is existence of 
 fields with only {\it negative} kinetic terms (with their rest mass equal to zero). Such fields automatically lead to the generation of {\it negative}
pressure which is one of the key characteristic features of the dark energy. The Abelian 3-form theory has been studied within the 
framework of the generalized BRST and BV formulation [33, 34] using the superspace technique. Furthermore, the BRST
analysis of the ABJM theory (connected with the M-theory) has also been performed [35]. It will be a nice future endeavor to look at these
works [33-35] in the light of our present investigation in terms of the constraints and conserved charges.\\

\noindent
{\bf Acknowledgments}\\

\noindent
One of us (AKR) gratefully acknowledges the financial support from the {\it BHU-fellowship program} of the Banaras Hindu University (BHU),
Varanasi, under which the present investigation has been carried out. Fruitful discussions with Dr. B. Chauhan, on 
the relevant issues of our present work, are thankfully acknowledged, too.
Useful and thought-provoking comments by our esteemed Reviewer are gratefully acknowledged which have been responsible for the improvement in the quality as well as clarity of our presentation.\\


\noindent
{\bf Data Availability}\\

\noindent
No data were used to support this study.\\

\noindent
{\bf Conflicts of Interest}\\

\noindent
The authors declare that there are no conflicts of interest.\\

\begin{center}
{\bf Appendix A: On the Equality of the Coupled Lagrangian Densities and the (Anti-)BRST Invariant CF-Type Restrictions}\\
\end{center}

\noindent
As pointed out earlier, the root-cause behind the existence of the {\it coupled} Lagrangian densities 
${\cal L}_{B}$ and ${\cal L}_{\bar B}$  [cf. Eqs. (5), (6)] is the set of {\it six} (anti-) BRST invariant CF-type restrictions 
(29) that incorporate into themselves  a set of {\it four} fermionic and {\it two} bosonic type  relationships. 
The purpose of our present Appendix is to demonstrate the existence of the {\it six} (anti-)BRST invariant CF-type restrictions {\it directly} from the 
equality of ${\cal L}_{B}$ and ${\cal L}_{\bar B}$. In other words, we plan to show that 
${\cal L}_{ B} - {\cal L}_{ \bar B} = 0$ provided {\it all} the {\it six}  CF-type restrictions are satisfied. 
Towards this goal in mind, first of all, we note that there is a {\it common} part  [${\cal L}^{(C)}$] that is present 
in the (anti-)BRST invariant Lagrangian densities ${\cal L}_{B}$ and ${\cal L}_{\bar B}$ 
[cf. Eqs. (5), (6)] as follows
\[
{\cal L}^{(C)} = {\cal L}_{S}^{(A)} 
+ \frac{m^2}{2} \bar C_{\mu\nu} C^{\mu\nu}
 +(\partial_\mu \bar C_{\nu\lambda} + \partial_\nu \bar C_{\lambda\mu} + 
\partial_\lambda \bar C_{\mu\nu}) (\partial^\mu C^{\nu\lambda}) 
\]
\[
~~~~~+ (\partial_\mu\, \bar C_\nu - \partial_\nu\, \bar C_\mu)\, (\partial^\mu \ C^\nu)
- \frac{1}{2}\, \Big[\pm m\, \bar \beta^\mu 
- \partial^\mu\, \bar\beta \Big]\, \Big[\pm m\,  \beta_\mu - \partial_\mu\, \beta \Big]
\]
\[
\pm \, m\, \bar C^\nu\, (\partial^\mu\, C_{\mu\nu})
-\,  (\partial_\mu \, \bar\beta_\nu - \partial_\nu\, \bar\beta_\mu)\, (\partial^\mu\,\beta^\nu) ~~~~~~~~~~~~~~~~~~~~~~~~~~
\]
\[
-   \partial_\mu \bar C_2 \partial^\mu  C_2
 - \, m^2\, \bar C_2\, C_2  \pm m (\partial_\mu \bar C^{\mu\nu}) C_\nu  
  -\, B\, B_2  -\, \frac{1}{2}\, B_1^2 ~~~~~~~~
\]
\[
+ [(\partial \cdot \bar\beta) \mp m\, \bar \beta]\, B
 - [(\partial \cdot \phi) \mp m\, \phi]\, B_1
 - [(\partial \cdot \beta) \mp m\, \beta]\,  B_2, ~~~
\eqno (A.1)
\]
where the superscript $(C)$, on the l.h.s. of the Lagrangian density, denotes  the {\it common} part
of the (anti-)BRST invariant Lagrangian densities ${\cal L}_{B}$ and ${\cal L}_{\bar B}$. It is obvious that 
 ${\cal L}^{(C)}$ will {\it cancel} out in the mathematical 
proof of ${\cal L}_{ B} - {\cal L}_{ \bar B} = 0$.

Against the backdrop of the above paragraph, we note that there are differences in ${\cal L}_{ B}$ and ${\cal L}_{ \bar B}$
$(i)$ in the {\it bosonic} sector of the {\it physical} gauge field $A_{\mu \nu \lambda }$, St$\ddot u$ckelberg field $\Phi_{\mu\nu}$
as well as the auxiliary fields associated with {\it them}, and $(ii)$ in the {\it fermionic} ghost-sector where the auxiliary fields 
$F_\mu, \; \bar F_\mu, \; f_\mu, \; \bar f_\mu, \; F, \; \bar F, \, f $ and $\bar f$ are present. First of all, we focus on the
{\it physically} important  bosonic sector of the coupled Lagrangian densities where we have: 
\[
{\cal L}_{ B}^{(b)} - {\cal L}_{ \bar B}^{(b)} =  (\partial_\mu A^{\mu\nu\lambda}) \, (B_{\nu\lambda} + \bar B_{\nu\lambda})
 + \frac{1}{2}\, (\bar B_{\mu\nu}\, \bar B^{\mu\nu} -  B_{\mu\nu}\, B^{\mu\nu})
\]
\[
~~~~~~ \mp \, \frac{m}{2} \Phi_{\mu\nu} \, (B^{\mu\nu} +  \bar B^{\mu\nu})
- \, (\partial_\mu\, \Phi^{\mu\nu})\, (B_\nu + \bar B_\nu) + \frac{1}{2}\, (\bar B_{\mu} \bar B^{\mu} -  B_{\mu} \, B^{\mu})
\]
\[
 + \frac{1}{2}[ \pm m\, \Phi^\mu - \partial^\mu \phi] (B_\mu - \bar B_\mu)
 + \frac{1}{2}( \partial_\mu  \phi_\nu - \partial_\nu \phi_\mu) (B^{\mu\nu} -  \bar B^{\mu\nu}).
\eqno (A.2)
\]
In the above, the superscript $(b)$ on the Lagrangian densities (on the l.h.s.) denote the {\it bosonic} sector of the coupled 
(but equivalent) Lagrangian densities (5) and (6). 
We would like to point out, at this juncture, that we have freedom to add/subtract the {\it total} spacetime derivative term(s)
to (A.2) because the {\it latter} is a part of the {\it physical} Lagrangian density. Thus, it is elementary to note that 
{\it first} and {\it fifth} terms of (A.2) can be written as 
$(\partial_\mu A^{\mu\nu\lambda})\,  [B_{\nu\lambda} + \bar B_{\nu\lambda} -\, (\partial_\nu \, \phi_\lambda 
- \partial_\lambda \, \phi_\nu)]  - \, (\partial_\mu\, \Phi^{\mu\nu})\, [B_\nu + \bar B_\nu + \partial_\nu\, \phi]$
 without  {\it changing} the corresponding action integral due to the celebrated Gauss's divergence theorem. Using the following 
simple and straightforward algebraic tricks, namely; 
\[
 \frac{1}{2}\, (\bar B_{\mu\nu} \,  \bar B^{\mu\nu} -  B_{\mu\nu}\, B^{\mu\nu}) =  
\frac{1}{2}\, (\bar B^{\mu\nu}-  B^{\mu\nu})\, (\bar B_{\mu\nu} +  B_{\mu\nu}), 
\]
\[
 \frac{1}{2}\, (\bar B_{\mu} \,  \bar B^{\mu} -  B_{\mu}\, B^{\mu}) =  
\frac{1}{2}\, (\bar B^{\mu}-  B^{\mu})\, (\bar B_{\mu} +  B_{\mu}), ~~~~~~~~~~
\eqno (A.3)
\]
the above expression for ${\cal L}_{ B}^{(b)} - {\cal L}_{ \bar B}^{(b)}$ [cf. Eq. (A.2)] can be re-expressed as
\[
{\cal L}_{ B}^{(b)} - {\cal L}_{ \bar B}^{(b)} = \big(\partial_\lambda A^{\lambda\mu\nu}\big) \, 
\Big[B_{\mu\nu} + \bar B_{\mu\nu} - (\partial_\mu \, \phi_\nu - \partial_\nu \, \phi_\mu)\Big]
\]
\[
 + \frac{1}{2}\, \big(\bar B^{\mu\nu}-  B^{\mu\nu}\big)\,\Big [B_{\mu\nu} + \bar B_{\mu\nu} - (\partial_\mu \, \phi_\nu - \partial_\nu \, \phi_\mu)\Big]
\]
\[
\pm \frac{m}{2} \Phi_{\mu\nu}  ( \partial^\mu  \phi^\nu - \partial^\nu \phi^\mu)
 - \, (\partial_\mu\, \Phi^{\mu\nu})\, [B_\nu + \bar B_\nu + \partial_\nu\, \phi]
\mp \frac{m}{2}\, \Phi_{\mu\nu} \, (\bar B^{\mu\nu} + B^{\mu\nu})
\]
\[
 \mp \frac{m}{2}\, \Phi_{\mu\nu} \, ( \partial^\mu \, \phi^\nu - \partial^\nu\, \phi^\mu)
+ \frac{1}{2}\, (\bar B^{\mu}-  B^{\mu})\, [B_{\mu} +  \bar B_{\mu} - (\pm m\, \phi_\mu - \partial_\mu\, \phi)],
\eqno (A.4)
\]
where we have added 
and subtracted [$\pm \frac{m}{2} \, (\Phi_{\mu \nu })\, \{ (\partial^\mu\, \phi^\nu - \partial^\nu\, \phi^\mu)\}$] in the above for the algebraic convenience. 
Re-arranging the whole equation, we have the following form of the r.h.s. of the above equation, namely;
\[
\big(\partial_\lambda A^{\lambda\mu\nu}\big) \, \Big[B_{\mu\nu} + \bar B_{\mu\nu} - (\partial_\mu \, \phi_\nu - \partial_\nu \, \phi_\mu)\Big]
\]
\[
~~~~+ \frac{1}{2}\, \big(\bar B^{\mu\nu}-  B^{\mu\nu}\big)\,\Big[B_{\mu\nu} + \bar B_{\mu\nu} - (\partial_\mu \, \phi_\nu - \partial_\nu \, \phi_\mu)\Big]
\]
\[
\mp \frac{m}{2}\, \Phi^{\mu\nu}   \, \Big[B_{\mu\nu} + \bar B_{\mu\nu} - (\partial_\mu \, \phi_\nu - \partial_\nu \, \phi_\mu)\Big] ~~~~~~~
\]
\[
- \, (\partial_\nu\, \Phi^{\nu\mu}) \, \Big[B_{\mu} +  \bar B_{\mu} - (\pm m\, \phi_\mu - \partial_\mu\, \phi)\Big]~~~~
\]
\[
+ \frac{1}{2}\, (\bar B^{\mu}-  B^{\mu})\, \Big[B_{\mu} +  \bar B_{\mu} - (\pm m\, \phi_\mu - \partial_\mu\, \phi)\Big], 
\eqno (A.5)
\]
where the explicit expressions for the bosonic CF-type restrictions appear [cf. Eq. (29)]. 
The above equation can be re-written, in a more compact form,  as: 
\[
\Big[(\partial_\lambda A^{\lambda\mu\nu}) + \frac{1}{2}\, (\bar B^{\mu\nu}-  B^{\mu\nu}) \mp \frac{m}{2}\, \Phi^{\mu\nu} \Big]\, 
\Big[B_{\mu\nu} + \bar B_{\mu\nu} - (\partial_\mu \, \phi_\nu - \partial_\nu \, \phi_\mu)\Big]
\]
\[
+ \Big[ \frac{1}{2}\, (\bar B^{\mu}-  B^{\mu}) - \, (\partial_\nu\, \Phi^{\nu\mu})  \Big] \, 
\Big[ B_{\mu} +  \bar B_{\mu} - (\pm m\, \phi_\mu - \partial_\mu\, \phi)   \Big]. ~~~~~~~~~~~~~
\eqno (A.6)
\]
Thus, we observe that ${\cal L}_{ B}^{(b)} - {\cal L}_{ \bar B}^{(b)} = 0$ provided we invoke the validity of the {\it two} bosonic type of 
CF-type restrictions that have been quoted in (29), namely;
\[
B_{\mu\nu} + \bar B_{\mu\nu} = \partial_\mu \, \phi_\nu - \partial_\nu \, \phi_\mu, \qquad 
 B_{\mu} +  \bar B_{\mu} = \pm m\, \phi_\mu - \partial_\mu\, \phi.
\eqno (A.7)
\]
We draw the conclusion that the bosonic (physical) sector of the Lagrangian 
densities ${\cal L}_{ B}$ and ${\cal L}_{ \bar B} $
are {\it equal} (and {\it equivalent} w.r.t. the (anti-)BRST symmetries [18]) 
on the submanifold of the Hilbert space where the specific {\it quantum} fields satisfy (A.7).

Let us focus on the {\it fermionic} (anti-)ghost part of the Lagrangian densities ${\cal L}_{ B}$ and ${\cal L}_{ \bar B} $.
It turns out that we have the following for  ${\cal L}_{ B}^{(f)} - {\cal L}_{ \bar B}^{(f)}$, namely;
\[
 -\, 2 \, F\, f - \, 2  \, \bar F\, \bar f + \frac{1}{2}\, (\partial \cdot C)\, [\bar f + F] \mp m\, C_1\, [F - \bar f]
 -\,  \frac{1}{2}\, (\partial \cdot \bar C)\, [ f + \bar F]
\]
\[
\pm m\, \bar C_1\, [f - \bar F] - \, 2\, F^\mu\,f_\mu - \, 2\, \bar F^\mu\,\bar f_\mu 
+ [(\partial_\nu \bar C^{\nu\mu}) \mp \frac{m}{2}\, \bar C^\mu]\, (f_\mu + \bar F_\mu)
\]
\[
- [(\partial_\nu  C^{\nu\mu}) \mp \frac{m}{2}  C^\mu] (\bar f_\mu + F_\mu)
+ (\partial^\mu \bar C_1)\, [f_\mu - \bar F_\mu] + (\partial^\mu C_1) [\bar f_\mu - F_\mu], 
\eqno (A.8)
\]
where the superscript $(f)$ on the Lagrangian densities ${\cal L}_{ B}^{(f)}$ and ${\cal L}_{ \bar B}^{(f)}$
denotes a part of the  {\it fermionic} ghost sector of the {\it coupled} Lagrangian densities. At this stage, let us, first of all, 
concentrate on the following part of the above equation, namely;
\[
-\, 2 \, F\, f - \, 2  \, \bar F\, \bar f + \frac{1}{2}\, (\partial \cdot C)\, [\bar f + F] 
 -\,  \frac{1}{2}\, (\partial \cdot \bar C)\, [ f + \bar F]  
 \]
 \[
 \mp m\, C_1\, [F - \bar f]
\pm m\, \bar C_1\, [f - \bar F].
\eqno (A.9)
\]
It can be seen that the substitutions of the CF-type restrictions
\[
f + \bar F = \pm m\, C_1, \qquad  \bar f + F = \pm m\, \bar C_1,
\eqno (A.10)
\]
into (A.9) leads to the following: 
\[
\pm \frac{m}{2}\, (\partial \cdot C )\, \bar C_1 \mp \frac{m}{2}\, (\partial  \cdot \bar C)\, C_1.
\eqno (A.11)
\]
We would like to point out that the substitutions of (A.10) into (A.9) results in the cancellation of the 
{\it first} two terms with the {\it last} two terms of the {\it latter}.

We are now in the position to concentrate on the remaining part of (A.8) which is:
\[
- \, 2\, F^\mu\,f_\mu - \, 2\, \bar F^\mu\,\bar f_\mu 
+ [(\partial_\nu \bar C^{\nu\mu}) \mp \frac{m}{2}\, \bar C^\mu]\, (f_\mu + \bar F_\mu)
-\, [(\partial_\nu  C^{\nu\mu}) 
\]
\[
\mp \frac{m}{2}\,  C^\mu]\, (\bar f_\mu + F_\mu)
+ (\partial^\mu\, \bar C_1)\, [f_\mu - \bar F_\mu] + (\partial^\mu\, C_1)\, [\bar f_\mu - F_\mu]. 
\eqno (A.12)
\]
The substitutions of the following CF-type restrictions
\[
f_\mu + \bar F_\mu = \partial_\mu \, C_1, \qquad  \bar f_\mu + F_\mu = \partial_\mu \, \bar C_1,
\eqno (A.13)
\]
into the above equation lead to $(i)$ the  cancellation of the 
{\it first} two terms with the {\it last} two terms, and $(ii)$  the observations that 
$(\partial_\nu  \bar C^{\nu\mu})\, \partial_\mu \, C_1 \equiv \partial_\mu\, [(\partial_\nu  \bar C^{\nu\mu})\, C_1]$ and 
$(\partial_\nu   C^{\nu\mu})\, \partial_\mu \, \bar C_1 \equiv \partial_\mu\, [(\partial_\nu  C^{\nu\mu})\,\bar C_1]$ 
are the {\it total} spacetime derivatives that do {\it not} play any significant role in the dynamics of our theory. Hence, they can be ignored 
in the action integral due to the Gauss divergence theorem. Ultimately, we obtain [from (A.12)] the following 
\[ 
\mp \frac{m}{2}\, \bar C^\mu\, \partial_\mu\, C_1 \pm \frac{m}{2}\,  C^\mu\, \partial_\mu\, \bar C_1,
\eqno (A.14)
\]
which cancels with (A.11) modulo a couple of  total spacetime derivatives.

We end this Appendix with the concluding remarks that the straightforward requirement of the 
{\it equality} between ${\cal L}_{ B}$ and ${\cal L}_{ \bar B} $ demonstrates that these Lagrangian densities are 
equivalent on the submanifold of the quantum Hilbert space of fields where,  modulo some total spacetime derivatives, the 
CF-type restrictions (29) are satisfied. 
The proof of the {\it two} bosonic CF-type restrictions:
$B_{\mu\nu} + \bar B_{\mu\nu} = \partial_\mu \, \phi_\nu - \partial_\nu \, \phi_\mu$ and 
 $B_{\mu} +  \bar B_{\mu} = \pm m\, \phi_\mu - \partial_\mu\, \phi$  has been demonstrated quite explicitly in
the equality ${\cal L}_{B}^{(b)} $ - ${\cal L}_{ \bar B}^{(b)} = 0$  [cf. Eqs. (A.6), (A.7)] which is valid for the {\it bosonic} sector.  
 However, the {\it fermionic} four CF-type restrictions: 
 $f + \bar F = \pm m\, C_1, \;  \bar f + F = \pm m\, \bar C_1, \; 
f_\mu + \bar F_\mu = \partial_\mu \, C_1$ and   $\bar f_\mu + F_\mu = \partial_\mu \, \bar C_1$ are {\it hidden} in a {\it subtle} manner 
when we prove:  ${\cal L}_{ B}^{(f)} - {\cal L}_{ \bar B}^{(f)} = 0$ for the coupled 
Lagrangian densities in their {\it fermionic} sector. This is why there is cancellation between (A.11) and (A.14). However, there
is also an explicit way to demonstrate the existence of the above {\it four}  fermionic CF-type restrictions by re-arranging terms
in (A.8) and throwing away all the total spacetime derivative terms. Ultimately, we can show that 
${\cal L}_{ B}^{(f)} - {\cal L}_{ \bar B}^{(f)}$, modulo some total spacetime derivative terms,  is equal to the following
explicit expression:
\[
\Big[\frac{1}{2}\, (\partial\cdot C) 
\pm \, m\, C_1 -\, 2\, \bar F \Big]\, \big(\bar f + F \mp\, m\, \bar C_1 \big)
\]
\[
-\, \Big [\frac{1}{2}\, (\partial\cdot \bar C) 
\mp \, m\, \bar C_1 + 2\, F \Big]\, \big(f + \bar F \mp\, m\, C_1 \big) ~~~~~~~~~~~~~
\]
\[
+ \big [\partial_\nu\, \bar C^{\nu\mu} \mp\, (\frac{m}{2})\, \bar C^\mu  + \partial^\mu\, \bar C_1 - 2\, F^\mu  \big]\, \big(f_\mu + \bar F_\mu - \partial_\mu\, C_1  \big)
\]
\[
- \big [\partial_\nu\, C^{\nu\mu} \mp \, (\frac{m}{2})\, C^\mu  - \partial^\mu\, C_1 + 2\, \bar F^\mu  \big]\, \big(\bar f_\mu + F_\mu - \partial_\mu\, \bar C_1  \big).  
\eqno (A.15)
\]
The above equation demonstrates that we have the equality of the fermionic sector (${\cal L}_{ B}^{(f)} - {\cal L}_{ \bar B}^{(f)} = 0$)
of the coupled Lagrangian densities if and only if {\it all} the four fermionic CF-type restrictions of (29) are satisfied.\\

\begin{center}
{\bf Appendix B: On the Absolute Anticommutativity of the Nilpotent (Anti-)BRST Charges: CF-Type Restrictions}\\
\end{center}

\noindent
The purpose of this Appendix is to establish the existence of a set of 
{\it six} CF-type restrictions on our theory from the requirement of the 
absolute anticommutativity property of the off-shell nilpotent versions of the
(anti-) BRST charges [cf. Eqs. (62), (53)]. In other words, the theoretical tricks we apply for this purpose are
as follows
\[
s_b\, Q_{ab}^{(1)}  = -\, i\, \{Q_{ab}^{(1)}, \, Q_{b}^{(1)} \}  = 0 
\Longrightarrow Q_{ab}^{(1)}\, Q_{b}^{(1)} + Q_{b}^{(1)}\, Q_{ab}^{(1)} = 0,
\]
\[
 s_{ab}\, Q_{b}^{(1)}  = -\, i\, \{Q_{b}^{(1)}, \, Q_{ab}^{(1)} \}  = 0 
 \Longrightarrow  Q_{b}^{(1)}\, Q_{ab}^{(1)} + Q_{ab}^{(1)}\, Q_{b}^{(1)} = 0,
\eqno (B.1)
\]
where one can find the  proper conditions under which the absolute anticommutativity property between the nilpotent 
versions of the BRST and anti-BRST charges (B.1) are precisely  satisfied. Towards, this goal in mind, 
we shall focus on the explicit computation of the l..h.s. of the above equations by the direct applications of the 
(anti-)BRST symmetry transformations [cf. Eqs. (9), (7)] on the explicit expressions for the nilpotent 
versions of the BRST charge [cf. (53)] and the anti-BRST charge [cf. Eq. (62)], respectively.

Now we concentrate on the application of the BRST symmetry transformations on the explicit expression for the 
nilpotent anti-BRST charge $Q_{ab}^{(1)}$. This leads to the following:  
\[
s_b\, Q_{ab}^{(1)}  = \int  d^{D-1}\, x \Big[ 
\big(\partial^0\, \bar B^{ij} + \partial^i\, \bar B^{j0} + \partial^j\, \bar B^{0i}\big)\, B_{ij} 
- \big(\partial^0\, B^{ij} + \partial^i\,  B^{j0} 
\]
\[
~~~~~~+ \partial^j\, B^{0i}\big)\, \bar B_{ij}
- \, 2\,  \big(\partial^0\, \bar C^{ij} + \partial^i\, \bar C^{j0} 
+ \partial^j\, \bar C^{0i}\big)\, \big(\partial_i\, \bar F_j\big) + \big(\pm m\, \bar F^0 - \partial^0\, \bar F\big)\, F
\]
\[
 + 2\,  \big(\partial^0\, C^{ij} + \partial^i\,  C^{j0} + \partial^j\, C^{0i}\big)\, \big(\partial_i\, F_j\big) 
 - \big(\partial^0 \bar F^0 - \partial^i\, \bar F^0\big)\, \bar f_i ~~~~~~~~~~~~~~~
\]
\[
 -2\, \big(\partial^0\, \bar F^i - \partial^i\, \bar F^0\big)\, F_i 
+ 2\,\big (\partial^0\, \beta^i - \partial^i\, \beta^0\big)\, \big(\partial_i\, B_2\big)~~~~~~~~~~~~~~~~~~~~~~~~~~
\]
\[
~~~~~~\mp m\,\big [\pm \, m\, C^{0i} - \big(\partial^0 \,  C^i - \partial^i\, C^0\big)\big]\, F_i 
+ \big[\pm m\,  C^{0i} - \big(\partial^0\, C^i - \partial^i \,  C^0\big)\big]\, (\partial_i\, F) 
\]
\[
\pm m\, \big(\partial^0\, \bar F^i - \partial^i\, \bar F^0\big)\, \bar C_1 \mp \, m \bar B^{0i}\, B_i
\pm\, m\, \big(\partial^0\, \bar f^i - \partial^i\, \bar f^0\big)\, \bar C_i ~~~~~~~~~~~~~~~
\]
\[
+ \big(\partial^0\, \bar B^i - \partial^i\, \bar B^0\big)\, B_i \pm m\, B^{0i}\, \bar B_i - \big(\partial^0\, B^i - \partial^i\, B^0\big)\, \bar B_i ~~~~~~~~~~~~~~~~~~~~~
\]
\[
~~~~~+ \big[\pm m\, \bar C^{0i} - \big(\partial^0\, \bar C^i - \partial^i\, \bar C^0\big)\big]\, (\partial_i\, f)\,
\mp \, m\, \big[\pm m\, \bar C^{0i} - \big(\partial^0\, \bar C^i - \partial^i\, \bar C^0\big)\big]\, f_i
\]
\[
~~~~~\mp\, m \big(\pm m\, \beta^0 - \partial^0 \beta\big)\, B_2 + B\, \dot B_2 
-\, 2\, \dot B\, B_2 + B_1\, \dot B_1 - \frac{1}{2} \big(\pm\, m\, f^0 
- \partial^0\, f\big)\, \bar f
\]
\[
- \bar B^{0i}\, \big(\partial_i\, B_1\big) \pm \frac{m}{2}\, \bar B^0\, B_1 
- \big(\partial^0\, F^i - \partial^i\, F^0\big)\, \bar F_i + \big(\partial^0\, \bar\beta^i - \partial^i\, \bar\beta^0\big)\, \partial_i\, B 
\]
\[
+ \frac{1}{2}\,\big [\pm m\, F^0 - \partial^0\, F\big]\, \bar F
\mp\, \frac{m}{2}\, \big(\pm m\, \bar\beta^0 - \partial^0\, \bar\beta\big)\, B\Big].~~~~~~~~~~~~~~~~~~~~~~~
\eqno (B.2)
\]
The stage is now set  to apply the Gauss divergence theorem and the appropriate EL-EoMs 
from the Lagrangian densities ${\cal L}_{B}$ and ${\cal L}_{\bar B}$, respectively, at appropriate 
places so that the above computation  of the operation of the BRST symmetry transformations $(s_b)$ on the 
nilpotent version of the anti-BRST charge $[Q_{ab}^{(1)}]$ can be completed 
and the necessary conditions can be found so that the absolute anticommutativity 
property is satisfied.

It is very interesting to note that {\it all} the terms containing  the auxiliary fields $(B, \, B_1, \, B_2)$
and the derivatives on them can be explicitly expressed as follows
\[
\int d^{D-1}\,x\, \Big[
(\partial^0\, \bar\beta^i - \partial^i\, \bar\beta^0)\, (\partial_i\, B)  
+ 2\, (\partial^0\,\beta^i - \partial^i\, \beta^0)\, (\partial_i\, B_2)
\]
\[
-\, \bar B^{0i} \, (\partial_i\, B_1) + B\, \dot B_2 - 2\, \dot B\, B_2 \pm \frac{m}{2}\, \bar B^0\, B_1 + B_1\, \dot B_1
\]
\[
\mp m\, (\pm m\, \beta^0 - \partial^0\, \beta)\, B_2
\mp \frac{m}{2}\, (\pm m \, \bar\beta^0 - \partial^0\, \bar\beta)\, B
\Big],
\eqno (B.3)
\]
where we can apply the Gauss divergence theorem on the {\it first} three 
terms of (B.3) and drop the total space derivative terms. 
This mathematical operation leads to the following form of the above terms 
that are present in the integral, namely;  
\[
\int d^{D-1}\,x\, \Big[
 -\, \partial_i\, (\partial^0\, \bar\beta^i - \partial^i\, \bar\beta^0)\, B  
- 2\, \partial_i\, (\partial^0\,\beta^i - \partial^i\, \beta^0)\,  B_2
\]
\[
+ (\partial_i\, \bar B^{0i}) \, B_1  + B\, \dot B_2 - 2\, \dot B\, B_2 \pm \frac{m}{2}\, \bar B^0\, B_1 + B_1\, \dot B_1
\]
\[
\mp m\, (\pm m\, \beta^0 - \partial^0\, \beta)\, B_2
\mp \frac{m}{2}\, (\pm m \, \bar\beta^0 - \partial^0\, \bar\beta)\, B
\Big].
\eqno (B.4)
\]
Using the following EL-EoMs from ${\cal L}_B$, namely; 
\[
\partial_\mu \, (\partial^\mu\, \beta^\nu - \partial^\nu\, \beta^\mu) - \partial^\nu\, B 
-\, \frac{m^2}{2}\, \beta^\nu \pm \frac{m}{2}\, \partial^\nu\, \beta = 0, 
\]
\[
\partial_\mu \, (\partial^\mu\, \bar \beta^\nu - \partial^\nu\, \bar \beta^\mu) + \partial^\nu\, B_2 
-\, \frac{m^2}{2}\, \bar \beta^\nu \pm \frac{m}{2}\, \partial^\nu\, \bar \beta = 0, 
\]
\[
\partial_\mu\, \bar B^{\mu\nu} - \partial^{\nu} \, B_1 \mp \frac{m}{2}\, \bar B^\nu = 0, ~~~~~~~~~~~~~~~~~~~~~~~~~~~~~
\eqno (B.5)
\]
it is straightforward to prove that all the terms inside the integral (B.3)
produce {\it zero} result. Hence, we note that, in (B.2), all the terms containing
the auxiliary fields $(B, \, B_1, \, B_2)$ and/or derivatives on them contribute
to zero as far as the explicit computation of $s_b\, Q_{ab}^{(1)}$  (i.e. $s_b\, Q_{ab}^{(1)} 
= -\, i\,\{ Q_{ab}^{(1)}, Q_{b}^{(1)} \} = 0$) is concerned.

We focus now on all the terms that do {\it not} contain any kind 
(i.e. bosonic/fermionic) of the (anti-)ghost fields. These terms of the {\it non-ghost} sector of (B.2)  are: 
\[
\int d^{D-1}\,x \Big[
\big(\partial^0\, \bar B^{ij} + \partial^i\, \bar B^{j0} + \partial^j\, \bar B^{0i}\big)\, B_{ij} 
- \big(\partial^0\, B^{ij} + \partial^i\,  B^{j0} + \partial^j\, B^{0i}\big)\, \bar B_{ij}
\]
\[
\pm \, m\, B^{0i}\, \bar B_i
 - \, \big(\partial^0\, B^i - \partial^i\, B^0\big)\, \bar B_i
 \mp \, m\, \bar B^{0i}\, B_i
 + \, \big(\partial^0\,\bar  B^i - \partial^i\,\bar B^0\big)\,  B_i
\Big].
\eqno (B.6)
\]
The above terms can be re-written as follows
\[
\int d^{D-1}\,x \Big[
\partial^0\, \big[B^{ij} + \bar B^{ij} - \big(\partial^i\, \phi^j - \partial^j\, \phi^i\big)\big]B_{ij} 
\]
\[
+ \partial^i\, \big[B^{j0} + \bar B^{j0} - \big(\partial^j\, \phi^0 - \partial^0\, \phi^j\big)\big]B_{ij} ~~~~~~~~~~
\]
\[
+ \partial^j\, \big[B^{0i} + \bar B^{0i} - \big(\partial^0\, \phi^i - \partial^i\, \phi^0\big)\big]B_{ij} ~~~~~~~~~~
\]
\[
~~~~~~~~~~~~~~~- \big(\partial^0\, B^{ij} + \partial^i\,  B^{j0} + \partial^j\, B^{0i}\big)\,
\big[B_{ij} + \bar B_{ij} - \big(\partial_i \, \phi_j - \partial_j\, \phi_i\big)\big]
\]
\[
~~~~~~~~~~~~~~~~~~~~~~~~~+ \partial^0\, \big[B^i + \bar B^i \mp \, m\, \phi^i + \partial^i\, \phi\big]\, B_i
- \partial^i\, \big[B^0 + \bar B^0 \mp \, m\, \phi^0 + \partial^0\, \phi\big]\, B_i 
\]
\[
\mp\, m\, \big[B^{0i} + \bar B^{0i} - \big(\partial^0\, \phi^i - \partial^i\, \phi^0\big)\big]\, B_{i} ~~~~~~~~~~~~
\]
\[
-\,  \big(\partial^0\, B^i - \partial^i\, B^0\big)\,\big[B_i + \bar B_i \mp \, m\, \phi_i + \partial_i\, \phi\big] ~~~~
\]
\[
\pm\, m\, B^{0i}\, \big[B_i + \bar B_i \mp \, m\, \phi_i + \partial_i\, \phi\big]
\Big],~~~~~~~~~~~~~~
\eqno (B.7)
\]
where we have exploited the Gauss divergence theorem at appropriate 
places and used the following EL-EoMs that are derived from ${\cal L}_B$, namely; 
\[
\partial_\mu\, H^{\mu\nu\lambda\xi} + m^2\, A^{\nu\lambda\xi}
\mp\, m\, \Sigma^{\nu\lambda\xi}
+ (\partial^\nu\, B^{\lambda\xi} 
+ \partial^\lambda \,  B^{\xi\nu} + \partial^\xi\, B^{\nu\lambda}) = 0, 
\]
\[
\partial_\mu\, \Sigma^{\mu\nu\lambda}  \mp m\, (\partial_\mu\, A^{\mu\nu\lambda}) - \frac{1}{2}\, (\partial^\nu\,  B^\lambda - \partial^\lambda\, B^\nu)
\mp  \, \frac{m}{2}\, B^{\nu\lambda}  = 0,~~~~~~~
\eqno (B.8)
\]
where $\Sigma_{\mu\nu\lambda} = \partial_\mu\, \Phi_{\nu\lambda} + 
\partial_\nu\, \Phi_{\lambda\mu} + \partial_\lambda\, \Phi_{\mu\nu}$
is a totally antisymmetric tensor defined in terms of the sum of the cyclic derivatives [cf. Eq. (3)] on the St$\ddot u$ckelberg field $\Phi_{\mu\nu}$
(which is {\it also} antisymmetric: $\Phi_{\mu\nu} = -\, \Phi_{\nu\mu}$). 
Ultimately, we note that the {\it non-ghost} sector of the terms in (B.2)
(i.e. $s_{b}\, Q_{ab}^{(1)}$) lead to the existence of the {\it bosonic} CF-type restrictions: 
$B_{\mu\nu} + \bar B_{\mu\nu} - (\partial_\mu \, \phi_\nu - \partial_\nu\, \phi_\mu) = 0$ and  
$B_\mu + \bar B_\mu \mp m\, \phi_\mu + \partial_\mu\, \phi = 0$ in equation (B.7) if we demand that the 
{\it non-ghost} sector of (B.2) {\it must} be equal to zero on its own so that the {\it first} entry of (B.1) could be satisfied 
(i.e. $s_{b}\, Q_{ab}^{(1)} = -\, i\,\{ Q_{ab}^{(1)}, Q_{b}^{(1)} \} = 0$).

We now concentrate on terms in (B.2) which contain bosonic/fermionic (anti-)ghost 
fields. These terms in the integral (B.2) are as follows: 
\[
\int d^{D-1}\,x \Big[
\big(\partial^0\, \bar C^{ij} + \partial^i\, \bar C^{j0} 
+ \partial^j\, \bar C^{0i}\big)\, \big(\partial_i\, \bar F_j - \partial_j\, \bar F_i\big)
\]
\[
 -  \big(\partial^0\, C^{ij} + \partial^i\,  C^{j0} + \partial^j\, C^{0i}\big)\, \big(\partial_i\, F_j - \partial_j\, F_i\big) 
\]
\[
+ \big[\pm m\, \bar C^{0i} - \big(\partial^0\, \bar C^i - \partial^i\, \bar C^0\big)\big]\, (\partial_i\, f)\,
\]
\[
+ \big[\pm m\,  C^{0i} - \big(\partial^0\, C^i - \partial^i \,  C^0\big)\big]\, (\partial_i\, F)
\]
\[
+ (\pm m\, \bar F - \partial^0\, \bar F)\, F - (\partial^0\, \bar F^i - \partial^i\, \bar F^0)\, \bar f_i 
\]
\[
- 2\, (\partial^0\, \bar F^i - \partial^i\, \bar F^0)\, F_i ~~~~~~~~~~~~~~~~~~~~~~~~
\]
\[
\mp \, m\, \big[\pm m\, C^{0i} - \big(\partial^0\, C^i - \partial^i\, C^0\big)\big]\, F_i
\]
\[
\mp \, m\, \big[\pm m\, \bar C^{0i} - \big(\partial^0\, \bar C^i - \partial^i\, \bar C^0\big)\big]\, f_i ~~
\]
\[
\pm \,m\, [\partial^0\, (f^i + \bar F^i)] - \partial^i\, (f^0 + \bar F^0)]\, \bar C_i
\]
\[
- \frac{1}{2}\, (\pm m\, f^0 - \partial^0\, f)\, \bar f
- (\partial^0 \, F^i - \partial^i \, F^0)\, \bar F_i
\]
\[
+ \frac{1}{2}\, (\pm m\, F^0 - \partial^0\, F)\, \bar F
 \Big].~~~~~~~~~~~~~~~~~~~
\eqno (B.9)
\]
We can apply the Gauss divergence theorem on the first {\it four} terms and drop the total space derivative terms. This theoretical operation leads to the following explicit form (for the first four terms), namely; 
\[
\int d^{D-1}\,x \Big[
-\,  2\, \partial_i\,   \big(\partial^0\, C^{ij} + \partial^i\,  C^{j0} + \partial^j\, C^{0i}\big)\, F_j 
\]
\[
+ 2\, \partial_i\, \big(\partial^0\, \bar C^{ij} + \partial^i\, \bar C^{j0} 
+ \partial^j\, \bar C^{0i}\big)\,\bar F_j
\]
\[
- \partial_i\, \big[\pm m\, \bar C^{0i} - \big(\partial^0\, \bar C^i - 
 \partial^i\, \bar C^0\big)\big]\,  f
 \]
 \[
- \partial_i\, \big[\pm m\,  C^{0i} - \big(\partial^0\, C^i - \partial^i \,  C^0\big)\big]\, F
\Big].
\eqno (B.10)
\]
Using the following EL-EoMs from ${\cal L}_B$, namely; 
\[
\partial_\mu\, \big [\partial^\mu\, \bar C^{\nu\lambda}  + 
\partial^\nu\, \bar C^{\lambda\mu} + \partial^\lambda\, \bar C^{\mu\nu}\big]
\pm \frac{m}{2}\, (\partial^\nu\, \bar C^\lambda - \partial^\lambda\, \bar C^\nu)
\]
\[
+ \frac{1}{2}\, (\partial^\nu \, F^\lambda - \partial^\lambda \, F^\nu)
-\, \frac{m^2}{2}\, \bar C^{\nu\lambda} = 0, ~~~~~~~~~~~~~~~~~~~~
\]
\[
\partial_\mu\, \big [\partial^\mu\, C^{\nu\lambda}  + 
\partial^\nu\,  C^{\lambda\mu} + \partial^\lambda\,  C^{\mu\nu}\big]
\pm \frac{m}{2}\, (\partial^\nu\,  C^\lambda - \partial^\lambda\, C^\nu)
\]
\[
+ \frac{1}{2}\, (\partial^\nu \, f^\lambda - \partial^\lambda \, f^\nu)
-\, \frac{m^2}{2}\,  C^{\nu\lambda} = 0,~~~~~~~~~~~~~~~~~~~~~~~~
\]
\[
\partial_\mu\, [\partial^\mu\, \bar C^\nu - \partial^\nu \, \bar C^\mu]
-\, \frac{1}{2}\, \partial^\nu \, F \mp \, m\, (\partial_\mu\, \bar C^{\mu\nu})
\pm \frac{m}{2}\, F^\nu = 0, 
\]
\[
\partial_\mu\, [\partial^\mu\, C^\nu - \partial^\nu \,  C^\mu]
-\, \frac{1}{2}\, \partial^\nu \, f \mp \, m\, (\partial_\mu\, C^{\mu\nu})
\pm \frac{m}{2}\, f^\nu = 0, 
\eqno (B.11)
\]
we observe that (B.10) reduces to the following:
\[
\int d^{D-1}\,x \Big[
 \pm m\, \big[\pm m\,  C^{0i} - \big(\partial^0\,  C^i - \partial^i\,  C^0\big)\big]\, F_i
 -\, \frac{1}{2}\, (\pm m\, \bar f^0 - \partial^0\, \bar f)
 \]
\[
 + \frac{1}{2}\, (\pm\, m\, f^0 - \partial^0\, f)\, F
\mp\, m\,  \big[\pm m\, \bar C^{0i} - \big(\partial^0\, \bar C^i - \partial^i\, \bar C^0\big)\big]\,  \bar F_i 
\]
\[
+ (\partial^0\, F^i - \partial^i\, F^0)\, \bar F_i + (\partial^0\, \bar F^i - \partial^i\, \bar F^0)\, F_i. ~~~~~~~~~~~~~~~~~~~~~~~~~
\eqno (B.12)
\]
 It is interesting to point out that the {\it first} term of the above equation 
 cancels out with the first term in the {\it fourth} row of (B.9). 
 A careful observation of {\it all} the terms of (B.12) and (B.9)
 as well as use of the Gauss divergence theorem at appropriate places along 
 with the utility of EL-EoMS (B.11), ultimately, lead to the 
following explicit expression for the ghost-sector of (B.2), namely;
\[
\int d^{D-1}\,x \Big[
\pm\, m\, \big[ \partial^0\, \big(f^i + \bar F^i - \partial^i\, C_1 \big) 
- \partial^i\, \big(f^0 + \bar F^0 - \partial^0\, C_1 \big) \big]\, f_i
\]
\[
\mp m\, \big[ \pm m\, \bar C^{0i} - \big(\partial^0\, \bar C^i - \partial^i\, \bar C^0\big) \big]\,  \big(f_i + \bar F_i - \partial_i\, C_1 \big) ~~~~~~~~~~~~~~~~
\]
\[
~~~~~~~~~~~~-\, \big(\partial^0\, \bar F^i - \partial^i\, \bar F^0\big)\, 
\big(\bar f_i +  F_i - \partial_i\, \bar C_1 \big) 
-\, \frac{1}{2}\, (\pm m\,  f^0 - \partial^0\,  f) \, (\bar f + F \mp\, m \, \bar C_1)
\]
\[
+ \big[ \pm m\, \big(f^0 + \bar F^0 - \partial^0\, C_1 \big) - 
\partial^0\, (f + \bar F \mp\, m \,  C_1)   \big] \, F~~~~~~~~~~~~~~~
\]
\[
+ \frac{1}{2}\, \big[\pm m\, \big(\bar f^0 +  F^0 - \partial^0\, \bar C_1 \big) 
- \, \partial^0\, (\bar f + F \mp\, m \, \bar C_1)   \big] \, \bar F ~~~~~~~~~~~
\]
\[
-\, \frac{1}{2}\, (\pm m\,  \bar f^0 - \partial^0\,  \bar f) \, (f + \bar F \mp\, m \, C_1)
\Big].~~~~~~~~~~~~~~~~~~~~~~~~~~~~~~~
\eqno (B.13)
\]
A close look at the above equation demonstrate that it can be zero if we exploit the validity of the {\it four} fermionic CF-type restrictions:
$f_\mu + \bar F_\mu -\, \partial_\mu\, C_1 = 0, \, 
\bar f_\mu + F_\mu -\, \partial_\mu\, \bar C_1 = 0, \, 
f + \bar F \mp\, m\, C_1 = 0, \, $ and $\bar f +  F \mp\, m\,\bar C_1 = 0$. 
In other words, if we demand the absolute anticommutativity of the nilpotent 
(anti-)BRST charges [$Q_{(a)b}^{(1)}$] from the condition: 
$s_b\, Q_{ab}^{(1)} = 0$, we observe that {\it all} the {\it four} 
set of fermionic CF-type restrictions emerge out.

Before we end this Appendix, we briefly comment on the appearance of the 
CF-type restrictions in the application  of $s_{ab}\, Q_b^{(1)}$. 
For this purpose, we now apply the anti-BRST symmetry transformations (9) on the 
off-shell nilpotent version of the BRST charge [cf. Eq. (53)] so that the explicit form of the 
l.h.s. of the second entry of (B.1) is as follows: 
\[
s_{ab}\, Q_{b}^{(1)} = \int d^{D-1}\,x \Big[
\big(\partial^0\, \bar B^{ij} + \partial^i\, \bar B^{j0} + \partial^j\, \bar B^{0i}\big)\, B_{ij} 
- \big(\partial^0\, B^{ij} + \partial^i\,  B^{j0} 
\]
\[
+ \partial^j\, B^{0i}\big)\, \bar B_{ij}
 - \,  \big(\partial^0\, \bar C^{ij} + \partial^i\, \bar C^{j0} 
+ \partial^j\, \bar C^{0i}\big)\, \big(\partial_i\, \bar F_j - \partial_j\, \bar F_i\big) 
~~~~~~~~~~~~~~~~~~~
\]
\[
+   \big(\partial^0\, C^{ij} + \partial^i\,  C^{j0} + \partial^j\, C^{0i}\big)\, \big(\partial_i\, F_j - \partial_j\, F_i\big) ~~~~~~~~~~~~~~~~~~~~~~~~~~~~~~~~~~~~~
\]
\[
+ 2\, \dot B_2\, B - \dot B\, B_2 
\mp m\, \big[\partial^0 \, \{\bar f^i + F^i \} - \partial^i \, \{\bar f^0 + F^0 \}\big]\, C_i \pm m\, B^{0i}\, \bar B_i ~~~~~~~~
\]
\[
+ \, \big[\pm \, m\, \bar C^{0i} - \big(\partial^0 \,  \bar C^i - \partial^i\, \bar C^0\big)\big]\, \big(\pm m\, \bar F_i - \partial_i\, \bar F\big) 
-\, 2\, \big(\partial^0\, F^i - \partial^i\, F^0\big)\, \bar F_i 
\]
\[
+ 2\, \big(\partial^0\, \bar\beta^i - \partial^i \, \bar\beta^0\big)\, \partial_i\, B - B_1\, \dot B_1 
- \, \big[\pm\, m\, \bar B^{0i} - \big(\partial^0\, B^i - \partial^i\, B^0\big)\big]\, B_i~~~~~~ 
\]
\[
+  \big(\partial^0\, \beta^i - \partial^i \, \beta^0\big)\, \partial_i\, B_2
- \, \big[\pm \, m\,  C^{0i} - \big(\partial^0 \,   C^i - \partial^i\,  C^0\big)]\, \big(\pm m\, \bar f_i - \partial_i\, \bar f\big) 
\]
\[
~~~\mp \, \frac{m}{2}\, B^0\, B_1 + \frac{1}{2}\, \big(\pm \, m\, \bar F^0 - \partial^0\, \bar F\big)\, F \pm \frac{m}{2}\, \big(\pm \, m\, \beta^0 - \partial^0\, \beta\big)\, B_2 + B^{0i}\, \partial_i\, B_1
\]
\[
- \big(\partial^0\, F^i - \partial^i\, F^0\big)\, f_i + \big(\pm \, m\, 
 F^0 - \partial^0\,  F\big)\, \bar F - \frac{1}{2}\, \big(\pm \, m\, \bar f^0 - \partial^0\, \bar f\big)\, f ~~~~~~~
\]
\[
~~~~~ \mp \, m\, \big[\pm \, m\, \bar \beta^0 -  \partial^0 \, \bar \beta\big]\,B
 -\,  \big(\partial^0\, \bar F^i - \partial^i\, \bar F^0\big)\, F_i
 - \, \big(\partial^0\, B^i - \partial^i\, B^0\big)\, \bar B_i
\Big]. 
\eqno (B.14)
\]
To simplify the above expression, first of all, we select {\it all} the terms 
that contain the auxiliary fields $B, \, B_1$ and $B_2$. These are as follows:
\[
\int d^{D-1}\,x \Big[
 2\, \dot B_2\, B - \dot B\, B_2 - B_1\, \dot B_1   \pm \frac{m}{2}\, \big(\pm \, m\, \beta^0 - \partial^0\, \beta\big)\, B_2 
\]
\[
 \mp \, m\, \big[\pm \, m\, \bar \beta^0 -  \partial^0 \, \bar \beta\big]\,B 
\mp \, \frac{m}{2}\, B^0\, B_1
+ B^{0i}\, \partial_i\, B_1
\]
\[
+  \big(\partial^0\, \beta^i - \partial^i \, \beta^0\big)\, \partial_i\, B_2
+ 2\, \big(\partial^0\, \bar\beta^i - \partial^i \, \bar\beta^0\big)\, \partial_i\, B 
\Big].
\eqno (B.15)
\]
In the last {\it three} terms of the above equation, we can apply the Gauss divergence theorem 
and drop the total space derivative terms to obtain 
the following [from (B.15)], namely; 
\[
\int d^{D-1}\,x \Big[
 2\, \dot B_2\, B - \dot B\, B_2 - B_1\, \dot B_1  \mp \, \frac{m}{2}\, B^0\, B_1 \pm \frac{m}{2}\, \big(\pm \, m\, \beta^0 - \partial^0\, \beta\big)\, B_2 
\]
\[
\mp \, m\, \big[\pm \, m\, \bar \beta^0 -  \partial^0 \, \bar \beta\big]\,B
- \big(\partial_i\,B^{0i}\big)\,  B_1
-  \partial_i\,\big(\partial^0\, \beta^i - \partial^i \, \beta^0\big)\, B_2
\]
\[
- 2\, \partial_i\,\big(\partial^0\, \bar\beta^i - \partial^i \, \bar\beta^0\big)\, B 
\Big].~~~~~~~~~~~~~~~~~~~~~~~~~~~~~~~~~~~~~~~~~~~~~~~
\eqno (B.16)
\]
At this stage, we use the following EL-EoMs, from ${\cal L}_B$: 
\[
\partial_\mu\, (\partial^\mu\, \bar\beta^\nu - \partial^\nu\, \bar\beta^\mu) + \partial^\nu\, B_2 - \frac{m^2}{2}\, \bar \beta^\nu \pm \frac{m}{2}\, \partial^\nu\, \bar\beta = 0, 
\]
\[
\partial_\mu\, B^{\mu\nu} - \partial^\nu\, B_1 \mp \frac{m}{2}\, B^\nu = 0, ~~~~~~~~~~~~~~~~~~~~~~~~~~
\]
\[
\partial_\mu\, (\partial^\mu\, \beta^\nu - \partial^\nu\, \beta^\mu) - \partial^\nu\, B 
- \frac{m^2}{2}\,  \beta^\nu \pm \frac{m}{2}\, \partial^\nu\, \beta = 0.  
\eqno (B.17)
\]
The substitutions of these values into (B.16) shows that all the terms, 
containing auxiliary fields $B, \, B_1, \, B_2$, cancel out and the 
whole integral (B.16) turns out to be {\it zero}.

At this juncture, we now concentrate on the {\it non-ghost} sector  of the terms 
that are present in (B.14). A close look at these terms demonstrate that they are 
{\it same} as (B.6). Hence, the integral with these terms can be expressed as 
(B.7) provided we use the EL-EoMs (B.8). Thus, we shall have validity of the
two {\it bosonic} CF-type restrictions provided we assume that non-ghost sector contribute to {\it zero} in $s_{ab}\, Q_b^{(1)}$ [cf. Eq. (B.18)]. 
We concentrate now on the {\it ghost-sector} of the terms that are 
present in (B.14) which necessarily contain bosonic/fermionic (anti-)ghost
fields. These are:  
\[
\int d^{D-1}\,x \Big[
 \big(\partial^0\, C^{ij} + \partial^i\,  C^{j0} + \partial^j\, C^{0i}\big)\, \big(\partial_i\, F_j - \partial_j\, F_i\big) ~~~~~~~~~~~~~~~~~~~
\]
\[
- \,  \big(\partial^0\, \bar C^{ij} + \partial^i\, \bar C^{j0} 
+ \partial^j\, \bar C^{0i}\big)\, \big(\partial_i\, \bar F_j - \partial_j\, \bar F_i\big) 
~~~~~~~~~~~~~~~~~~~~~~~~~~~~~~~~~
\]
\[
\mp m\, \big[\partial^0 \, \{\bar f^i + F^i \} - \partial^i \, \{\bar f^0 + F^0 \}\big]\, C_i
 -\, \big(\partial^0\, \bar F^i - \partial^i\, \bar F^0\big)\, F_i ~~~~~~~~~~~~~~
\]
\[
~~~~+ \, \big[\pm \, m\, \bar C^{0i} - \big(\partial^0 \,  \bar C^i - \partial^i\, \bar C^0\big)\big]\, \big(\pm m\, \bar F_i - \partial_i\, \bar F\big) 
-\, 2\, \big(\partial^0\, \bar F^i - \partial^i\, \bar F^0\big)\, F_i 
\]
\[
~~~~~- \, \big[\pm \, m\,  C^{0i} - \big(\partial^0 \,   C^i - \partial^i\, C^0\big)\big]\, \big(\pm m\,  f_i - \partial_i\, f \big) 
+ \frac{1}{2}\, \big(\pm m\,  \bar F^0 - \partial^0\, \bar F\big)\, F 
\]
\[
-\,  \big(\partial^0\,  F^i - \partial^i\,  F^0\big)\, f_i 
+  \big(\pm m\,   F^0 - \partial^0\, F\big)\, \bar F 
-\, \frac{1}{2}\, \big(\pm m\,  \bar f^0 - \partial^0\, \bar f \big)\, f 
\Big].
\eqno (B.18)
\]
We apply (i) the Gauss divergence theorem at appropriate places, and (ii)
the appropriate EL-EoMs from ${\cal L}_B$ and ${\cal L}_{\bar B}$, respectively.
It turns out that the final form of the above integral 
(B.18) appears as follows:
\[
\int d^{D-1}\,x \Big[
\pm m \, \big[\pm \, m\, C^{0i} - \big(\partial^0 \,  C^i - \partial^i\, C^0\big)\big]\, (\bar f_i + F_i - \, \partial_i\,  \bar C_1) 
\]
\[
\mp \, m\, \big[ \partial^0\, \big(f^i + \bar F^i - \partial^i\, C_1 \big) 
- \partial^i\, \big(f^0 + \bar F^0 - \partial^0\, C_1 \big) \big]\, C_i ~~~~~~~~~~
\]
\[
~~~~~~~~~~~~~-\, \big(\partial^0\,  F^i - \partial^i\, F^0\big)\, 
\big( f_i + \bar F_i - \partial_i\, C_1 \big) 
+  \frac{1}{2}\, (\pm m\,  F^0 - \partial^0\,  F) \, (f + \bar F \mp\, m \, C_1)
\]
\[
+ \frac{1}{2}\,  \big[ \pm m\, \big(f^0 + \bar F^0 - \partial^0\, C_1 \big) - 
\partial^0\, (f + \bar F \mp\, m \, C_1)   \big] \, F~~~~~~~~~~~~~
\]
\[
-\, \frac{1}{2}\, (\pm m\,  f^0 - \partial^0\,  f) \, (\bar f + F \mp\, m \, \bar C_1) ~~~~~~~~~~~~~~~~~~~~~~~~~~~~~~~~~~~~
\]
\[
-\, \frac{1}{2}\, \big[ \pm m\,  \big(\bar f^0 + F^0 - \, \partial^0\, \bar C_1 \big) -  \partial^0\,(\bar f + F \mp\, m \, \bar C_1)\big]\, f
\Big].~~~~~~~~~~
\eqno (B.19)
\]
Hence, the contributions of the terms in the {\it ghost-sector}
of (B.14) will be equal to {\it zero} provided all the {\it four} 
fermionic CF-type restrictions: 
$f_\mu + \bar F_\mu -\, \partial_\mu\, C_1 = 0, \, 
\bar f_\mu + F_\mu -\, \partial_\mu\, \bar C_1 = 0, \, 
f + \bar F \mp\, m\, C_1 = 0, \, $ and $\bar f +  F \mp\, m\,\bar C_1 = 0$ are satisfied.

We end this Appendix with the final remark that the existence of the 
CF-type restrictions is the hallmark of a BRST-quantized theory. 
We have shown, in our present endeavor, the existence of a set of {\it six} 
CF-type restrictions (i) from the EL-EoMs [cf. Eq. (29)], 
(ii) from the equality of the Lagrangian density ${\cal L}_{B}$
and ${\cal L}_{\bar B}$ [cf. Appendix A], and (iii) from the requirement of absolute 
anticommutativity of the nilpotent (anti-)BRST charges.\\

\vskip 1.5cm

\begin{center}
{\bf Appendix C: On the Standard BRST Algebra}\\
\end{center}

\vskip 0.5cm
\noindent
We have seen that the standard Noether conserved (anti-)BRST charges [i.e. $Q_{(a)b}]$ are {\it not} nilpotent of  order two
(i.e. $Q_{(a)b}^2 \neq 0$). Hence, they do not participate in the construction of the {\it standard} BRST algebra where the 
nilpotency of the (anti-)BRST charges is one of the key ingredients. In fact, the nilpotency property plays a decisive role 
in the discussion of the BRST cohomology (see, e.g. [27] for details) where the {\it original} state and the 
{\it gauge} transformed state {\it both} are cohomologically equivalent w.r.t. the nilpotent (anti-)BRST charges. Thus, 
in the construction and derivations of the standard BRST algebra, the off-shell nilpotent versions of the 
(anti-) BRST  charges $[Q_{(a)b}^{(1)}]$ (that have been derived in an explicit forms in (62) and (53), respectively) play a crucial role. 
In the process of the derivation of the standard BRST algebra, we take into consideration the celebrated relationship 
between the continuous symmetry transformations and their generators as the conserved charges
[cf. Eq. (49)]. In other words, we note the 
following explicit relationships, namely;
\[
s_b\, Q_b^{(1)}\,\;  = \;  -\, i\, \{ Q_b^{(1)}, \, Q_b^{(1)} \} \; = \; 0 \quad \Longrightarrow \quad [Q_b^{(1)}]^2 = 0, ~~~~~~~~~~~~~~~~~~
\]
\[
s_{ab}\, Q_{ab}^{(1)}\,\;  = \;  -\, i\, \{ Q_{ab}^{(1)}, \, Q_{ab}^{(1)} \} \; = \; 0 \quad \Longrightarrow \quad [Q_{ab}^{(1)}]^2 = 0,~~~~~~~~~~~~~~~~~
\]
\[
s_b\, Q_{ab}^{(1)}\,\;  = \;  -\, i\, \{ Q_{ab}^{(1)}, \, Q_b^{(1)} \} \; = \; 0 \quad \Longrightarrow 
\quad  Q_{ab}^{(1)}\, Q_b^{(1)} + Q_b^{(1)}\, Q_{ab}^{(1)} = 0, 
\]
\[
s_{ab}\, Q_b^{(1)}\,\;  = \;  -\, i\, \{ Q_b^{(1)}, \, Q_{ab}^{(1)} \} \; = \; 0
 \quad \Longrightarrow \quad Q_b^{(1)}\, Q_{ab}^{(1)} + Q_{ab}^{(1)}\, Q_b^{(1)}  = 0,
\]
\[
s_g\, Q_b \,\;  = \;  -\, i\, [ Q_b, \, Q_{g}] \; = \; + \, Q_b, 
\qquad \quad s_g\, Q_g \,\;  = \;  -\, i\, [ Q_g, \, Q_{g}] \; = \; 0, 
\]
\[
s_g\, Q_{ab} \,\;  = \;  -\, i\, [ Q_{ab}, \, Q_{g}] \; = \; - \, Q_{ab}, ~~~~~~~~~~~~~~~~~~~~~~~~~~~~~~~~~~~~~~~~~
\eqno (C.1)
\]
where we have taken into account the general mathematical relationship (49).

We conclude this Appendix with the following remarks. First of all, we note that the absolute 
anticommutativity property between the (anti-)BRST charges is satisfied on a submanifold of the quantum Hilbert space 
of fields where the full set of all the {\it six} CF-type restrictions are satisfied [cf. Appendix B]. 
Second, the (anti-)BRST charges $Q_{(a)b}^{(1)}$ are off-shell nilpotent of order two and,  hence, they are 
fermionic in nature and they generate continuous symmetry transformations where the fermionic fields transform into 
bosonic fields and vice-versa. Third, the off-shell nilpotent (anti-)BRST charges are {\it not} like the $ \mathcal{N } = 2$
SUSY off-shell nilpotent supercharges because the {\it latter} set of charges do {\it not} absolutely anticommute. 
Fourth, the nilpotency property plays a crucial role in the discussion on the BRST cohomology (see, e.g. [27] for details). Fifth, we note that the 
ghost number of the BRST charge is $(+1)$ and that of the anti-BRST charge is $(-1)$, respectively. The ghost number
of the ghost charge is {\it zero} as is clear from the algebra (C.1). 
Finally, we obtain, in a nut-shell, the standard BRST algebra with the help of the nilpotent 
versions of the (anti-)BRST charges [$Q_{(a)b}^{(1)}$] and the ghost charge $Q_g$ as follows:  
\[
[Q_{(a)b}^{(1)}]^2 \; = \; 0, ~~~~~~~~~
\]
\[   i\, [Q_g, \, Q_b^{(1)}] = +\;  Q_b^{(1)}, 
\]
\[   i\, [Q_g, \, Q_{ab}^{(1)}] = -\;   Q_{ab}^{(1)},
\]
\[
~~~~~~~~~~~~~~~\{ Q_b^{(1)}, \, Q_{ab}^{(1)} \}  \equiv  \{ Q_{ab}^{(1)}, \, Q_{b}^{(1)} \} \;  = \; 0, 
\eqno (C.2)
\]
where the last entry (i.e. absolute anticommutativity property) is satisfied only on the submanifold of the 
Hilbert space of quantum fields where all the {\it six} CF-type restrictions are valid (see, e.g. Appendix B for details). \\

\vskip 0.3cm

\begin{center}
{\bf Appendix D: On the Glossary of Fields}
\end{center}

\noindent
In this Appendix, we provide the list of {\it all} the fields that are present in the coupled (but\\

\newpage

\noindent
 equivalent) BRST and anti-BRST invariant Lagrangian densities ${\cal L}_B$ and ${\cal L}_{\bar B}$, respectively [cf. Eqs. (5),(6)]. We lay emphasis on their bosonic/fermionic, basic/auxiliary, independent/restricted, tensor/vector/scalar, etc., nature and mention their ghost number(s), too.
We list them in the following tabulated form: 

\vskip 0.5cm

\begin{table}[h!]
\centering
\begin{tabular}{ |p{1.2cm}|p{1.9cm}|p{1.99cm}|p{2.5cm}|p{1.2cm}|p{4.2cm}| }
\hline
 \textbf{Fields} & \textbf{Bosonic/ Fermionic} & \textbf{Basic/Au- xiliary}
 & \textbf{Tensorial Nature} & \textbf{Ghost Number} & \textbf{Independent/Restr- icted} \\
 \hline
 $A_{\mu\nu\lambda}$  & Bosonic  & Basic & Totally Antisymmetric & 0 & Independent  \\
 $C_{\mu\nu} $ & Fermionic & Basic & Antisymmetric & +1 &  Independent \\
 $\bar C_{\mu\nu} $  & Fermionic & Basic & Antisymmetric & - 1 &  Independent\\
 $B_{\mu\nu}$ & Bosonic & Auxiliary & Antisymmetric & 0 & Restricted  (D.1) \\
 $\bar B_{\mu\nu}$  & Bosonic & Auxiliary & Antisymmetric & 0 & Restricted  (D.1)  \\
 $\phi_{\mu\nu} $ & Bosonic  &  Basic & Antisymmetric & 0 & Independent  \\
 $\beta_\mu$ & Bosonic & Basic & Vector & +2 & Independent \\
 $\bar \beta_\mu$ & Bosonic & Basic & Vector & - 2 & Independent \\
 $\beta$ & Bosonic & Basic & Scalar & +2 & Independent \\
 $\bar \beta$ & Bosonic & Basic & Scalar & - 2 & Independent \\
 $C_\mu$  & Fermionic  &  Basic & Vector & +1 & Independent  \\
 $\bar C_\mu$  & Fermionic  &  Basic & Vector & - 1 & Independent  \\
 $C_1$  & Fermionic  &  Basic & Scalar & +1 & Restricted (D.2), (D.5) \\
$\bar C_1$  & Fermionic  &  Basic & Scalar & - 1 & Restricted (D.3), (D.6)  \\
 $C_2$  & Fermionic  &  Basic & Scalar & +3 & Independent  \\
 $\bar C_2$  & Fermionic  &  Basic & Scalar & - 3 & Independent  \\
 $\phi_\mu$ & Bosonic  &  Basic & Vector & 0 & Restricted (D.4), (D.1)\\
 $\phi$ & Bosonic  &  Basic & Scalar & 0 & Restricted (D.4) \\
 $B_\mu$  & Bosonic  &  Auxiliary & Vector & 0 & Restricted (D.4)  \\
 $\bar B_\mu$ & Bosonic  &  Auxiliary  & Vector & 0 & Restricted (D.4)\\
 $B$ & Bosonic  &  Auxiliary & Scalar & +2 & Independent  \\
 $B_1$  & Bosonic  &  Auxiliary & Scalar & 0 & Independent  \\
 $B_2$  & Bosonic  &  Auxiliary & Scalar & - 2 & Independent  \\ 
 $f_\mu$ & Fermionic & Auxiliary & Vector & +1 & Restricted (D.2) \\
 $\bar f_\mu$  &  Fermionic & Auxiliary & Vector & - 1 & Restricted (D.3) \\
 $F_\mu$ &  Fermionic & Auxiliary & Vector & - 1 & Restricted (D.3) \\
 $\bar F_\mu$ & Fermionic & Auxiliary & Vector & +1 & Restricted (D.2) \\
 $f$  &  Fermionic & Auxiliary & Scalar & +1 & Restricted (D.5) \\
 $\bar f$ &  Fermionic & Auxiliary & Scalar & - 1 & Restricted (D.6)\\
 $F$ &  Fermionic & Auxiliary & Scalar & - 1 & Restricted (D.6) \\
 $\bar F$  &  Fermionic & Auxiliary & Scalar & +1 & Restricted (D.5)  \\
 \hline 
\end{tabular}
\caption{Tower of fields and their specifications}
\end{table}
In the above, the the equation numbers from $(D.1)$ to $(D.6)$ correspond to the 
celebrated (anti-)BRST invariant CF-type restrictions [cf. Eq. (29)]  that are
present on our theory. These restrictions, in their explicit form,  are as follows:
\[
B_{\mu\nu} + \bar B_{\mu\nu} = \partial_\mu \phi_\nu - \partial_\nu \phi_\mu,
 ~~~~~~~~~~~~~~~~~~~~~~~~~~~~~~
\eqno (D.1)
\]
\[
f_\mu +  \bar F_\mu = \partial_\mu C_1,
 ~~~~~~~~~~~~~~~~~~~~~~~~~~~~~~
\eqno (D.2)
\]
\[
\bar f_\mu +  F_\mu = \partial_\mu \bar C_1,
 ~~~~~~~~~~~~~~~~~~~~~~~~~~~~~~
\eqno (D.3)
\]
\[
B_\mu + \bar B_\mu = \pm m\, \phi_\mu - \partial_\mu \, \phi,
 ~~~~~~~~~~~~~~~~~~~~~~~~~~~~~~
\eqno (D.4)
\]
\[
f + \bar F = \pm m\, C_1,
 ~~~~~~~~~~~~~~~~~~~~~~~~~~~~~~
\eqno (D.5)
\]
\[
\bar f + F = \pm \, m\, \bar C_1.
 ~~~~~~~~~~~~~~~~~~~~~~~~~~~~~~
\eqno (D.6)
\]
The above equations, demonstrate that there are many fields in the theory which are {\it not} independent because they
are restricted by the CF-type restrictions from $(D.1)$ to $(D.6)$. \\

\vskip 1.5cm

\begin{center}
{\bf Appendix E: On the Off-Shell Nilpotency and  Absolute Anticommutativity of the 
(Anti-)BRST Symmetries and Importance of the CF-Type Restrictions}\\
\end{center}

\vskip 0.5cm

\noindent
We have already noted that the absolute anticommutativity property of the (anti-)BRST symmetry transformations [cf. Eq. (11)] crucially
depends on the validity of the CF-type restrictions [cf. Eq. (29)]. This anticommutativity property 
between the BRST and anti-BRST transformations is very important because it
distinguishes the nilpotent (anti-)BRST symmetry transformations from the nilpotent $\mathcal{N} = 2 $ SUSY
transformations which do {\it not} anticommute (with each-other). The purpose of our present Appendix is to show that,
{\it even} at the level of the (anti-)BRST transformed fields, the CF-type restrictions are valid in the sense that the off-shell
nilpotency property of the (anti-)BRST symmetry transformations, at this stage, crucially depends upon them. 
We take here 
an example from the BRST  symmetry transformations [cf. Eq. (7)]  to corroborate our claim. Let us focus on 
the BRST transformation: $s_b \bar C_{\mu\nu} = B_{\mu\nu}$ which is nilpotent of order two (i.e. $s_b^2 = 0 $) because we observe that
$s_b^2 \bar C_{\mu\nu} = s_b B_{\mu\nu} = 0 $ due to the fact that $s_b B_{\mu\nu} = 0$ [cf. Eq. (7)].  We would like to point out that
the r.h.s. of the transformation: $s_b \bar C_{\mu\nu} = B_{\mu\nu}$ is {\it not} independent in the sense that it is restricted by the
CF-type restriction: $B_{\mu\nu} + \bar  B_{\mu\nu} = \partial_\mu \phi_\nu - \partial_\nu \phi_\mu$. Taking recourse 
to {\it this} CF-type restriction, we can re-express the {\it above} BRST transformation as:
\[
s_b \bar C_{\mu\nu} = - \, \bar B_{\mu\nu} +(\partial_\mu \phi_\nu - \partial_\nu \phi_\mu).
 ~~~~~~~~~~~~~~~~~~~~~~~~~~~~~~
\eqno (E.1)
\]
The question we address is whether the off-shell nilpotency is satisfied at this stage or not. Interestingly, we find that
the off-shell nilpotency is maintained because we observe that  
\[
s^2_b \bar C_{\mu\nu} = s_b \big [- \, \bar B_{\mu\nu} +(\partial_\mu \phi_\nu - \partial_\nu \phi_\mu) \big] = 0,
 ~~~~~~~~~~~~~~~~~~~~~~~~~~~~~~
\eqno (E.2)
\]
because of the fact that we have: $s_b \bar B_{\mu\nu} = \partial_\mu f_\nu -  \partial_\nu f_\mu$
and $s_b \phi_\mu = f_\mu $ [cf. Eq. (7)]. Thus, it is very clear that the off-shell nilpotency (i.e. $s_{(a)b}^2 = 0 $)
and absolute anticommutaivity [cf. Eq. (11)] properties
of the (anti-)BRST symmetry transformations and the 
(anti-)BRST invariant CF-type restrictions are intertwined together in a very meaningful fashion.

To complete our discussion, let us take an example from the anti-BRST symmetry transformations (9) 
and focus on the application
of the anti-BRST symmetry transformation on a bosonic field. For instance, let us choose 
$s_{ab} \phi_\mu = \bar f_\mu $. It is 
clear that the off-shell nilpotency (i.e. $s_{ab}^2 = 0 $) is {\it true}, at this stage, 
because we observe that $s_{ab} \bar f_\mu  = 0 $
[cf. Eq. (9)]. However, we note that the r.h.s. of this transformation is {\it not} 
independent because $\bar f_\mu$ is restricted
by the CF-type restriction: $ \bar f_\mu + F_\mu = \partial_\mu \bar C_1$. As 
a consequence, we can replace the r.h.s.
of  $s_{ab} \phi_\mu = \bar f_\mu $ (due to the above CF-type restriction) as follows:

\[
s_{ab} \phi_\mu = - \, F_\mu + \partial_\mu \bar C_1.
 ~~~~~~~~~~~~~~~~~~~~~~~~~~~~~~
\eqno (E.3)
\]
At this stage, we would like to check whether the off-shell nilpotency $s_{ab}^2 = 0 $ is maintained or not
after the application of the CF-type restriction. It turns out, interestingly,
that the off-shell nilpotency (i.e. $s_{ab}^2 \phi_\mu = 0 $) is maintained 
due to the fact that we have the following anti-BRST transformations [cf. Eq. (9)]:
\[
s_{ab} F_\mu = - \,  \partial_\mu B_2, \qquad s_{ab} \bar C_1 = -\, B_2\quad  \Longrightarrow
\quad s_{ab}^2 \phi_\mu =  s_{ab} \big [ - \, F_\mu + \partial_\mu \bar C_1] = 0.
 ~~~~~~~~~~~~~~~~~~~~~~~~~~~~~~
\eqno (E.4)
\]
Thus, we conclude that the nilpotency is maintained {\it even} at the level of the anti-BRST transformed fields (provided we exploit the
appropriate variety of the CF-type restriction).

We end this Appendix with the following concluding remarks. First of all, we note that the absolute anticommutativity 
property (i.e. $\{s_b, \; s_{ab} \} = 0 $) of the (anti-)BRST symmetry transformations [cf. Eq. (11)] crucially depends on
the existence of the CF-type restrictions [cf. Eq. (29)] on our theory. Second, we note that the straightforward equality of 
the BRST and anti-BRST invariant Lagrangian densities ${\cal L}_B$ and  ${\cal L}_{\bar B}$ leads
to the derivation of the CF-type restrictions (cf. Appendix A). In other words,  Lagrangian densities ${\cal L}_B$ and  ${\cal L}_{\bar B}$
are equivalent on the submanifold of the fields where the CF-type restrictions are satisfied. Third, at the level of the
(anti-)BRST transformed fields, it turns out that the uses of the appropriate CF-type restrictions are responsible for the
off-shell nilpotency of the (anti-)BRST symmetry transformations.  Finally, it is worthwhile 
to point out that we have chosen only {\it two} examples to show 
the sanctity of the CF-type restrictions at the level of the (anti-)BRST transformed fields. However,
this observation is {\it true} for {\it other} similar examples that are present in the (anti-)BRST transformations 
[cf. Eqs. (7),(9)], too.\\


\vskip 0.9cm









\begin{thebibliography}{99}
\bibitem{RPM1}      M. B. Green, J. H. Schwarz, E. Witten, {\it Superstring Theory}, Vols. 1 and 2,\\
                    Cambridge University Press, Cambridge (1987)
\bibitem{RPM2}      J. Polchinski, {\it String Theory}, Vols. 1 and 2,\\
                    Cambridge University Press, Cambridge (1998)   
\bibitem{RPM3}      D. Lust, S. Theisen, {\it Lectures in String Theory}, Springer-Verlag, New York (1989) 
\bibitem{RPM4}      K. Becker,  M. Becker, J. H. Schwarz, {\it String Theory and M-Theory},\\ 
                    Cambridge University Press, Cambridge (2007)   
\bibitem{RPM5}      D. Rickles, {\it A Brief History of String Theory From Dual Models to M-Theory},\\ Springer, Germany  (2014)                
\bibitem{RPM6}      R. Kumar, S. Krishna, A. Shukla, R. P. Malik,\\
                    {\it Int. J. Mod. Phys.} A {\bf 29}, 1450135 (2014) (a brief review)
\bibitem{RPM7}     S. Krishna, R. Kumar, R. P. Malik, {\it Annals of Physics} {\bf 414}, 168087 (2020)
\bibitem{RPM8}     B. Chauhan, S. Kumar, A. Tripathi, R. P. Malik, \\ {\it Advances in High Energy Physics} {\bf 2020}, 3495168 (2020)

\bibitem{RPM9}     T. Eguchi, P. B. Gilkey, A. Hanson, {\it Physics Reports} {\bf 66}, 213 (1980)    
\bibitem{RPM10}     S. Mukhi, N. Mukanda, {\it Introduction to Topology, Differential Geometry and Group Theory for Physicists}, 
                    Wiley Eastern Private Limited, New Delhi (1990)   
\bibitem{RPM11}     J. W. van Holten, {\it Phys. Rev. Lett.} {\bf 64}, 2863 (1990)   
\bibitem{RPM12}     S. Deser, A. Gomberoff, M. Henneaux, C. Teitelboim, {\it Phys. Lett.} B {\bf 400}, 80 (1997)   
\bibitem{RPM13}     P. J. Steinhardt, N. Turok, {\it Science} {\bf 296}, 1436 (2002)    
\bibitem{RPM14}     J. L. Lehners, {\it Physics Reports} {\bf 465}, 223 (2008)  
\bibitem{RPM15}     M. Novello, S. E. P. Bergliaffa, {\it Physics Reports} {\bf 463}, 127 (2008) 
\bibitem{RPM16}     V. M. Zhuravlev, D. A. Kornilov, E. P. Savelova,\\
                    {\it General Relativity and Gravitation} {\bf 36}, 1736 (2004)
\bibitem{RPM17}     Y. Aharonov, S. Popescu, D. Rohrlich, L. Vaidman, \\ {\it Physical Review} A {\bf 48}, 40844090 (1993)
\bibitem{RPM18}     A. K. Rao, R. P. Malik, {\it  Nucl. Phys.} B, {\bf 983},  115926 (2022) 
\bibitem{RPM19}     G. Curci, R. Ferrari, {\it Phys. Lett.} B {\bf 63}, 91 (1976)  
\bibitem{RPM20}     L. Bonora, R. P. Malik, {\it Phys. Lett.} B {\bf 655}, 75 (2007)
\bibitem{RPM21}     L. Bonora, R. P. Malik, {\it J. Phys.} A: {\it Math. Theor.} {\bf 43}, 375403 (2010) 
\bibitem{RPM22}     A. K. Rao, A. Tripathi, B. Chauhan, R. P. Malik, {\it Universe} {\bf 8},  566 (2022)
\bibitem{RPM23}     P. A. M. Dirac, {\it Lectures on Quantum Mechanics} (Belfer Graduate
                    School of Science), Yeshiva University Press, New York (1964)
\bibitem{RPM24}     K. Sundermeyer, {\it Constraint Dynamics, Lecture notes in Physics}, \\Vol. 169, 
                    Springer-Verlag, Berlin (1982)
\bibitem{RPM25}     M.  Henneaux,  C. Teitelboim, {\it Quantization of Gauge System}, \\Princeton University, New Jersey (1992)
\bibitem{RPM26}     D. M. Gitman, I. V. Tyutin, {\it Quantization of Fields with Constraints},\\
                    Springer-Verlag, Berlin Heidelberg (1990)                     
\bibitem{RPM27}     S. Weinberg, {\it The Quantum Theory of Fields, Volume 2: 
                      Modern Applications}, \\Cambridge University Press, Cambridge (1996)
\bibitem{RPM28}     P. Mitra, R. Rajaraman, {\it Annals of Physics} {\bf 203}, 137 (1990)
\bibitem{RPM29}     P. Mitra,  R. Rajaraman, {\it Annals of Physics} {\bf 203}, 157  (1990)
\bibitem{RPM30}    N. Nakanishi, I. Ojima,  {\it Covariant Operator Formalism of Gauge Theories 
                    and Quantum Gravity},  World Scientific, Singapore (1996) 
\bibitem{RPM31}     K. Nishijima, {\it Czechoslovak Journal of Physics} {\bf 46}, 140 (1996)
\bibitem{RPM32}     B. Chauhan,  A. K. Rao,  R. P.  Malik,  {\it Nucl. Phys. } B  {\bf 996}, 116366  (2023)
\bibitem{RPM33}     S. Upadhyay, B. P. Mandal, {\it  Euro. Phys. J.} C {\bf 72},  2059 (2012) 
\bibitem{RPM34}     S. Upadhyay, M. K. Dwivedi, B. P. Mandal, {\it Int. J. Mod. Phys.} A {\bf 28}, 1350033 (2013)
\bibitem{RPM35}     M. Faizal,   {\it Phys. Rev.} D {\bf 84},   106011 (2011)




\end{thebibliography}
\end{document}